\documentclass[a4paper,11pt]{article}
\usepackage{jheppub}
\usepackage{lineno}
\usepackage{graphicx}
\usepackage{tikz-feynman}
\usepackage{tikz-cd}
\usepackage{slashed}
\usepackage{multirow}
\usepackage{array}
\usepackage{xcolor}
\usepackage{float}
\usepackage{placeins}
\usepackage{bbding,amssymb,pifont}
\usepackage[justification=centering,singlelinecheck=false]{subcaption}
\usepackage[colorinlistoftodos]{todonotes}
\usepackage{hyperref}
\nolinenumbers
\title{Mono-Z signal for dark matter searches at future lepton colliders and cosmological interpretation}
\author[a]{Anupam Ghosh,}
\author[b]{Partha Kumar Paul,}
\author[a,c]{Abhik Sarkar,}
\author[d]{Rituparna Ghosh,}
\author[e,f,g]{Rachit Sharma,}
\author[a]{and, Subhaditya Bhattacharya}

\affiliation[a]{Department of Physics, Indian Institute of Technology Guwahati, North Guwahati, 781039, India.}
\affiliation[b]{Department of Physics, Indian Institute of Technology Hyderabad, Kandi, Telangana-502285, India.}
\affiliation[c]{Laboratoire de Physique Subatomique et de Cosmologie, Université Grenoble-Alpes, CNRS/IN2P3, 53 Avenue des Martyrs, 38026 Grenoble, France}
\affiliation[d]{Indian Institute of Technology Kanpur, Kalyanpur, Kanpur 208016, Uttar Pradesh, India.}
\affiliation[e]{Indian Institute of Science Education and Research Thiruvananthapuram, Vithura, Kerala 695 551, India.}
\affiliation[f]{School of Physics and Institute for Collider Particle Physics, University of the Witwatersrand, Wits, 2050, Johannesburg, South Africa.}
\affiliation[g]{SGTB Khalsa College, University of Delhi, Delhi, India-110007.}

\emailAdd{anupamg@rnd.iitg.ac.in}
\emailAdd{ph22resch11012@iith.ac.in}
\emailAdd{asarkar@lpsc.in2p3.fr}
\emailAdd{rituparnaoffc@gmail.com}
\emailAdd{rachit21@iisertvm.ac.in}
\emailAdd{subhab@iitg.ac.in}

\abstract{We investigate mono-$Z$ signature at the future electron-positron collider within the dark matter effective field theory (DMEFT) framework, considering operators up to dimension six for scalar, fermionic, and vector dark matter (DM) candidates. The operators are classified according to their underlying production mechanisms. We identify the phenomenologically viable parameter space by imposing the observed DM relic abundance together with direct and indirect detection constraints, Higgs invisible decay limits, and LHC mono-$Z$ bounds obtained through a recasting analysis. We study the discovery prospects of the surviving DM operators at the future $e^-e^+$ collider including initial and final state radiation, detector simulation, and the reconstruction of both leptonic and hadronic $Z$-boson decay modes. We exploit missing energy, missing transverse momentum, recoil mass, $Z$-boson pseudorapidity, together with beam polarization, which play a significant role to distinguish the signal from the Standard Model background. We find that fermionic dipole, scalar leptophilic, vector Higgs-portal, and gauge-portal DM operators offer robust mono-$Z$ discovery potential, probing multi-TeV effective scales.}
\begin{document}
\makeatletter
\gdef\@fpheader{}
\makeatother
\maketitle
\flushbottom
\section{Introduction}
\label{sec:intro}
The existence of dark matter (DM)~\cite{Zwicky:1933gu,Rubin:1970zza,Clowe:2006eq,Planck:2018vyg} remains one of the most compelling mysteries in modern physics. Although the Standard Model (SM) has achieved remarkable success in describing the interactions of elementary particles and accurately predicting a wide range of phenomena across current and future experiments, it fails to provide a viable particle candidate that satisfies the necessary properties of the DM. Consequently, the existence of DM constitutes one of the strongest pieces of evidence for physics beyond the SM (BSM), making its identification a central objective of contemporary particle physics. Although all confirmed evidence for DM is gravitational in origin~\cite{Zwicky:1933gu,Rubin:1970zza,Clowe:2006eq}, extensive experimental efforts have been undertaken to uncover its non-gravitational interactions through complementary search strategies. These include direct detection (DD) experiments~\cite{LZ:2024zvo,XENON:2025vwd,PandaX:2024qfu,DARWIN:2016hyl}, indirect detection (ID) experiments~\cite{Fermi-LAT:2009ihh,Fermi-LAT:2015att,Fermi-LAT:2025gei,MAGIC:2014zas,MAGIC:2021mog,HESS:2014zqa,HESS:2018kom}, and collider searches at the Large Hadron Collider (LHC)~\cite{Abdallah:2015ter,Abercrombie:2015wmb,Arina:2025zpi}. Direct-detection experiments probe the scattering of Galactic DM particles with nuclei, whereas indirect-detection searches constrain DM annihilation or decay through photons, charged particles, and neutrinos produced in astrophysical environments. A viable DM candidate must not only evade existing experimental constraints, but also reproduce the observed cosmological relic abundance~\cite{Planck:2018vyg},
\begin{equation}
	\Omega_{\rm DM}h^2 \approx 0.12 \pm 0.0012\,,
\end{equation}
where $\Omega=\frac{\rho}{\rho_c}$ represents the cosmological density, and $h$ represents the reduced Hubble constant. 

The DM relic and its detectability depend crucially on the DM-SM interaction, which can be represented by operators of the form $\mathcal{O}_{\rm DM-SM}\sim\mathcal{O}_{\rm DM}.\mathcal{O}_{\rm SM}$, where $\mathcal{O}_{\rm DM}$ represents operators consisting of DM particles, and $\mathcal{O}_{\rm SM}$ represents operators consisting of SM particles. In the absence of a preferred ultraviolet (UV) completion, dark matter effective field theory (DMEFT) offers a systematic and model-independent framework for describing interactions between the SM and the dark sector in terms of gauge-invariant higher-dimensional operators ($\mathcal{O}_{\rm DM-SM}$) suppressed by appropriate powers of effective scale $\Lambda$, where heavy mediators are assumed to be integrated out. The DMEFT approach is especially useful for comparing different experimental probes within a common parameter space, specified by the DM mass and the scale of the effective interaction. The complete set of operator bases for DMEFT up to a given mass dimension has been developed in the literature~\cite{Duch:2014xda,Criado:2021trs,Song:2023jqm}, and the corresponding effective operators can be matched onto a wide variety of UV-complete models.

Owing to the DM-SM interaction envisaged by $\mathcal{O}_{\rm DM-SM}$, collider experiments offer a search strategy to uncovering the particle nature of DM. At high-energy colliders, DM particles, or even heavier dark-sector states, can be produced directly from SM initial states. Since DM particles escape the detector without depositing energy, collider searches rely on the recoil of the invisible system against visible radiation, leading to characteristic mono-$X$~\cite{Belyaev:2016lok,ATLAS:2021kxv,CMS:2021far,ATLAS:2018bvc,ATLAS:2020uiq,ATLAS:2021shl,CMS:2025monophoton,Bhattacharya:2025wef} signatures, as well as final states with energetic leptons~\cite{Ghosh:2024boo,CMS:2023wjr,Ghosh:2025dcv,Ghosh:2025agw,ATLAS:2023cjo} or boosted fat jets~\cite{Ghosh:2021noq,Ghosh:2022rta,Bhardwaj:2020llc,Ghosh:2023xhs,ATLAS:2021yqv, Ghosh:2024nkj}. The latter arise when DM particles are produced in decays of heavier dark-sector states, and experimental and phenomenological studies of such signatures at the LHC can be found in Refs.~\cite{Ghosh:2021noq,Ghosh:2022rta,Bhardwaj:2020llc,Ghosh:2023xhs,Ghosh:2024nkj,ATLAS:2020gxq,ATLAS:2021yqv}.
Despite the extensive LHC searches, the non-observation of DM signal has placed stringent constraints on a wide class of DM models and significantly restricted their viable parameter space. However, it must be stated that LHC primarily being a QCD machine, has not been able to constrain the Electroweak extension of the SM to a significant extent, which is particularly applicable to the class of $\mathcal{O}_{\rm DM-SM}$ operators that we consider here.

In this work, we investigate the mono-$Z$ signature in association with missing DM particles at the future International Linear Collider (ILC) operating at a center-of-mass (CM) energy of $\sqrt{s}=1$ TeV within the framework of the DMEFT. We consider the complete set of DMEFT operators up to dimension-6 that can give rise to this signature and assess their discovery prospects at the ILC.  Although different DMEFT operators lead to the same experimental final state, they predict different dependences on beam polarization and distinct distributions in missing energy ($\slashed{E}$), missing transverse momentum ($p_T^{\rm miss}$), recoil mass ($m_{\rm rec}$), and $Z$-boson pseudorapidity ($\eta_Z$), as detailed in the analysis. Therefore, we separately optimize for each of these operators for their collider analysis. Furthermore, lepton colliders offer an exceptionally clean environment for DM searches owing to the absence of overwhelming QCD backgrounds and the precise knowledge of the initial-state kinematics, which enables an accurate reconstruction of the missing four-momenta. Beam polarization plays an important role because it modifies the signal rate according to the Electroweak structure of the effective interaction and can strongly suppress the $t$-channel $W$-exchange SM background.   

Consequently, mono-$X$ searches have received considerable attention at future $e^+e^-$ colliders, with the mono-$\gamma$ channel being the most extensively studied~\cite{Fox:2011fx,Yu:2013aca,Essig:2013vha,Kadota:2014mea,Yu:2014ula,Freitas:2014jla,Dutta:2017ljq,Choudhury:2019sxt,Horigome:2021qof,Barman:2021hhg,Kundu:2021cmo,Bhattacharya:2022qck,Ge:2023wye,Ma:2022cto,Barman:2024nhr,Barman:2024tjt}. Among the various mono-$X$ channels, the mono-$Z$ signature offers several advantages over the mono-$\gamma$ channel for probing the Electroweak structure of DM interactions. While mono-$\gamma$ searches are highly sensitive to electromagnetic couplings and certain dipole-type interactions, they are largely blind to the $SU(2)_L$ structure of the underlying operator. The mono-$Z$ channel, on the other hand, directly probes the coupling of DM to the neutral Electroweak gauge boson and is sensitive to both hypercharge and weak-isospin components of the interaction. Moreover, the $Z$ boson can be reconstructed in both leptonic and hadronic decay modes, providing multiple handles to control backgrounds and to exploit kinematic observables such as the recoil mass and missing transverse momentum. This makes the mono-$Z$ channel a promising one for probing the nature of DM. This also enables us to connect the collider prone parameter space to its cosmological interpretation.

The remainder of this paper is organized as follows. In Sec.~\ref{sec:dmeft}, we introduce the DMEFT operators relevant for the mono-$Z$ signal and classify their production topologies. In Sec.~\ref{sec:dpheno}, we study the relic-density, direct-detection, and indirect-detection constraints. The limits from invisible Higgs decays are discussed in Sec.~\ref{sec:higgs}, while the LHC recasting analysis is presented in Sec.~\ref{sec:lhc-recast}. In Sec.~\ref{sec:cpheno}, we investigate the mono-$Z$ discovery prospects at a future lepton collider. Finally, we summarize our results and conclude in Sec.~\ref{sec:conc}.

\section{DMEFT Operators}
\label{sec:dmeft}
There is no DM in the SM. Therefore, to incorporate DM-SM interaction, one necessitates to assume a DM candidate in the form of either a
scalar, or fermion, or a vector boson particle, not present in the SM. 
In the absence of any established UV completion for DM, DMEFT provides a systematic and model-independent framework to parametrize the low-energy interactions between the SM and the dark sector. The effects of heavy mediator particles are encoded in a series of higher-dimensional operators constructed from SM and DM fields, with each operator weighted by its corresponding Wilson coefficient and suppressed by an appropriate power of the cutoff scale $\Lambda$. The DMEFT Lagrangian can be written as
\begin{equation}
    \mathcal{L}_{\rm DMEFT} = \mathcal{L}_{\rm SM} + \mathcal{L}_{\rm DM} + \sum_{i,d} \frac{C_i^{(d)}\mathcal{O}_i^{(d)}}{\Lambda^{d-4}},
\end{equation}
where $\mathcal{L}_{\rm SM}$ and $\mathcal{L}_{\rm DM}$ denote the SM and DM Lagrangians, respectively. Here, $\mathcal{O}_i^{(d)}$ represents a gauge-invariant operator of mass dimension $d$, $C_i^{(d)}$ is the corresponding Wilson coefficient encoding the strength of the interaction, and $\Lambda$ denotes the cutoff scale above which the effective description is expected to break down. Throughout this work, we fix all Wilson coefficients to unity, i.e., $C_i^{(d)}=1$\footnote{One can easily reparametrise the Wilson coefficient in terms of the appropriate rescaling of the NP scale $\Lambda$, as the collider production cross section or that of the DM relic density is directly proportional to $\frac{C_i^{(d)}}{\Lambda^{d-4}}$, subject to the effective operator validity, $\Lambda > \sqrt{s}$.}. Consequently, the cutoff scale $\Lambda$ and the DM mass, $m_{\rm DM}$, constitute the only free parameters of the effective theory. These parameters determine the phenomenology of the effective interactions and are treated as the parameters of interest throughout the subsequent analyses.

Throughout this work, we consider the minimal scenario in which the DM field is a singlet under the SM gauge group, while all SM fields are neutral under the dark-sector symmetry responsible for stabilizing the DM candidate. Under these assumptions, every DMEFT operator can be factorized into a product of an SM operator and a dark-sector operator,
\begin{equation}
    \mathcal{O}_i^{(d)} \sim \mathcal{O}_{\rm SM}\,\mathcal{O}_{\rm DM},
\end{equation}
where $\mathcal{O}_{\rm SM}$ is invariant under the SM gauge symmetry and $\mathcal{O}_{\rm DM}$ is invariant under the dark symmetry. This factorized structure provides a convenient classification of all possible interactions between the visible and dark sectors in a model-independent manner.

In this work, we focus on the subset of DMEFT operators that contribute to the mono-$Z$ signature at an $e^+e^-$ collider. The representative operators for bosonic DM are presented in table~\ref{tab:dim-l1}, where we consider both real scalar ($\phi$) and vector ($X_\mu$) DM candidates. The corresponding operators for fermionic DM, namely Dirac ($\chi$) and Majorana ($\psi$) fermions, are summarized in table~\ref{tab:dim-l2}. The final column in each table indicates the dominant production topology responsible for the mono-$Z$ signature, where I and F denote initial-state radiation (ISR) and final-state emission (FS), respectively. In these operators, $B_{\mu\nu}$ and $W_{\mu\nu}$ denote the $U(1)_Y$ and $SU(2)_L$ field-strength tensors, respectively, $\ell_L$ ($e_R$) represents the left-handed lepton doublet (right-handed charged-lepton singlet), $H$ is the SM Higgs doublet, and $D_\mu$ denotes the SM gauge-covariant derivative. Throughout this work, we perform a detailed phenomenological analysis only for the scalar and Dirac fermion DM scenarios. The vector and Majorana cases possess analogous operator structures and collider signatures to the scalar and Dirac cases, respectively, differing primarily in their Lorentz structures and numerical coefficients. Corresponding analyses follow in a straightforward manner and not presented separately\footnote{For vector and majorana DM collider production, see appendix~\ref{app:vector}.}. For Higgs-portal operators, we study both the scalar and vector DM cases.

\begin{table}[htb!]
\centering
\begin{minipage}[t]{0.475\textwidth}
\centering
\renewcommand{\arraystretch}{1.25}
\begin{tabular}{>{\centering\arraybackslash}p{0.75cm}
                >{\centering\arraybackslash}p{3cm}
                >{\centering\arraybackslash}p{2.5cm}}
\hline\hline
\multicolumn{3}{c}{\textbf{Scalar ($\phi$)}} \\
\hline\hline
$\mathcal{O}^{(6)}_{L\phi}$ & $(\overline{\ell}_{L}He_{R})\phi^{2}$ & $\phi\phi Z$ (I) \\
$\mathcal{O}^{(6)}_{W\phi}$ & $(W^{I}_{\mu\nu}W^{\mu\nu I})\phi^{2}$ & $\phi\phi Z$ (F) \\
$\mathcal{O}^{(6)}_{B\phi}$ & $(B_{\mu\nu}B^{\mu\nu})\phi^{2}$ & $\phi\phi Z$ (F) \\
$\mathcal{O}^{(6)}_{D\phi}$ & $(H^{\dagger}H)(\partial_{\mu}\phi\,\partial^{\mu}\phi)$ & $Zh(\phi\phi)$ \\
\hline\hline
\end{tabular}
\end{minipage}
\hfill
\begin{minipage}[t]{0.475\textwidth}
\centering
\renewcommand{\arraystretch}{1.25}
\begin{tabular}{>{\centering\arraybackslash}p{0.75cm}
                >{\centering\arraybackslash}p{3cm}
                >{\centering\arraybackslash}p{2.5cm}}
\hline\hline
\multicolumn{3}{c}{\textbf{Vector ($X$)}} \\
\hline\hline
$\mathcal{O}^{(6)}_{LX}$ & $(\overline{\ell}_{L}He_{R})X^{\mu}X_{\mu}$ & $XXZ$ (I) \\
$\mathcal{O}^{(6)}_{WX}$ & $(W^{I}_{\mu\nu}W^{\mu\nu I})X^{\mu}X_{\mu}$ & $XXZ$ (F) \\
$\mathcal{O}^{(6)}_{BX}$ & $(B_{\mu\nu}B^{\mu\nu})X^{\mu}X_{\mu}$ & $XXZ$ (F) \\
$\mathcal{O}^{(6)}_{DX}$ & $(H^{\dagger}H)(X_{\mu\nu}X^{\mu\nu})$ & $Zh(XX)$ \\
\hline\hline
\end{tabular}
\end{minipage}
\caption{Representative DMEFT operators corresponding to bosonic DM (scalar and vector) contributing to the mono-$Z$ signature at $e^{+}e^{-}$ colliders. The last column indicates the dominant production topology, where I and F denote initial-state radiation and final-state emission, respectively.}
\label{tab:dim-l1}
\end{table}

\begin{table}[htb!]
\centering
\begin{minipage}[t]{0.475\textwidth}
\centering
\renewcommand{\arraystretch}{1.25}
\begin{tabular}{>{\centering\arraybackslash}p{0.75cm}
                    >{\centering\arraybackslash}p{3cm}
                    >{\centering\arraybackslash}p{2.5cm}}
\hline \hline
\multicolumn{3}{c}{\textbf{Dirac Fermion ($\chi$)}} \\ \hline \hline
$\mathcal{O}^{(5)}_{B\chi}$ & $(\overline{\chi} \sigma_{\mu\nu}\chi) B^{\mu\nu}$ & $\chi\overline{\chi} Z$(I, F) \\ \hline
$\mathcal{O}^{(6)}_{\ell \chi}$ & $(\overline{\ell}_{L}\gamma^{\mu} \ell_{L}) (\overline{\chi}\gamma_{\mu}\chi)$ & $\chi\overline{\chi} Z$(I) \\ 
$\mathcal{O}^{(6)}_{e\chi}$ & $(\overline{e}_{R}\gamma^{\mu}e_{R}) (\overline{\chi}\gamma_{\mu}\chi)$ & $\chi\overline{\chi} Z$(I) \\ 
$\mathcal{O}^{(6)}_{D\chi}$ & $(H^{\dagger} i \overleftrightarrow{D_{\mu}} H) (\overline{\chi} \gamma^{\mu} \chi)$ & $\chi\overline{\chi} Z$(I, F) \\
\hline \hline
\end{tabular}
\end{minipage}
\hfill
\begin{minipage}[t]{0.475\textwidth}
\centering
\renewcommand{\arraystretch}{1.25}
\begin{tabular}{>{\centering\arraybackslash}p{0.75cm}
                >{\centering\arraybackslash}p{3cm}
                >{\centering\arraybackslash}p{2.5cm}}
\hline \hline
\multicolumn{3}{c}{\textbf{Majorana Fermion ($\psi$)}} \\ \hline \hline
$\mathcal{O}^{(5)}_{B\psi}$ & $(\overline{\psi} \sigma_{\mu\nu} \gamma^{5}\psi) B^{\mu\nu}$ & $\psi\overline{\psi} Z$(I, F) \\ \hline
$\mathcal{O}^{(6)}_{\ell \psi}$ & $(\overline{\ell}_{L}\gamma^{\mu} \ell_{L}) (\overline{\psi}\gamma_{\mu}\gamma^{5}\psi)$ & $\psi\overline{\psi} Z$(I) \\ 
$\mathcal{O}^{(6)}_{e\psi}$ & $(\overline{e}_{R}\gamma^{\mu}e_{R}) (\overline{\psi}\gamma_{\mu}\gamma^{5}\psi)$ & $\psi\overline{\psi} Z$(I) \\ 
$\mathcal{O}^{(6)}_{D\psi}$ & $(H^{\dagger} i \overleftrightarrow{D_{\mu}} H) (\overline{\psi} \gamma^{\mu}\gamma^{5} \psi)$ & $\psi\overline{\psi} Z$(I, F) \\
\hline \hline
\end{tabular}
\end{minipage}
\caption{Representative DMEFT operators for fermion DM (Dirac and Majorana type) contributing to the mono-$Z$ signature at $e^{+}e^{-}$ colliders.}
\label{tab:dim-l2}
\end{table}

\subsection{Mono-$Z$ Signature}
The characteristic collider signature considered in this work is the production of a $Z$ boson in association with a pair of invisible DM particles, $e^{+}e^{-}\rightarrow Z+\slashed{E}$,
where the missing energy ($\slashed{E}$) originates from the undetected DM pair. Missing energy is defined as ,
\begin{equation}
    \slashed{E}=\sqrt{s}-E_{\rm vis}\,,
\end{equation}
where $E_{\rm vis}$ accommodates all detected particles. Note here, that $\slashed{E}$ can only be defined for lepton colliders,
where the CM energy of the hard scattering ($\sqrt{s}$) is known. For hadron colliders, where the subprocess CM energy is not known, one needs to rely only on the missing transverse momentum ($p_T^{\rm miss}$), 
\begin{equation}
\vec{p}_T^{\,\rm miss}
\equiv
-\sum_{i\in{\rm visible}} \vec{p}_{T,i}\,,
\qquad
p_T^{\rm miss} \equiv \big|\vec{p}_T^{\,\rm miss}\big|\,,
\end{equation}
defined as the negative vector sum of the transverse momenta of all reconstructed visible objects in the plane perpendicular to the beam axis. 
We further note that missing energy 
carries the information of DM mass, which $p_T^{\rm miss}$ does not accommodate, for details, see~\cite{Bhattacharya:2022qck}.

Another powerful observable at an $e^+e^-$ collider is the recoil mass of the invisible system against the visible $Z$ decay products. For the process $e^{+}e^{-}\rightarrow Z+\slashed{E}$, the recoil mass is defined as
\begin{equation}
	m_{\rm rec}^2
	=
	\left(p_{\rm ini}-p_Z\right)^2
	=
	s+M_{12}^2-2\sqrt{s}\,E_Z,
	\label{eq:recoil-mass}
\end{equation}
where $p_{\rm ini}=(\sqrt{s},\mathbf{0})$ is the initial-state four-momentum, and $M_{12}$ and $E_Z$ are the invariant mass and energy of the reconstructed visible pair from the $Z$ decay, respectively. At the parton level, $m_{\rm rec}$ coincides with the invariant mass of the invisible system. In realistic event samples, initial-state radiation, beamstrahlung, detector resolution, and reconstruction effects smear $m_{\rm rec}$, broadening the peak and shifting the kinematic endpoint. Nevertheless, $m_{\rm rec}$ retains substantial discriminating power for a massive DM pair: the signal recoil spectrum is pushed to higher values and, in contrast, to the lighter invisible systems of the SM backgrounds.

The same experimental signature can also arise from SM processes in which a $Z$ boson is produced in association with neutrinos. Since neutrinos interact only weakly and escape the detector without depositing energy, they manifest as missing energy, thereby constituting an irreducible background to the mono-$Z$ signal. This, however, must be contrasted with the huge QCD background in the jets $+\;p_T^{\rm miss}$ channel at the LHC.  The dominant SM background to the final state jets $+\;p_T^{\rm miss}$ arises from $Z+\text{jets}$ and $W^\pm+\text{jets}$ production processes, where the $Z$ decays invisibly and the $W^\pm$ decays leptonically with the charged lepton being mistagged, thereby generating large $p_T^{\rm miss}$. The higher-order production cross sections for these processes are of order $\mathcal{O}(10^4\,\text{pb})$~\cite{Catani:2009sm, Balossini:2009sa}. Consequently, it is extremely challenging to distinguish a signal with a small cross section relative to such overwhelming backgrounds at the LHC. At an $e^{+}e^{-}$ collider, the clean experimental environment, together with the precisely known initial-state four-momentum, enables an accurate reconstruction of the event kinematics using the visible decay products of the $Z$ boson. Consequently, the mono-$Z$ channel provides a sensitive probe of DM interactions at future lepton colliders encoded here via the DMEFT framework. 

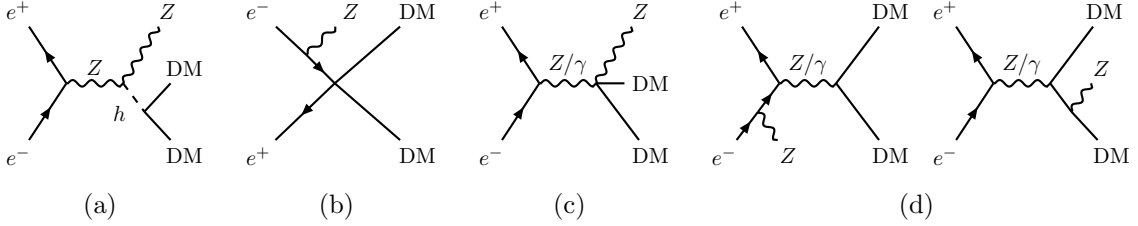
\begin{figure}[htb!]
    \centering
    \subfloat[]{\begin{tikzpicture}[baseline={(current bounding box.center)},style={scale=0.75, transform shape}]
	\begin{feynman}
		\vertex(a);
        \vertex[above left = 1cm and 0.5cm of a](a1){$e^{+}$};
        \vertex[below left = 1cm and 0.5cm of a](a2){$e^{-}$};
        \vertex[right = 1cm of a](b);
        \vertex[above right = 1cm and 0.5cm of b](b1){$Z$};
        \vertex[below right = 0.5cm and 0.375cm of b](b2);
        \vertex[below right = 0.5cm and 0.25cm of b2](b3){DM};
        \vertex[above right = 0.5cm and 0.25cm of b2](b4){DM};
        
		\diagram*{
			(a) -- [thick, fermion, arrow size=1pt] (a1),
            (a2) -- [thick, fermion, arrow size=1pt] (a),
            (a) -- [thick, boson, edge label = $Z$] (b),
            (b) -- [thick, boson] (b1),
            (b) -- [thick, scalar, edge label' = $h$] (b2),
            (b2) -- [thick, plain] (b3),
            (b2) -- [thick, plain] (b4),
		};
    \end{feynman}
    \end{tikzpicture}\label{fig:a}}
    \hspace{0.25cm}
    \subfloat[]{\begin{tikzpicture}[baseline={(current bounding box.center)},style={scale=0.75, transform shape}]
	\begin{feynman}
		\vertex(a);
		\vertex[above left = 0.5cm and 0.5cm of a] (a1);
        \vertex[above right =0.5cm and 0.5cm of a1] (a4){$Z$};
        \vertex[above left =0.5cm and 0.5cm of a1] (a3){$e^{-}$};
		\vertex[below left =1cm and 1cm of a] (a2){$e^{+}$};
		\vertex[above right = 1cm and 1cm of a] (b1) {DM};
		\vertex[below right = 1cm and 1cm of a] (b2){DM};
		\diagram*{
			(a) -- [thick, fermion, arrow size=1pt] (a2),
			(a1) -- [thick, fermion, arrow size=1pt] (a),
            (a3) -- [thick, plain, arrow size=1pt] (a1),
            (a1) -- [thick, boson, arrow size=1pt] (a4),
			(a) -- [thick, plain, arrow size=0.7pt] (b1),
			(b2) -- [thick, plain, arrow size=0.7pt] (a)
		};
	\end{feynman}
	\end{tikzpicture}\label{fig:b}}
    \hspace{0.25cm}
    \subfloat[]{\begin{tikzpicture}[baseline={(current bounding box.center)},style={scale=0.75, transform shape}]
	\begin{feynman}
		\vertex(a);
        \vertex[above left = 1cm and 0.5cm of a](a1){$e^{+}$};
        \vertex[below left = 1cm and 0.5cm of a](a2){$e^{-}$};
        \vertex[right = 1cm of a](b);
        \vertex[above right = 1cm and 0.5cm of b](b1){$Z$};
        \vertex[below right = 1cm and 0.5cm of b](b3){DM};
        \vertex[right = 0.5cm of b](b4){DM};
        
		\diagram*{
			(a) -- [thick, fermion, arrow size=1pt] (a1),
            (a2) -- [thick, fermion, arrow size=1pt] (a),
            (a) -- [thick, boson, edge label = $Z/\gamma$] (b),
            (b) -- [thick, boson] (b1),
            (b) -- [thick, plain] (b3),
            (b) -- [thick, plain] (b4),
		};
    \end{feynman}
    \end{tikzpicture}\label{fig:c}}
    \hspace{0.25cm}
    \subfloat[]{\begin{tikzpicture}[baseline={(current bounding box.center)},style={scale=0.75, transform shape}]
	\begin{feynman}
		\vertex(a);
        \vertex[above left = 1cm and 0.5cm of a](a1){$e^{+}$};
        \vertex[below left = 0.5cm and 0.375cm of a](a2);
        \vertex[below right = 0.5cm and 0.25cm of a2](a3){$Z$};
        \vertex[below left = 0.5cm and 0.25cm of a2](a4){$e^{-}$};
        \vertex[right = 1cm of a](b);
        \vertex[below right = 1cm and 0.5cm of b](b3){DM};
        \vertex[above right = 1cm and 0.5cm of b](b4){DM};
        
		\diagram*{
			(a) -- [thick, fermion, arrow size=1pt] (a1),
            (a2) -- [thick, fermion, arrow size=1pt] (a),
            (a4) -- [thick, fermion, arrow size=1pt] (a2),
            (a2) -- [thick, boson] (a3),
            (a) -- [thick, boson, edge label = $Z/\gamma$] (b),
            (b) -- [thick, plain] (b3),
            (b) -- [thick, plain] (b4),
		};
    \end{feynman}
    \end{tikzpicture}
	\begin{tikzpicture}[baseline={(current bounding box.center)},style={scale=0.75, transform shape}]
	\begin{feynman}
		\vertex(a);
        \vertex[above left = 1cm and 0.5cm of a](a1){$e^{+}$};
        \vertex[below left = 1cm and 0.5cm of a](a2){$e^{-}$};
        \vertex[right = 1cm of a](b);
        \vertex[below right = 0.5cm and 0.375cm of b](b1);
        \vertex[above right = 0.5cm and 0.25cm of b1](b2){$Z$};
        \vertex[below right = 0.5cm and 0.25cm of b1](b3){DM};
        \vertex[above right = 1cm and 0.5cm of b](b4){DM};
        
		\diagram*{
			(a) -- [thick, fermion, arrow size=1pt] (a1),
            (a2) -- [thick, fermion, arrow size=1pt] (a),
            (a) -- [thick, boson, edge label = $Z/\gamma$] (b),
            (b) -- [thick, plain] (b1),
            (b1) -- [thick, boson] (b2),
            (b1) -- [thick, plain] (b3),
            (b) -- [thick, plain] (b4),
		};
    \end{feynman}
    \end{tikzpicture}\label{fig:d}} 
    \caption{Feynman diagrams corresponding to mono-$Z$ signal at $e^{+}e^{-}$ colliders.}
    \label{fig:dm1}
\end{figure}

The DMEFT operators listed in Tabs.~\ref{tab:dim-l1} and \ref{tab:dim-l2} contribute to the mono-$Z$ final state through four distinct production topologies, illustrated in figure~\ref{fig:dm1}. The correspondence between these topologies and the relevant effective operators is summarized in table~\ref{tab:ops}. The first topology, diagram~(\ref{fig:a}), corresponds to associated Higgs production, $e^{+}e^{-}\rightarrow Zh$, followed by the invisible Higgs decay $h\rightarrow{\rm DM\,DM}$. This topology arises from the (derivative) Higgs-portal operators, respectively. The second topology, diagram~(\ref{fig:b}), originates from contact interactions between the initial-state leptons and the DM pair. In this case, the observed $Z$ boson is emitted from the incoming electron or positron, corresponding to initial-state radiation. This topology receives contributions from the leptophilic operators of all DM classes. In the third topology, diagram~(\ref{fig:c}), the virtual Electroweak gauge boson produced in the $s$-channel simultaneously couples to the outgoing $Z$ boson and the DM pair, resulting in $Z$ boson being emitted from the same outgoing vertex as the DMs. These interactions are generated by the gauge field-strength operators and may be interpreted as final-state emission of the $Z$ boson. Finally, the fourth topology, diagram~(\ref{fig:d}), contributes exclusively to fermionic DM. In this case, the $Z$ boson can be emitted either from the initial-state electron or from one of the final-state DM particles. Consequently, the dipole operators, together with the Higgs-current operators, receive contributions from both ISR and final-state emission diagrams.

\begin{table}[htb!]
    \centering
    \renewcommand{\arraystretch}{1.25}{
    \begin{tabular}{>{\centering\arraybackslash}p{3cm}
                    >{\centering\arraybackslash}p{2.5cm}
                    >{\centering\arraybackslash}p{2.5cm}
                    >{\centering\arraybackslash}p{2.5cm}
                    >{\centering\arraybackslash}p{2.5cm}}
    \hline \hline
    \multirow{2}*{Diagrams} & \multicolumn{4}{c}{Operators} \\ \cline{2-5}
    & Scalar ($\phi$) & Dirac ($\chi$) & Majorana ($\psi$) & Vector ($X$) \\ \hline \hline
    Diagram~\eqref{fig:a} & $\mathcal{O}^{(6)}_{D\phi}$ & $-$ & $-$ & $\mathcal{O}^{(6)}_{DX}$ \\ \hline
    Diagram~\eqref{fig:b} & $\mathcal{O}^{(6)}_{L\phi}$ & $\mathcal{O}^{(6)}_{\ell\chi}, \mathcal{O}^{(6)}_{e\chi}$ & $\mathcal{O}^{(6)}_{\ell\psi}, \mathcal{O}^{(6)}_{e\psi}$ & $\mathcal{O}^{(6)}_{LX}$ \\ \hline
    Diagram~\eqref{fig:c} & $\mathcal{O}^{(6)}_{W\phi}, \mathcal{O}^{(6)}_{B\phi}$ & $-$ & $-$ & $\mathcal{O}^{(6)}_{WX}, \mathcal{O}^{(6)}_{BX}$ \\ \hline
    Diagram~\eqref{fig:d} & $-$ & $\mathcal{O}^{(5)}_{B\chi}, \mathcal{O}^{(6)}_{D\chi}$ & $\mathcal{O}^{(5)}_{B\psi}, \mathcal{O}^{(6)}_{D\psi}$ & $-$ \\
    \hline \hline
    \end{tabular}}
    \caption{Feynman diagrams and corresponding operators responsible for those processes.}
    \label{tab:ops}
\end{table}

Although all of the above operators lead to the same experimental signature, a reconstructed $Z$ boson accompanied by missing energy, the different production mechanisms result in distinct kinematic characteristics. In particular, the distributions of missing energy, missing transverse momentum, and recoil mass differ significantly among the four topologies. These differences in kinematic distributions provide an effective means of discriminating between the various operator classes, as well as separating them from the SM backgrounds.

\section{Dark Matter Phenomenology}
\label{sec:dpheno}
Before turning to collider phenomenology, we first examine the constraints imposed on the operators listed in table \ref{tab:ops} set by cosmological observations. We evaluate relic density, direct detection, and indirect detection constraints for each type of DM candidate. Requiring consistency with these observations selects regions of the $\Lambda-m_{\rm DM}$ plane that are both theoretically motivated and experimentally viable, and we will use these regions as benchmarks for the collider study.

\subsection{Relic Density}
In this work, we assume the DM as Weakly Interacting Massive Particle (WIMP), which remains in thermal bath in the radiation dominated epoch, and freezes out later, as the Universe expands and cools down.
The thermal relic density of DM is set by the freeze-out of various interaction rates in the early universe. The cosmological evolution of the DM number density is governed by the following Boltzmann equation (BEQ), catering to DM depletion via $2 \to 2$ process,
\begin{equation}
\frac{dn_{\rm DM}}{dt}+3\mathcal{H}n_{\rm DM}=-\langle\sigma{v}\rangle\left(n_{\rm DM}^2-(n_{\rm DM}^{\rm eq})^2\right),
\end{equation}
where $n_{\rm DM}$ is the number density of DM and $n_{\rm DM}^{\rm eq}$ is its corresponding equilibrium value. $\mathcal{H}=1.66\sqrt{g_*}T^2/M_{\rm pl}$ is the Hubble parameter, $g_*$ is the number of relativistic degrees of freedom, $M_{\rm pl}=1.22\times10^{19}$ GeV is the Planck mass, $T$ defines the temperature of the thermal bath, and $\langle\sigma{v}\rangle$ is the thermally averaged annihilation cross section. We use \texttt{micrOMEGAs} \cite{Alguero:2023zol} to compute the DM relic density. For each DM candidate, we scan over the mass $m_{\rm DM}$ and the cutoff scale $\Lambda$, identifying the contours in this plane where the computed abundance matches the cosmic microwave background radiation observation by PLANCK ($\Omega_{\rm DM}h^2\sim0.12$) \cite{Planck:2018vyg}. 

\begin{figure}[htb!]
\centering
\includegraphics[width=0.48\textwidth]{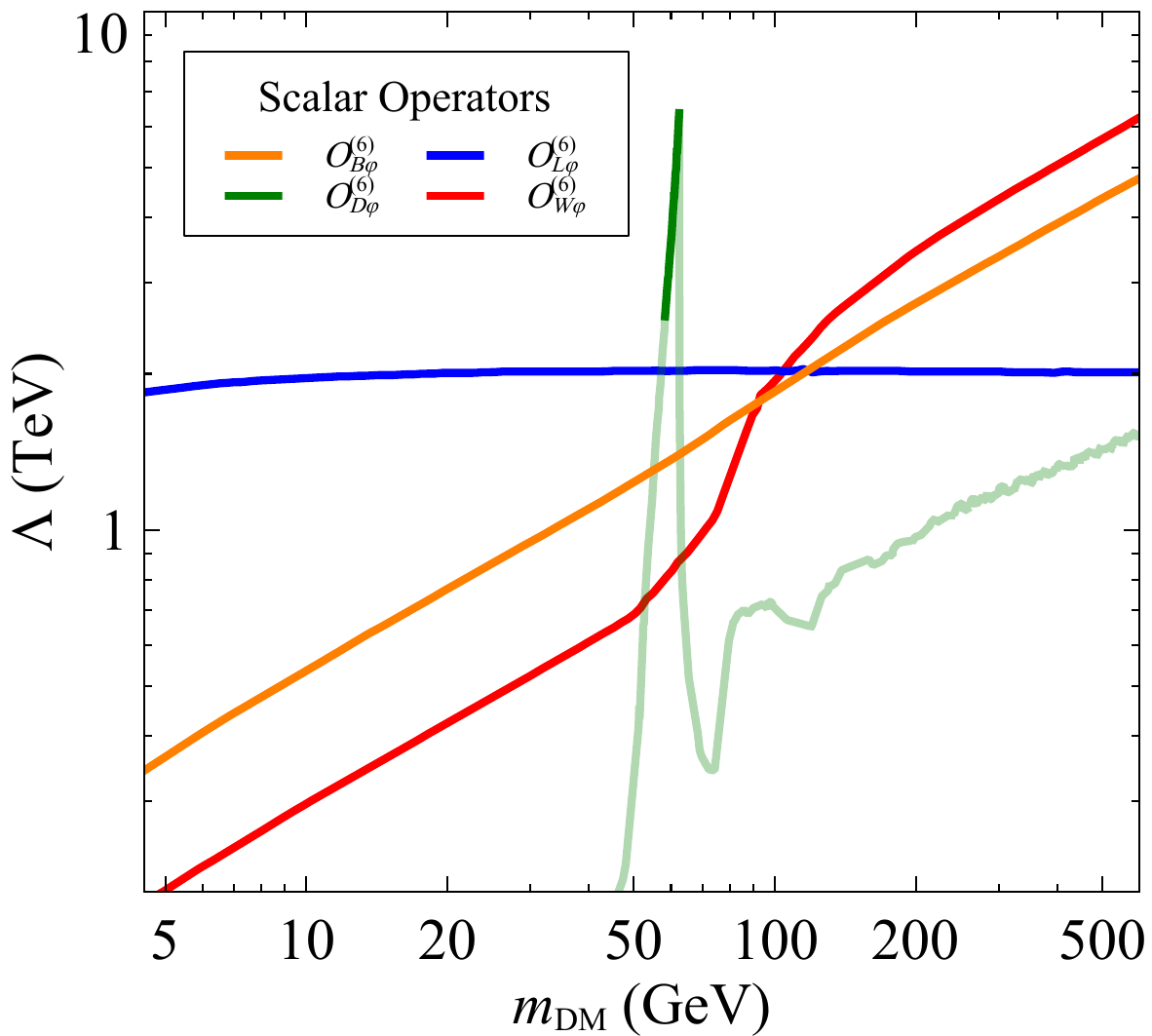}
\includegraphics[width=0.48\textwidth]{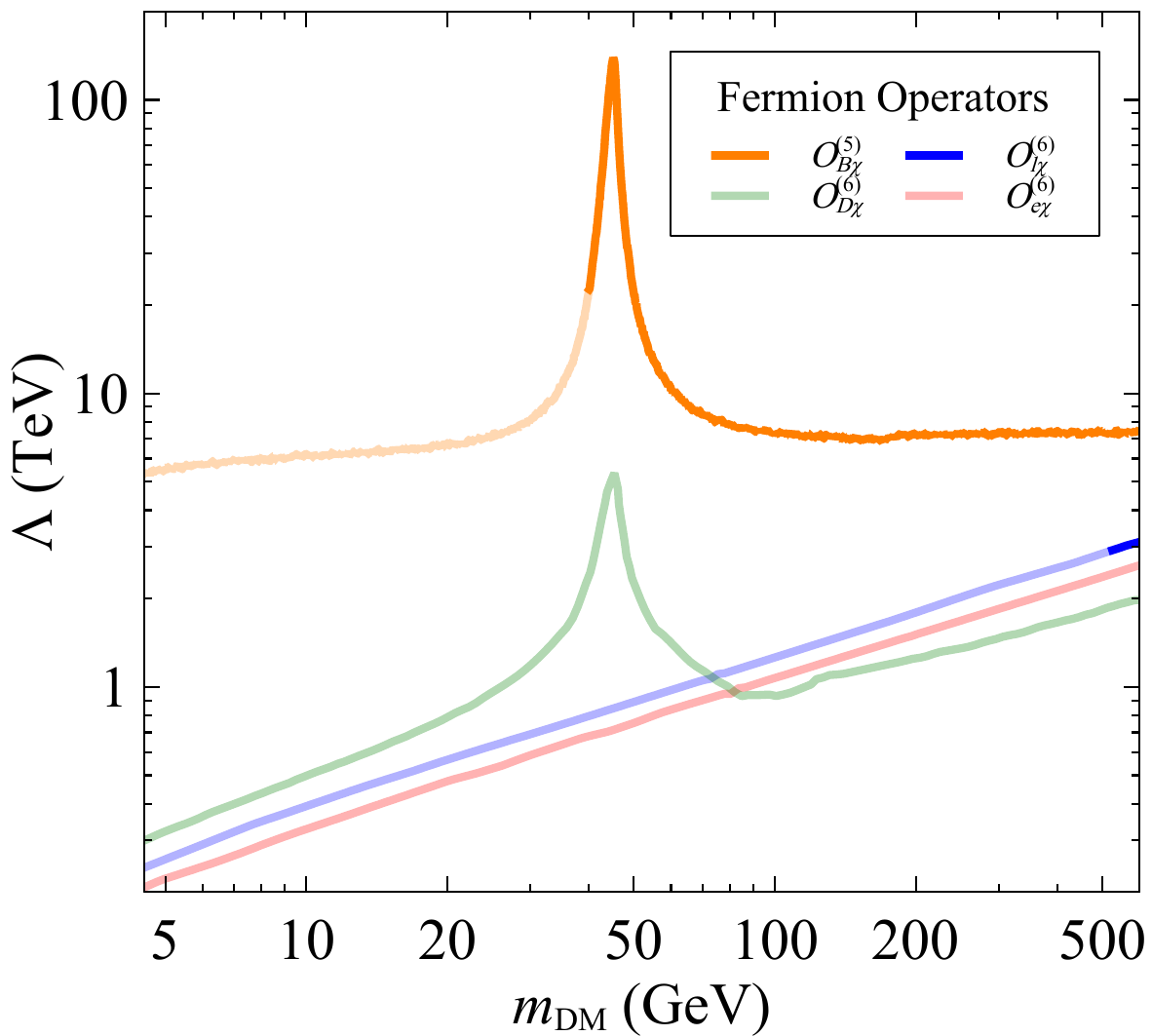}
\caption{The parameter space satisfying correct relic of DM and present direct detection limit from LZ \cite{LZ:2024zvo} are shown in the plane of $\Lambda$ vs $m_{\rm DM}$ for DMEFT operators as mentioned in the inset of the figures. The dark colored lines represent the allowed parameter space from relic and direct detection (LZ). The light colored lines correspond to the excluded parameter space from direct detection for the particular operator. The scalar operators are shown in \textit{left} panel and fermion operators are shown in \textit{right} panel.}
\label{fig:relic}
\end{figure}

In figure~\ref{fig:relic}, we present the correct DM relic parameter space in the plane of $\Lambda$ vs $m_{\rm DM}$ for scalar operators (\textit{left} panel) and Dirac fermion operators (\textit{right} panel) with different colored lines as mentioned in the inset of the figures. Now, let's briefly discuss each operator.
\begin{center}
\textbf{Scalar operators}
\end{center}
\begin{itemize}

\item \underline{\textbf{$\mathcal{O}^{(6)}_{B\phi}$:}} In this case, the scalar DM, $\phi$ has the following channel through which it can freeze-out: \{ $\phi\phi\rightarrow\gamma\gamma,~\gamma Z,~ ZZ$ \}. For $m_\phi<m_Z/2$, only the $\gamma\gamma$ channel remains kinematically open. The thermally averaged cross section is $\langle\sigma v\rangle_{\phi\phi\rightarrow\gamma\gamma}\sim8\cos\theta_w^4 m_\phi^2/\pi\Lambda^4$, where $\cos\theta_w$ is the cosine of the Weinberg angle. For a fixed $\Lambda$, as DM mass increases, the relic decreases. To obtain the correct relic, $\Lambda$ must be increased. This behavior is clearly visible in the \textit{left} panel of figure~\ref{fig:relic}, shown by the orange solid line for this operator. As the DM mass exceeds $m_Z/2$, the $\gamma Z$ channel opens up. And for $m_\phi>m_Z$, the $ZZ$ channel becomes accessible.

\item \underline{\textbf{$\mathcal{O}^{(6)}_{D\phi}$:}} For this operator, DM has the following channel through which it can freeze-out: \{ $\phi\phi\rightarrow hh,~ W^+W^-,~ ZZ,~ f\bar{f}$ \}. For the small DM masses, the relic is dominated by \textit{s}-channel $\phi\phi\rightarrow f\bar{f}$ process, mediated by Higgs. At the Higgs resonance, the scale $\Lambda$ increases significantly, thereby reducing the effective coupling. As the DM mass exceeds the $W$, $Z$, and $h$ masses, the $WW$, $ZZ$, and $hh$ annihilation channels open up, respectively, and contribute to the relic abundance. The correct relic contour is shown by a solid green line in the \textit{left} panel of figure~\ref{fig:relic}.

\item \underline{\textbf{$\mathcal{O}^{(6)}_{L\phi}$:}} The DM $\phi$ has the following channels through which it freezes out for this operator: \{ $\phi\phi\rightarrow e^+e^-,~ \mu^+\mu^-,~ \tau^+\tau^-$ \}. In the limit of $m_\phi\gg m_f$, the thermally averaged cross section is $\langle\sigma v\rangle\sim3v_h^2/2\pi\Lambda^4$, independent of DM mass. As a result, we see that the correct relic contour in the $\Lambda-m_{\rm DM}$ plane remains almost independent of $m_{\rm DM}$, as shown with the blue solid line in the \textit{left} panel of figure~\ref{fig:relic}.

\item \underline{\textbf{$\mathcal{O}^{(6)}_{W\phi}$:}} In the \textit{left} panel of figure~\ref{fig:relic}, the correct relic parameter space for the $\mathcal{O}^{(6)}_{W\phi}$ operator is shown with red solid line. The available channels are \{ $\phi\phi\rightarrow\gamma\gamma,~\gamma Z,~ ZZ,~ W^+W^-$ \}. For $m_\phi<m_Z/2$, $\phi\phi\rightarrow\gamma\gamma$ is the dominant channel which decides the relic. The thermally averaged cross section for this process is $\langle\sigma v\rangle_{\phi\phi\rightarrow\gamma\gamma}\sim8\sin\theta_w^4m_\phi^2/\pi\Lambda^4$. Therefore, as the DM mass increases, $\Lambda$ must also increase to maintain the required cross section and obtain the correct relic abundance. When the DM mass becomes $>m_W$, the $\phi\phi\rightarrow W^+W^-$ channel opens up. This cross section scales as $\langle\sigma v\rangle_{\phi\phi\rightarrow W^+W^-}\sim 16m_\phi^2/\pi\Lambda^4$. Here, too, $\Lambda$ increases with $m_\phi$ to obtain the correct relic abundance. Note that this $\sigma v$ is larger than that for the $\gamma\gamma$ channel and therefore requires a larger value of $\Lambda$. This behavior is clearly visible in the \textit{left} panel of figure~\ref{fig:relic}.
\end{itemize}

\begin{center}
\textbf{Dirac fermion operators}
\end{center}

\begin{itemize}
\item \underline{\textbf{$\mathcal{O}^{(5)}_{B\chi}$:}} In the \textit{right} panel of figure~\ref{fig:relic}, the correct relic contour for the $\mathcal{O}^{(5)}_{B\chi}$ operator is shown with orange solid line. Here the channels which contribute relic are \{$\chi\bar{\chi}\rightarrow f\bar{f},~ \gamma\gamma,~ \gamma Z,~ ZZ,~ W^+W^-,~ ZH$\}. The \textit{s}-channel processes are mediated by $\gamma/Z$. At the $Z$ resonance, the effective coupling drops and $\Lambda$ increases, as can be seen in the figure. The dominant channel in this case is $\chi\bar{\chi}\rightarrow f\bar{f}$ for the entire mass range of DM.

\item \underline{\textbf{$\mathcal{O}^{(6)}_{D\chi}$:}} The correct relic contour for this operator is shown in the \textit{right} panel of figure~\ref{fig:relic} with a green solid line. Here, the channels that contribute to relic are \{$\chi\bar{\chi}\rightarrow f\bar{f},~ ZZ,~ W^+W^-,~ ZH$\}. For a fixed $\Lambda$, increasing $m_\chi$ reduces the relic density. Consequently, $\Lambda$ must be increased to achieve the correct relic density. At smaller DM masses, the $f\bar{f}$ channel dominates, while additional channels open up kinematically as the DM mass increases. These processes are mediated by the $Z$ boson, and the $Z$ resonance is clearly visible in the figure. 

\item \underline{\textbf{$\mathcal{O}^{(6)}_{l\chi}$:}} The correct relic parameter space for this operator is shown in the \textit{right} panel of figure~\ref{fig:relic} with a blue solid line. The relic is decided by the freeze-out of the following channels: \{$\chi\bar{\chi}\rightarrow e^+e^-,~ \mu^+\mu^-,~ \tau^+\tau^-,~ \nu_e\overline{\nu_e},~ \nu_\mu\overline{\nu_\mu},~ \nu_\tau\overline{\nu_\tau}$\}. The total cross section scales as $\langle\sigma v\rangle\sim6m_\chi^2/2\pi\Lambda^4$. As a result, the $\Lambda$ increases with $m_\chi$ to give the correct relic density.

\item \underline{\textbf{$\mathcal{O}^{(6)}_{e\chi}$:}} For this operator, DM relic is decided by the freeze-out of the following channels: \{$\chi\bar{\chi}\rightarrow e^+e^-,~\mu^+\mu^-,~\tau^+\tau^-$\}. The total cross section scales as $\langle\sigma v\rangle\sim3m_\chi^2/2\pi\Lambda^4$. As a result, the $\Lambda$ increases with $m_\chi$ to give the correct relic density. It is worth noting that, in this case, the cross section is smaller compared to the $\mathcal{O}^{(6)}_{L\chi}$ operator. As a result, a comparatively smaller $\Lambda$ is required to get the correct relic density. The correct relic parameter space for this operator is shown in the \textit{right} panel of figure~\ref{fig:relic} with a red solid line. 
\end{itemize}

\subsection{Direct Detection}
Among all the operators considered above, only $\mathcal{O}^{(6)}_{D\phi}$, $\mathcal{O}^{(5)}_{B\chi}$, and $\mathcal{O}^{(6)}_{D\chi}$ can yield tree-level spin-independent direct detection (SIDD) contributions. 
\begin{figure}[htb!]
\centering
\includegraphics[width=0.6\textwidth]{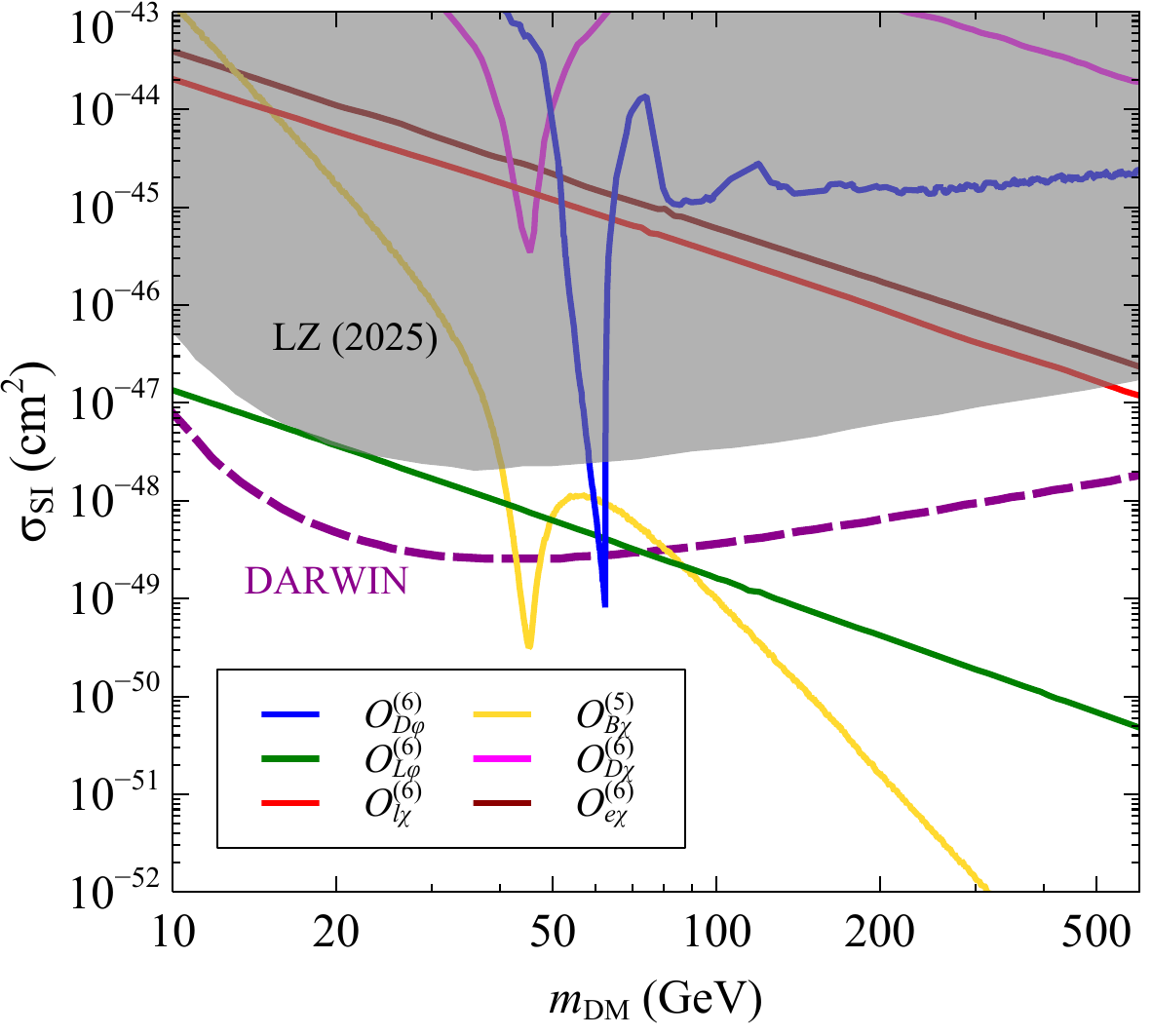}
\caption{Spin-independent DM-nucleon scattering cross-section as a function of DM mass for various operators as mentioned in the figure inset. The current exclusion limit from LZ \cite{LZ:2024zvo} is shown with gray shaded region. The DARWIN \cite{DARWIN:2016hyl} sensitivity
is shown with a dark magenta dashed line.}
\label{fig:relicDD}
\end{figure}
For the remaining operators, direct detection arises at one loop. We have computed the SIDD cross sections for each operator and shown them in figure~\ref{fig:relicDD}. The present constraint from the LZ experiment \cite{LZ:2024zvo} is shown with the gray shaded region. The DARWIN \cite{DARWIN:2016hyl} sensitivity
is shown with a dark magenta dashed line. It is clear from the figure that the operators $\mathcal{O}^{(6)}_{D\chi}$ and $\mathcal{O}^{(6)}_{e\chi}$ remain fully excluded for the DM mass range considered in the analysis by LZ experiment. We have also indicated the regions excluded by the LZ experiment \cite{LZ:2024zvo} in light color in the \textit{left} and \textit{right} panels of figure~\ref{fig:relic}.
\begin{table}[h]
\centering
\begin{tabular}{|c| c|c|c|} 
 \hline
Operators & DD& ID\\
 \hline
  $\mathcal{O}_{D\phi}^{(6)}$ & Allowed near resonance only&--\\
\hline
  $\mathcal{O}_{L\phi}^{(6)}$& Allowed & Allowed for $m_{\rm DM}>40$ GeV ($\tau^+\tau^-$)\\
\hline
  $\mathcal{O}_{W\phi}^{(6)}$& Allowed &Allowed\\
\hline
  $\mathcal{O}_{B\phi}^{(6)}$& Allowed&Allowed \\
\hline
  $\mathcal{O}_{D\chi}^{(6)}$& Excluded &Allowed\\
\hline
  $\mathcal{O}_{B\chi}^{(5)}$& Allowed for $m_{\rm DM}>40$ GeV&Allowed for $m_{\rm DM}>20$ GeV ($\tau^+\tau^-$)\\
\hline
  $\mathcal{O}_{l\chi}^{(6)}$& Allowed for $m_{\rm DM}>520$ GeV&Allowed for $m_{\rm DM}>20$ GeV ($\tau^+\tau^-$) \\
\hline
  $\mathcal{O}_{e\chi}^{(6)}$& Excluded & Allowed for $m_{\rm DM}>40$ GeV ($\tau^+\tau^-$) \\  
\hline
\end{tabular}
\caption{Summary of the scalar and Dirac fermion operators from relic, direct detection (DD), and indirect detection (ID).}\label{tab:ddid}
\end{table}
In other words, only the dark colored part of each line satisfies both the DM relic density and direct detection (LZ) constraints.

\subsection{Indirect Detection} 
\begin{figure}[htb!]
\centering
\includegraphics[width=0.475\textwidth]{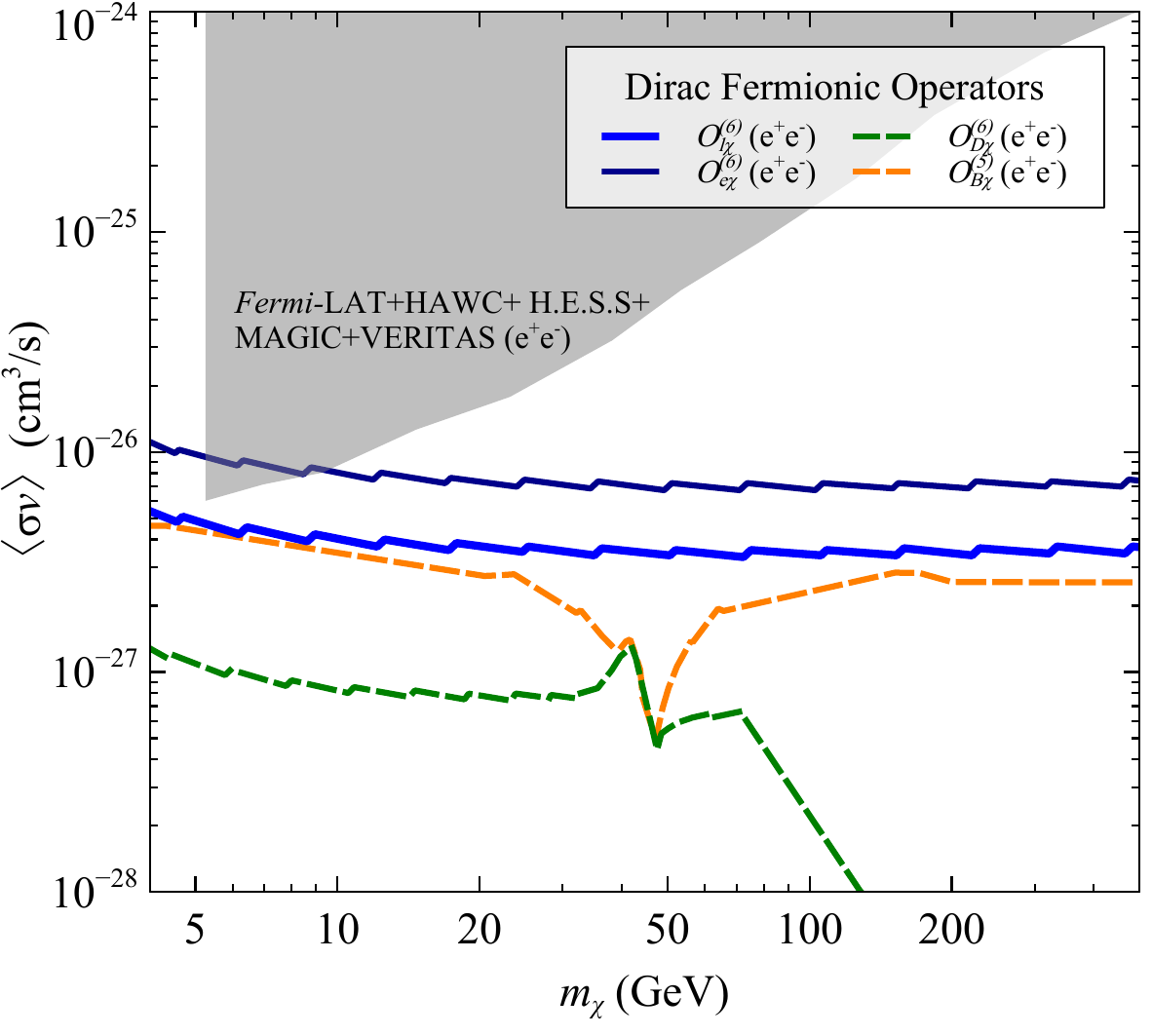}
\includegraphics[width=0.475\textwidth]{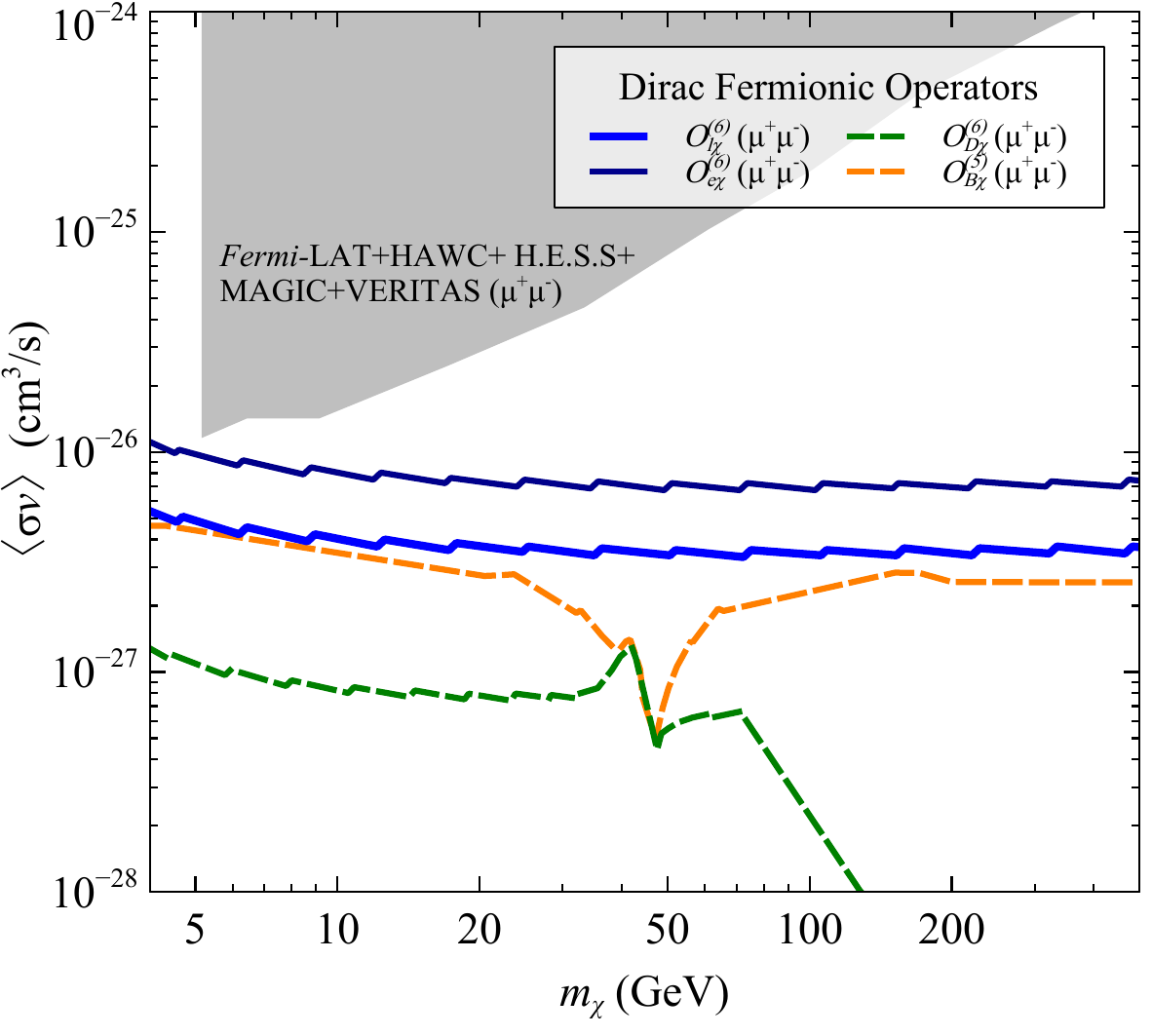} \\
\includegraphics[width=0.475\textwidth]{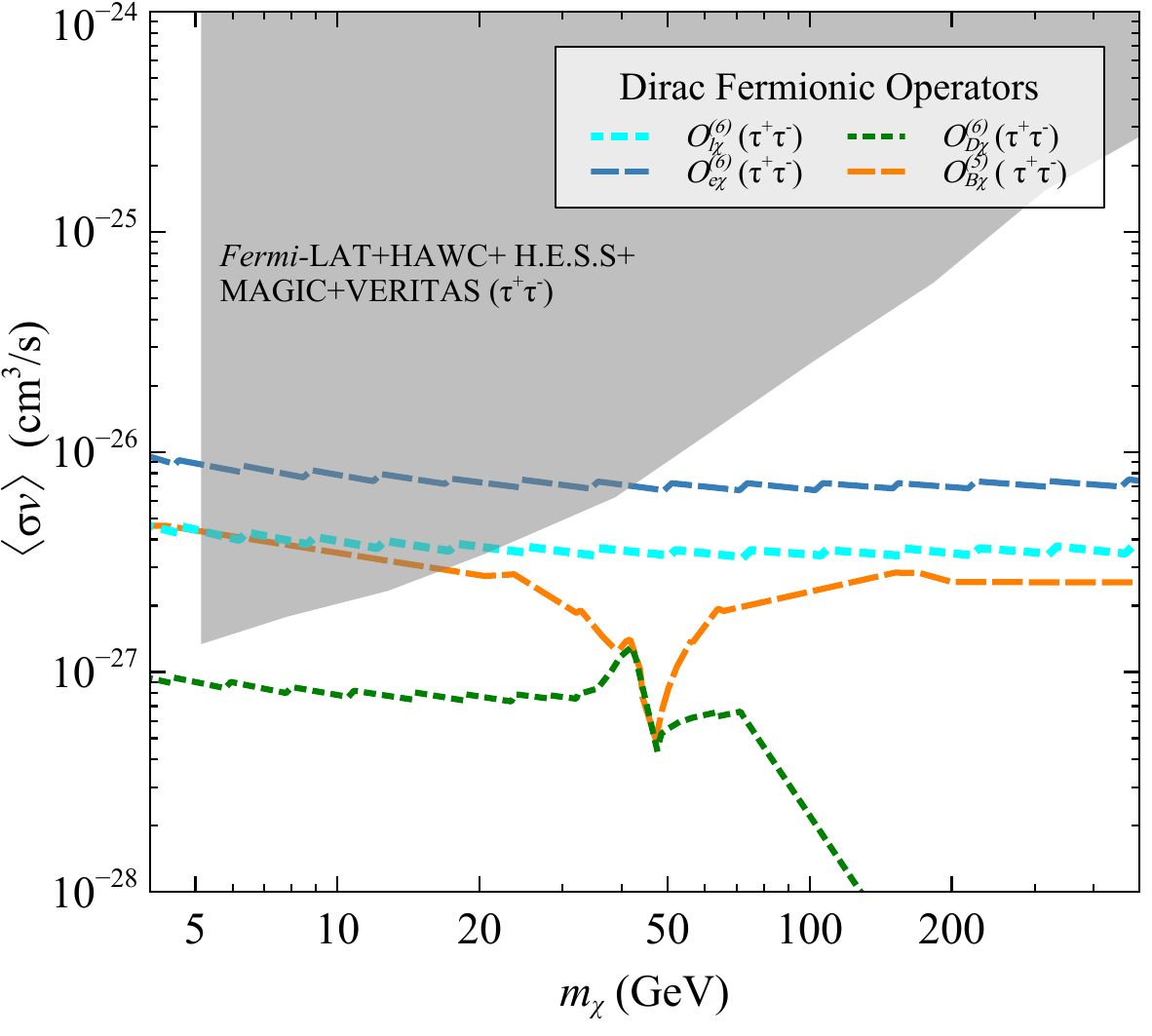}
\includegraphics[width=0.475\textwidth]{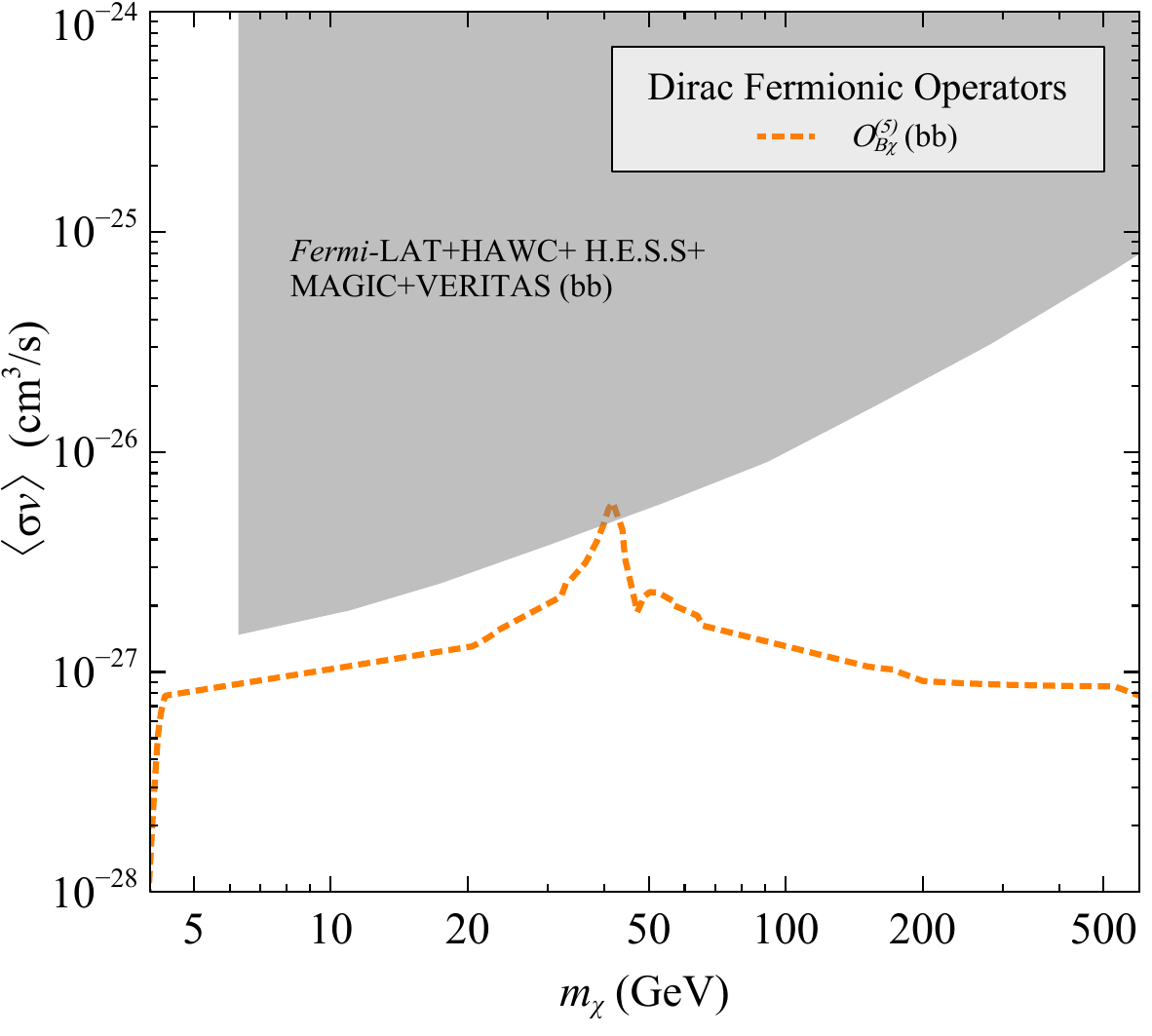}
\caption{DM annihilation cross section as a function of DM mass for the Dirac fermion operators.}
\label{fig:ID}
\end{figure}

In addition to direct detection, DM can also be probed via indirect detection experiments. These searches aim to identify SM particles generated through DM annihilation or decay at the centre of the galaxy. Among all possible final-state products, photons and neutrinos are especially important because their charge neutrality and stability allow them to propagate freely from the astrophysical source to the detector. Gamma rays, originating from electromagnetically charged final states, can be detected by space-based telescopes such as the Fermi Large Area Telescope (Fermi-LAT) \cite{Fermi-LAT:2009ihh,Fermi-LAT:2015att}, as well as by ground-based facilities like MAGIC \cite{MAGIC:2014zas,MAGIC:2021mog} and HESS \cite{HESS:2014zqa,HESS:2018kom}. By measuring the gamma-ray flux and combining it with standard astrophysical inputs, one can constrain possible DM annihilation channels into various SM final states, including $e^+e^-$, $\mu^+\mu^-$, $\tau^+\tau^-$, $b\bar{b}$, $W^+W^-$, etc. In figure~\ref{fig:ID}, we present the Dirac fermionic DM annihilating into various final states along with the most stringent combined constraints from \textit{Fermi}-LAT, HAWC, HESS, MAGIC, VERITAS \cite{Fermi-LAT:2025gei}.
\begin{figure}[h]
\centering
\includegraphics[width=0.475\textwidth]{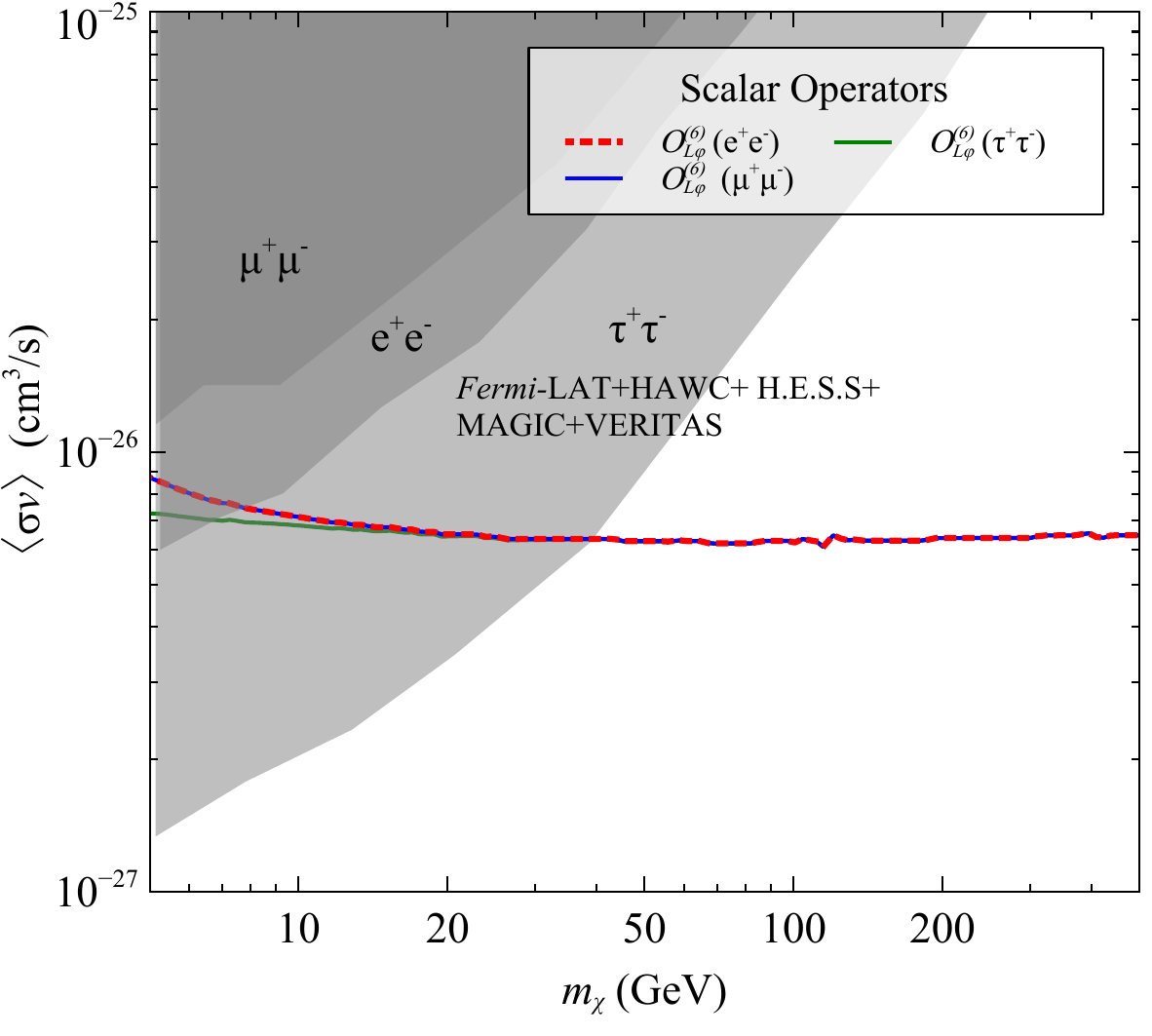}
\includegraphics[width=0.475\textwidth]{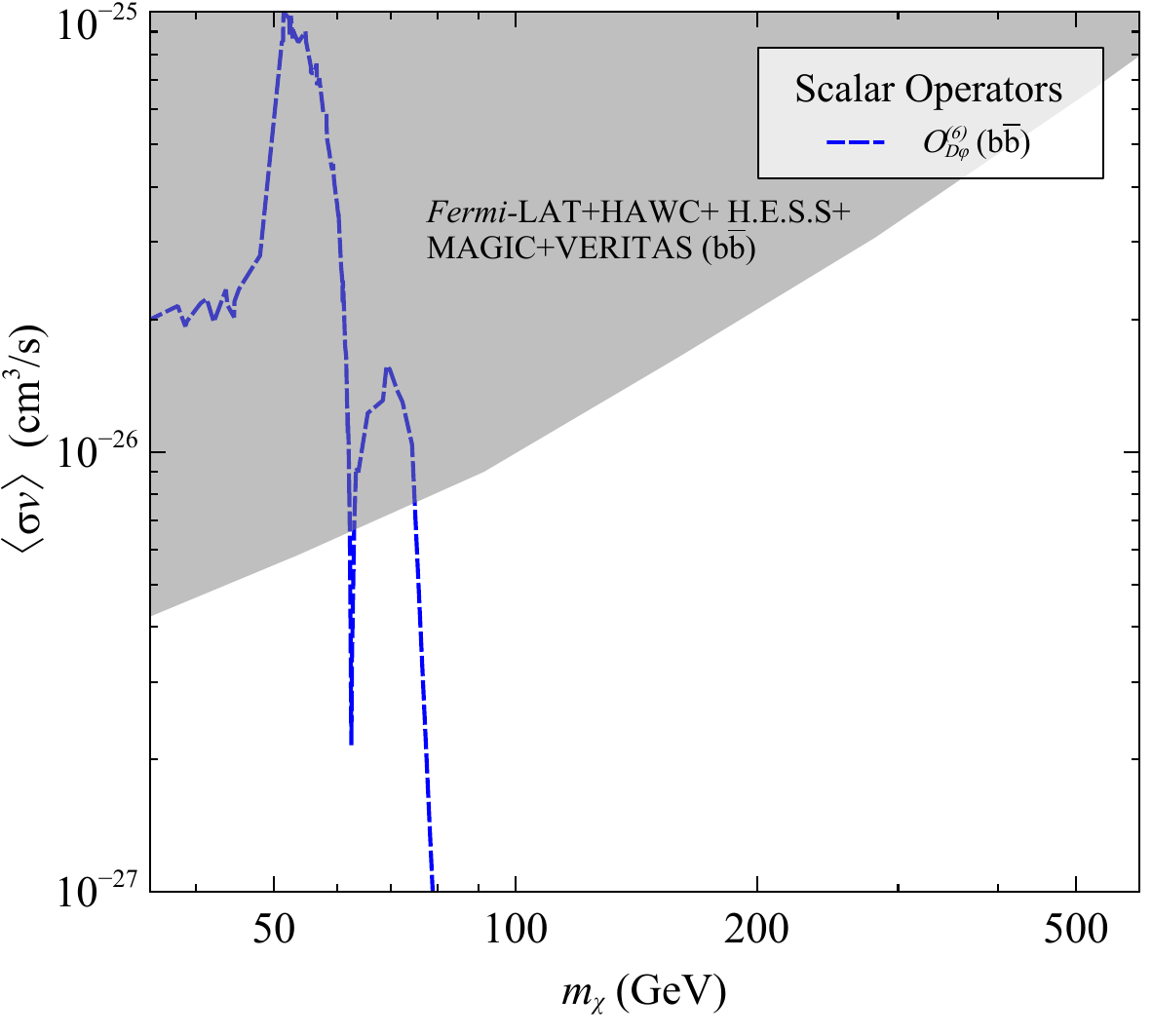} \\
\includegraphics[width=0.475\textwidth]{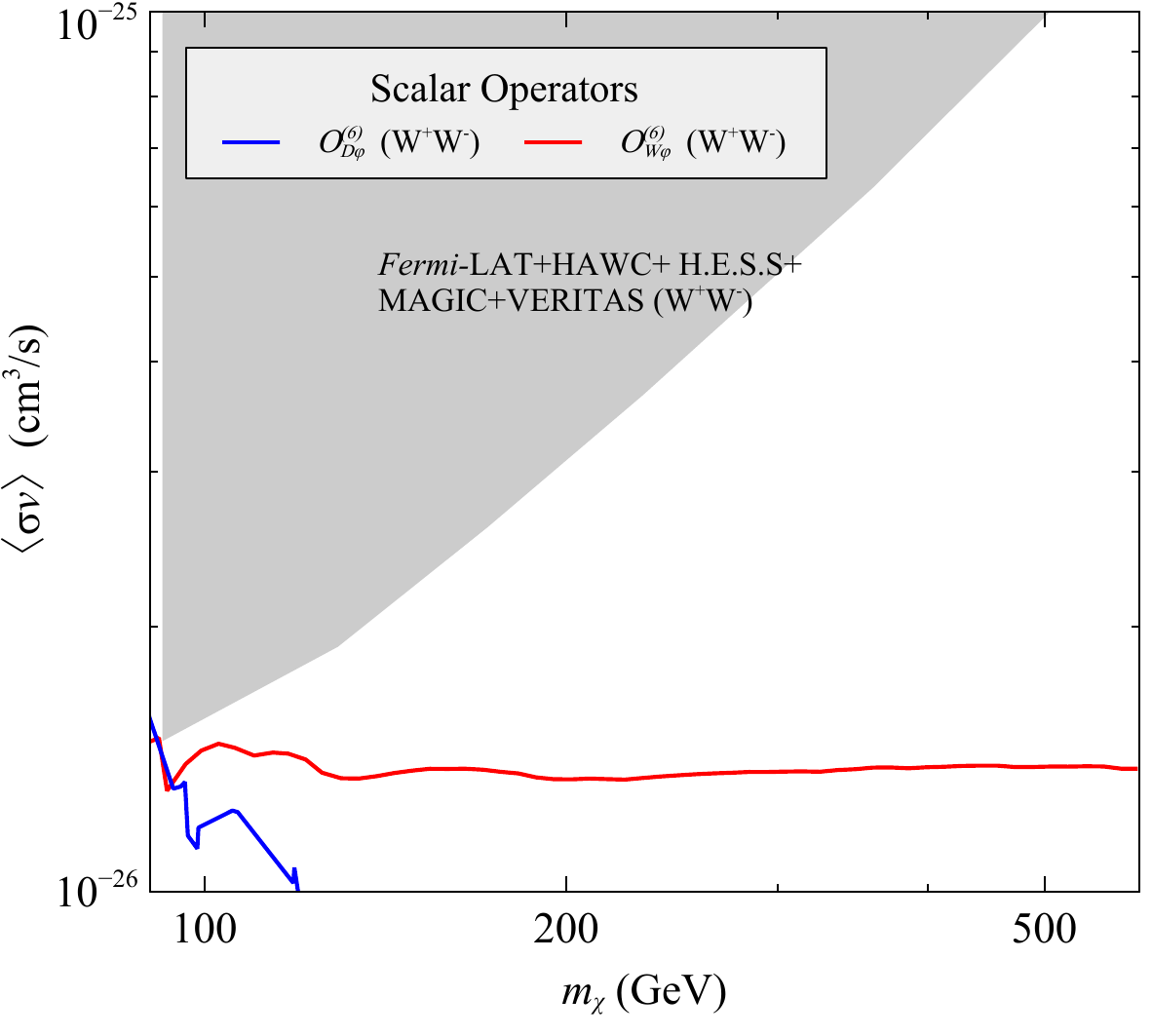}
\includegraphics[width=0.475\textwidth]{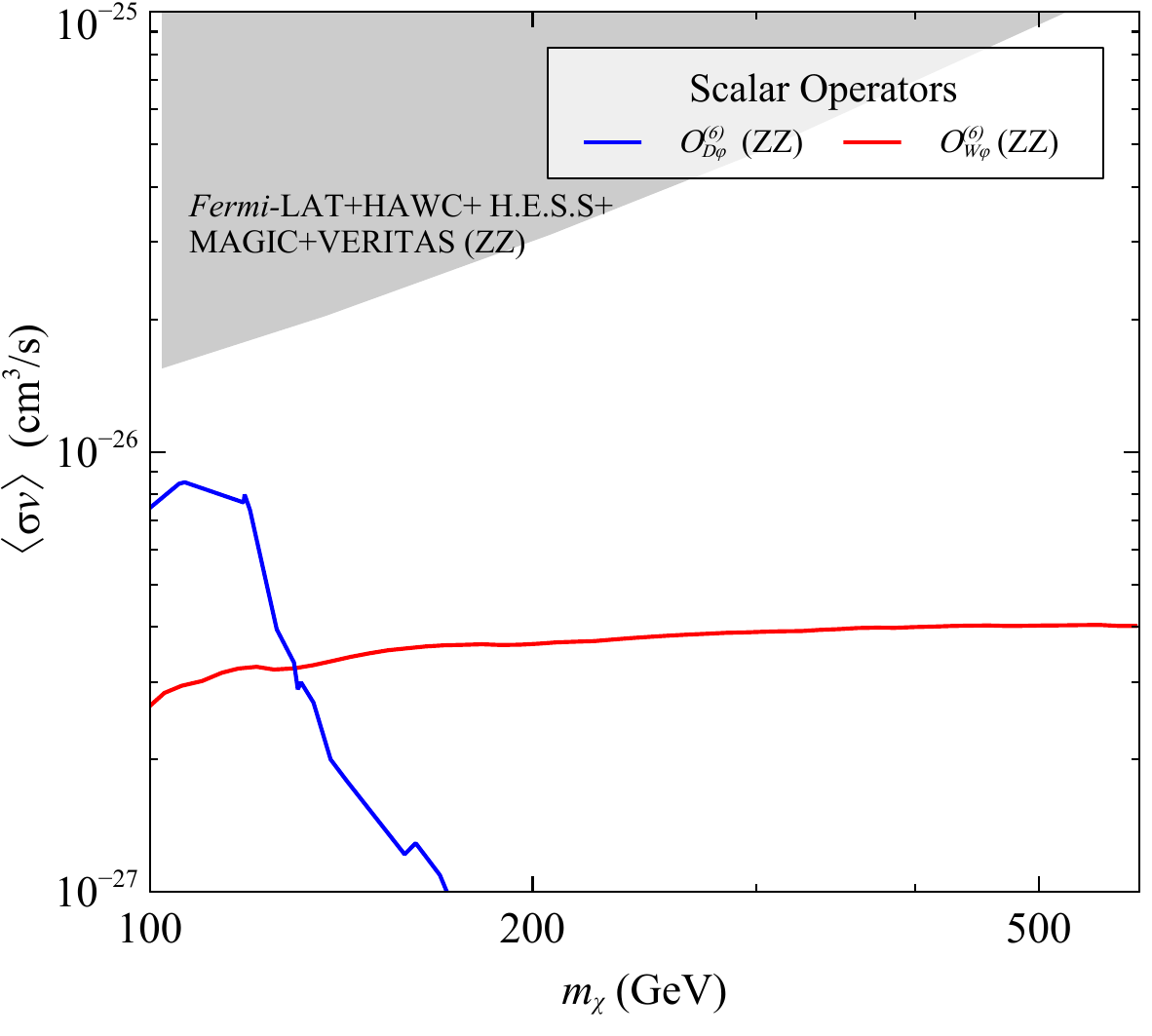}
\caption{DM annihilation cross section as a function of DM mass for the scalar operators.}
\label{fig:ID_scalar}
\end{figure}
Indirect detection constraints for the scalar operators are shown in figure~\ref{fig:ID_scalar}. We finally summarize the allowed parameter space from both direct detection and indirect detection in table \ref{tab:ddid}. The operators $\mathcal{O}_{D\chi}^{(6)}$ and $\mathcal{O}_{e\chi}^{(6)}$ remain fully excluded by the direct detection experiment for the DM mass range considered above. The other operators remain viable for the collider studies for the allowed mass ranges as mentioned in table \ref{tab:ddid}.

\section{Higgs Invisible Decay}
\label{sec:higgs}
\begin{figure}[htb!]
\centering
\includegraphics[width=0.6\textwidth,trim=0 0 0 0,clip]{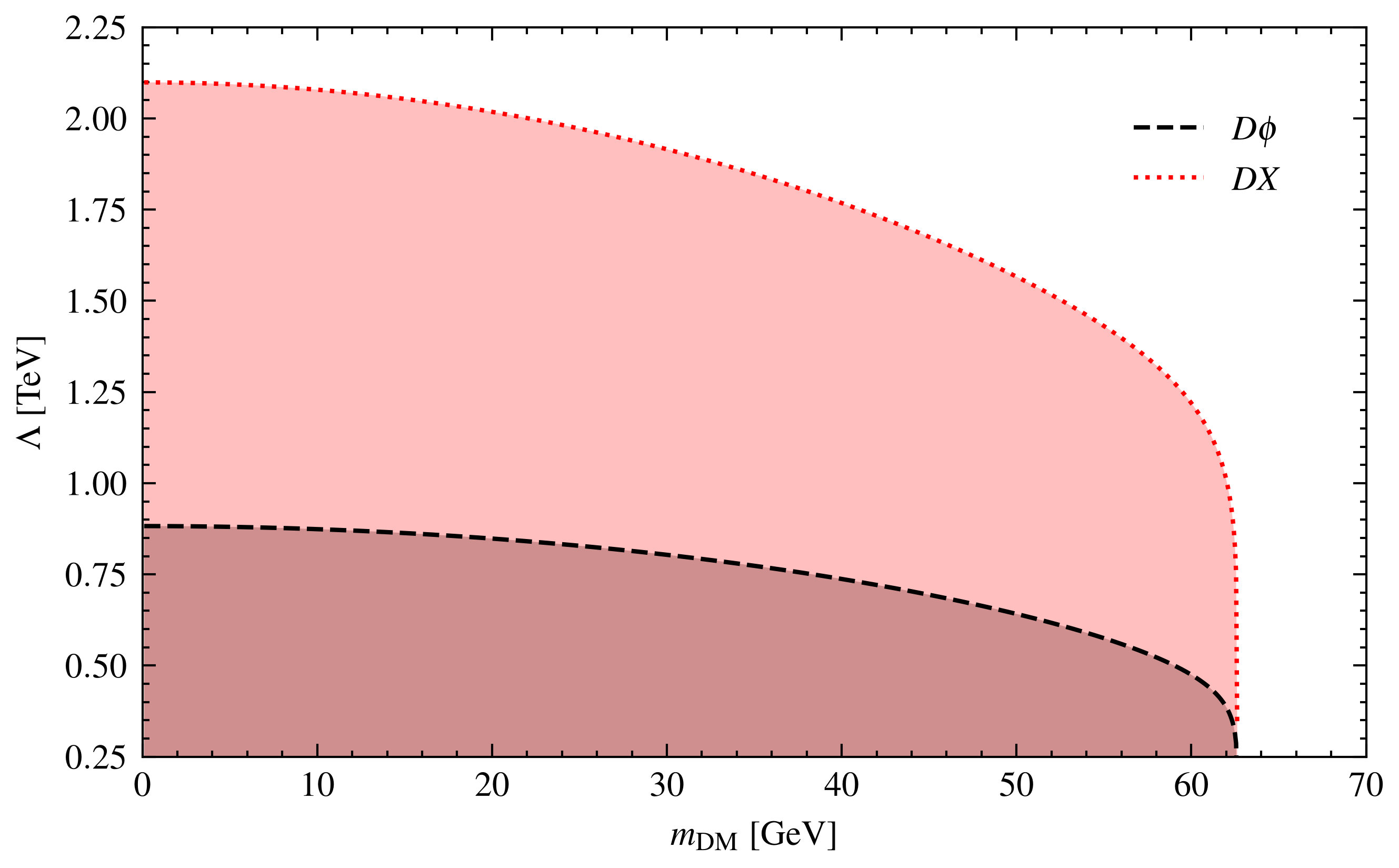}
\caption{Constraints on the Higgs-portal operators $\mathcal{O}^{(6)}_{D\phi}$ and $\mathcal{O}^{(6)}_{DX}$ arising from the current limit on the Higgs invisible branching fraction. The shaded region is excluded at the 95\% confidence level.}
\label{fig:higgs}
\end{figure}

Among the DMEFT operators considered in this work, the Higgs-portal operators,
$\mathcal{O}^{(6)}_{D\phi}$ and $\mathcal{O}^{(6)}_{DX}$, gives rise to an invisible decay of the Higgs boson whenever the bosonic DM mass satisfies the kinematic condition $m_{\rm DM} < m_h/2$. Since the Higgs boson has been extensively studied at the LHC, measurements of its invisible branching fraction provide stringent constraints on this operator, complementary to those obtained from relic abundance and DM direct detection searches. The partial decay width of the Higgs boson into a pair of bosonic DM particles through the $\mathcal{O}^{(6)}_{D\phi}$ and $\mathcal{O}^{(6)}_{DX}$ interaction is given by
\begin{equation}
\begin{split}
    \Gamma(h\rightarrow\phi\phi) &= \frac{v_h^2\left(m_h^2-2m_\phi^2\right)^2}{128\pi m_h\Lambda^4}\sqrt{1-\frac{4m_\phi^2}{m_h^2}}\,, \\
    \Gamma(h\rightarrow XX) &= \frac{v^2 m_h^3}{4 \pi^4 \Lambda^4}\sqrt{1-\frac{4m_X^2}{m_h^2}} \left(1 - \frac{4m_X^2}{m_h^2} + 6 \frac{m_X^4}{m_h^4}\right)
\end{split}
\end{equation}
where $m_h$ and $m_\phi(m_X)$ denote the Higgs boson and scalar(vector) DM masses, respectively, while $v_h=246$ GeV is the Electroweak vacuum expectation value. The corresponding invisible branching fraction is computed as
\begin{equation}
\mathcal{B}(h\rightarrow\mathrm{inv.})
=
\frac{\Gamma[h\rightarrow\phi\phi(XX)]}
{\Gamma_h^{\rm SM}+\Gamma[h\rightarrow\phi\phi(XX)]},
\end{equation}
where $\Gamma_h^{\rm SM}=4.07$ MeV is the total decay width of the SM Higgs. The latest ATLAS analyses~\cite{ATLAS:2023tkt} constrain the Higgs invisible branching fraction to
\begin{equation}
\mathcal{B}(h\rightarrow\mathrm{inv.})<10.7\%
\qquad
(95\%~\mathrm{C.L.})\,,
\end{equation}
which can be directly translated into a lower bound on the cutoff scale $\Lambda$ as a function of $m_{\rm DM}$. The resulting exclusion region is shown in figure~\ref{fig:higgs}. As expected, the constraint is applicable only for $m_{\rm DM} < m_h/2$, where the invisible decay is kinematically allowed. Near the threshold, the decay width is suppressed by the phase-space availability, leading to a gradual weakening of the bound. For $m_{\rm DM}>m_h/2$, the Higgs invisible decay channel is closed, and no constraint can be derived from the Higgs invisible branching fraction.

\section{Recasting LHC Limits}
\label{sec:lhc-recast}

\begin{figure}[htb!]
    \centering
    \begin{subfigure}[b]{0.33\textwidth}
        \centering
        \includegraphics[width=\textwidth]{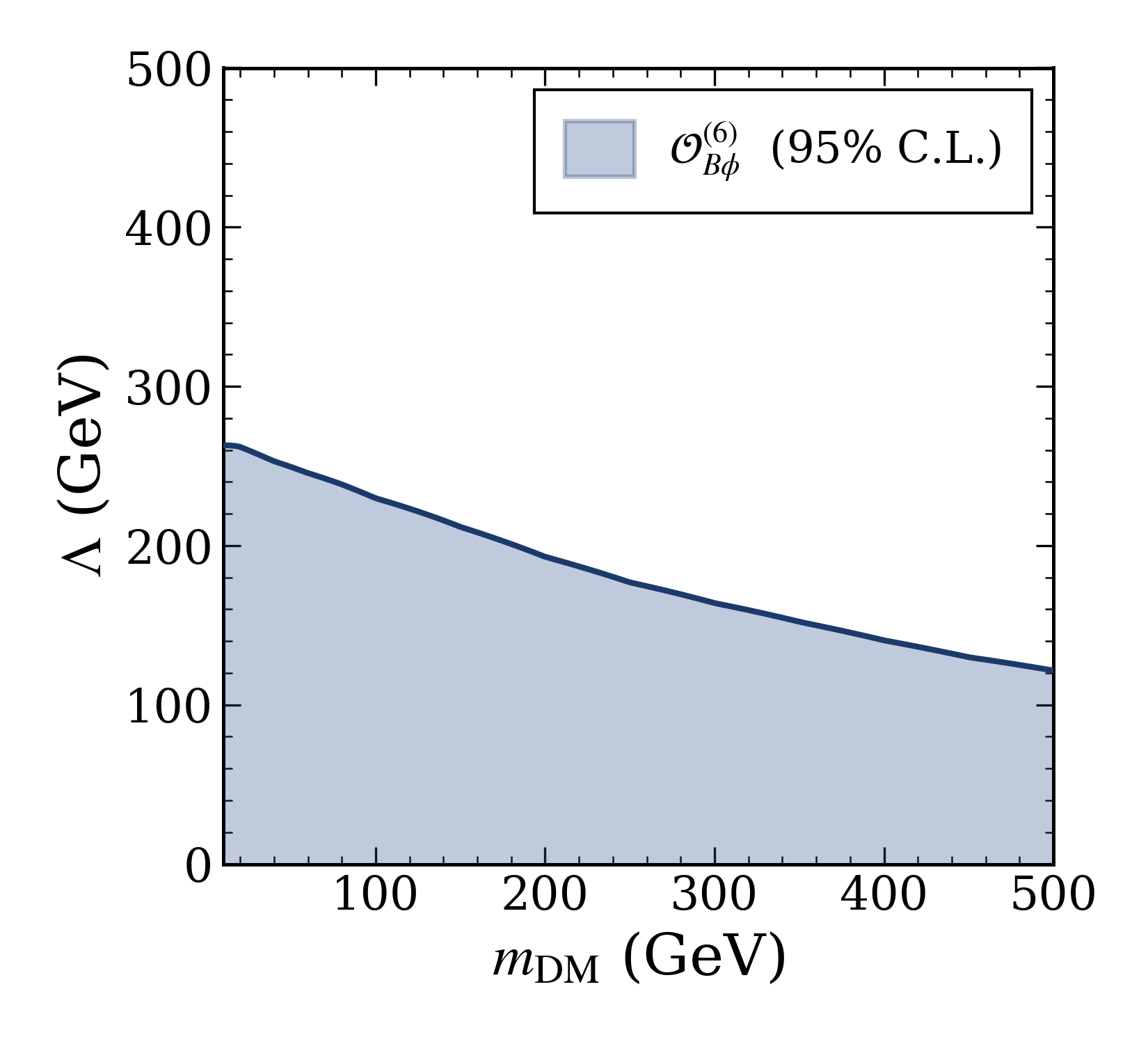}
        \caption{$\mathcal{O}_{B\phi}^{(6)}$}
        \label{fig:recast-Bphi}
    \end{subfigure}\hfill
    \begin{subfigure}[b]{0.33\textwidth}
        \centering
        \includegraphics[width=\textwidth]{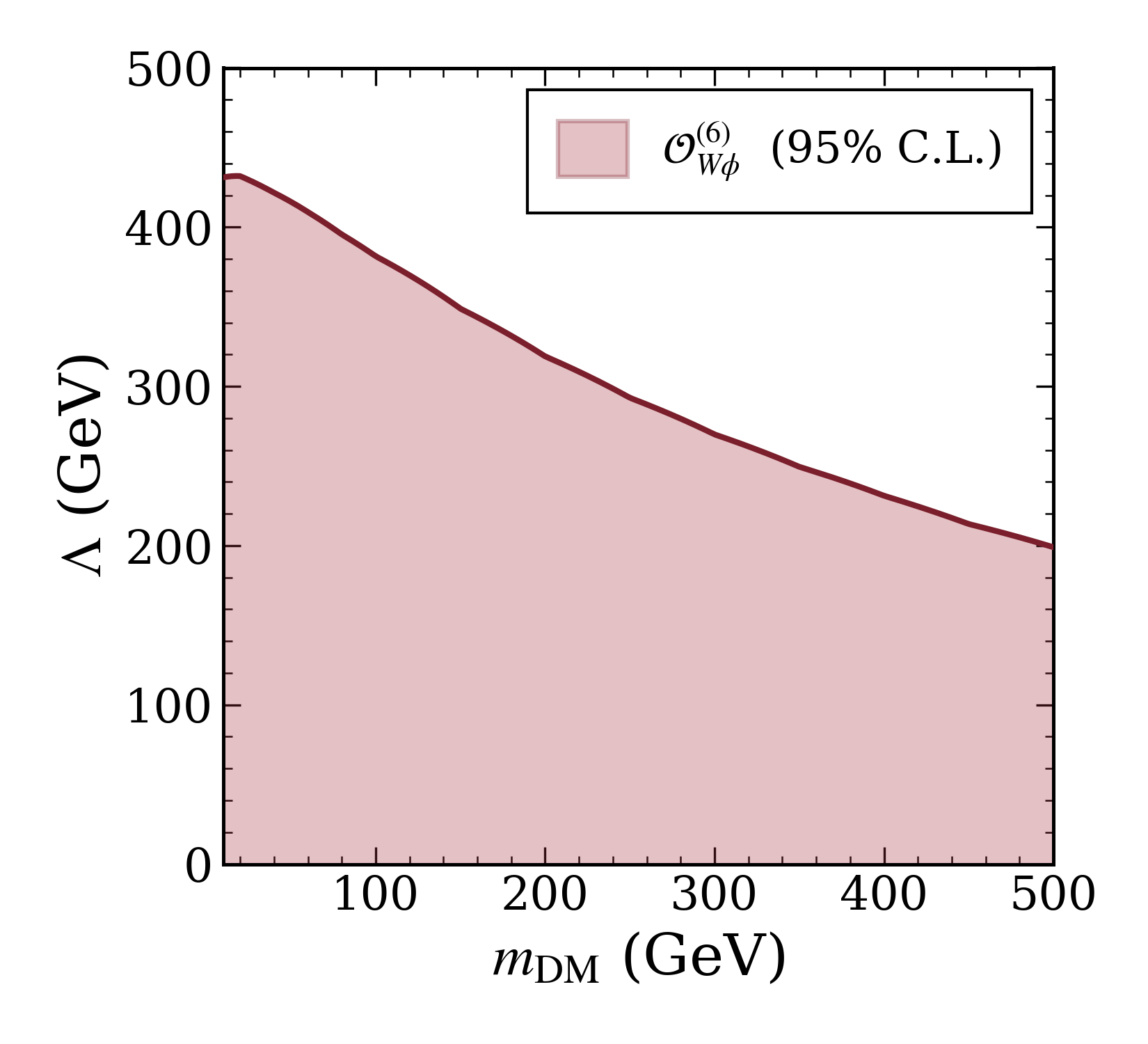}
        \caption{$\mathcal{O}_{W\phi}^{(6)}$}
        \label{fig:recast-Wphi}
    \end{subfigure}\hfill
    \begin{subfigure}[b]{0.33\textwidth}
        \centering
        \includegraphics[width=\textwidth]{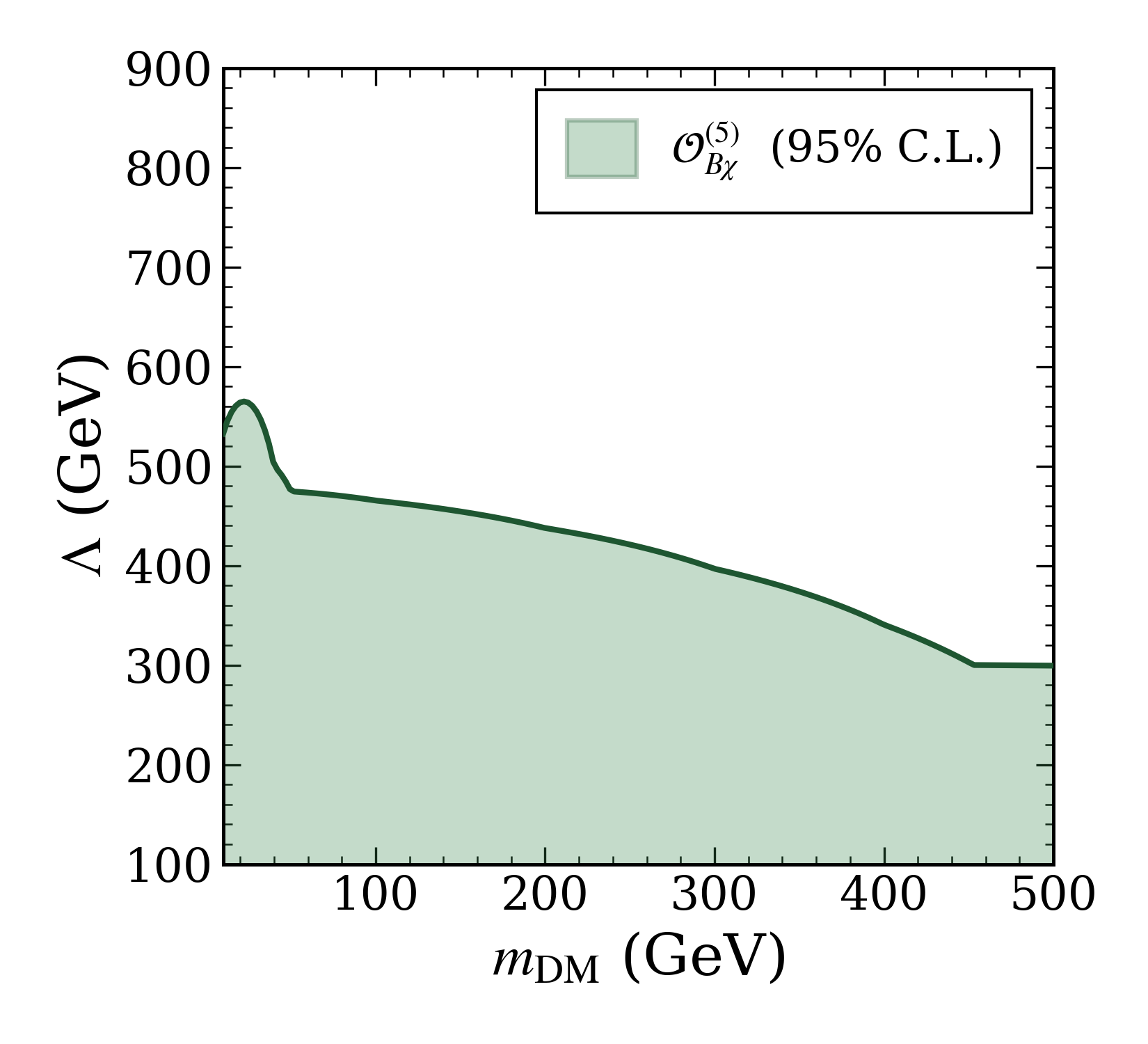}
        \caption{$\mathcal{O}_{B\chi}^{(5)}$}
        \label{fig:recast-Bchi5}
    \end{subfigure}
    \caption{95\% C.L.\ exclusion regions in the $(m_{\rm DM},\Lambda)$ plane
    obtained by recasting the ATLAS mono-$Z(\ell\ell)$~+~$p_T^{\rm miss}$
    search~\cite{ATLAS:2021monoZ} for the (a)~$\mathcal{O}_{B\phi}^{(6)}$,
    (b)~$\mathcal{O}_{W\phi}^{(6)}$, and (c)~$\mathcal{O}_{B\chi}^{(5)}$ operators. Shaded regions are excluded
    at 95\% confidence level.}
    \label{fig:recast-all}
\end{figure}
 
To constrain the Electroweak-field-strength-coupled DMEFT operators $\mathcal{O}_{B\phi}^{(6)}$, $\mathcal{O}_{W\phi}^{(6)}$, and $\mathcal{O}_{B\chi}^{(5)}$, we recast the ATLAS search for a $Z$ boson produced in association with DM candidates in the dilepton~+~$p_T^{\rm miss}$ final state~\cite{ATLAS:2021monoZ}, based on $139~\text{fb}^{-1}$ of $\sqrt{s}=13$~TeV proton--proton collision data.
 
Signal events for each operator were generated at leading order with \textsc{MG5\_aMC}~\cite{Alwall:2014hca}, using the default \texttt{NNPDF3.0} parton distribution set~\cite{Ball:2014uwa}, showered and hadronized with \textsc{Pythia\,8}~\cite{Sjostrand:2014zea}, and passed through a parameterized ATLAS detector simulation using \textsc{Delphes~3.5}~\cite{deFavereau:2013fsa}. We reproduce the ATLAS signal-region selection: exactly two same-flavor, opposite-sign leptons with transverse momentum $p_T>30\,(20)$~GeV for the leading (subleading) lepton, a dilepton invariant mass window $76<m_{\ell\ell}<106$~GeV, missing transverse energy $p_T^{\rm miss}>90$~GeV, angular separation $\Delta R_{\ell\ell}<1.8$, and a veto on events containing $b$-tagged jets.
 
For each operator the predicted signal yield scales as $N_{\rm sig}\propto (C/\Lambda^{\,n})^2$, with $n=1,2$ for the dimension-5, and dimension-6 operators respectively. We derive 95\% confidence-level exclusion limits on $\Lambda$ by requiring $N_{\rm sig}$ not exceed the maximum signal yield compatible with the observed ($N_{\rm obs}=6382$) and expected background ($N_{\rm bkg}=6385\pm81$) event counts in the ATLAS signal region. The resulting exclusion contours in the $(m_{\rm DM},\Lambda)$ plane are shown in figure~\ref{fig:recast-all}, which actually excludes $\Lambda \lesssim 500$ GeV.

For instance, at $m_{\rm DM}=100$~GeV ($200$~GeV), cutoff scale up to $\Lambda\approx230$~GeV ($195$~GeV) are excluded at 95\% C.L.\ for the $\mathcal{O}_{B\phi}^{(6)}$ operator, $\Lambda\approx345$~GeV ($315$~GeV) for $\mathcal{O}_{W\phi}^{(6)}$, and $\Lambda\approx460$~GeV ($420$~GeV) for $\mathcal{O}_{B\chi}^{(5)}$. In the subsequent collider analysis, we adhere to the obtained bounds.

\section{Collider Analysis}
\label{sec:cpheno}
We investigate the sensitivity of a high-energy $e^+e^-$ collider to the DMEFT operators introduced in Sec.~\ref{sec:dmeft} through the mono-$Z$ signature,
\begin{equation}
	e^+e^- \rightarrow Z + {\rm DM\,DM}.
\end{equation}
The DM particles escape detection and give rise to missing four-momentum. The visible $Z$ boson consequently provides the only handle for reconstructing the event kinematics.
We perform the analysis at $\sqrt{s}=1$~TeV for all scalar and Dirac-fermion operators listed in tables~\ref{tab:dim-l1} and \ref{tab:dim-l2}. The benchmark points are chosen from the parameter regions compatible with the cosmological constraints discussed in Sec.~\ref{sec:dpheno}. The operators involving the Higgs field, $\mathcal{O}^{(6)}_{D\phi}$ and $\mathcal{O}^{(6)}_{DX}$, are considered separately at $\sqrt{s}=250$~GeV in Subsec.~\ref{colli-Higgs}, where Higgsstrahlung provides the dominant production mechanism.

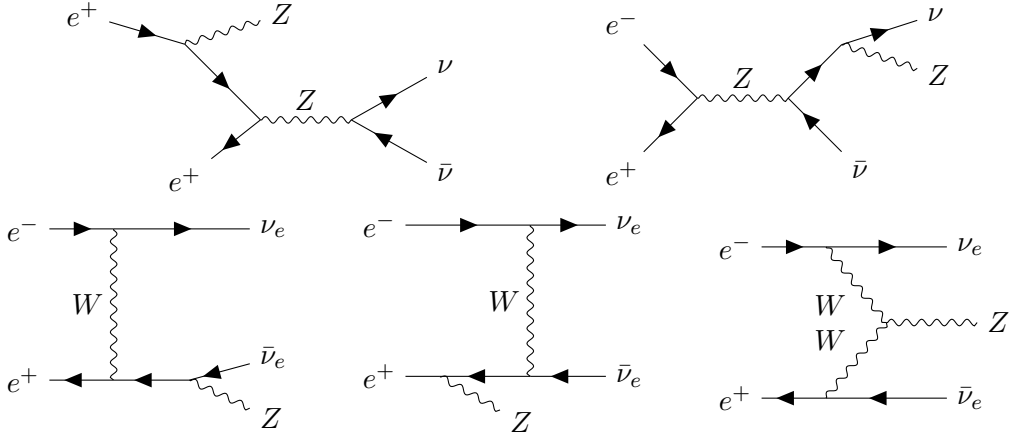
\begin{figure}[htbp!]
	\centering
	\begin{tikzpicture}[baseline=(current bounding box.center), >={Latex[open]}]
		\begin{feynman}
			\vertex (i1);
            \vertex [left=1cm of i1, yshift=0.4cm] (i3) {\(e^+\)};
            \vertex [right=1cm of i1, yshift=0.4cm] (i4) {\(Z\)};
			\vertex [below=1.5cm of i1] (i2) {\(e^+\)};
			\vertex [right=1cm of i1, yshift=-1cm] (v1);
			\vertex [right=1.2cm of v1] (v2);
			\vertex [right=1cm of v2, yshift=0.7cm] (v3) {\(\nu\)};
			\vertex [right=1cm of v2, yshift=-0.7cm] (f3) {\(\bar{\nu}\)};		
			\diagram* {
				(i1) -- [fermion] (v1),
                (i3) -- [fermion] (i1),
                (i4) -- [boson] (i1),
				(i2) -- [anti fermion] (v1),
				(v1) -- [boson, edge label=\(Z\)] (v2),
				(v2) -- [fermion] (v3),
				(v2) -- [anti fermion] (f3)
			};
		\end{feynman}
	\end{tikzpicture}
	\hspace{1.5cm}
	\begin{tikzpicture}[baseline=(current bounding box.center), >={Latex[open]}]
    
		\begin{feynman}
			\vertex (i1) {\(e^-\)};
			\vertex [below=2cm of i1] (i2) {\(e^+\)};
			\vertex [right=1cm of i1, yshift=-1cm] (v1);
			\vertex [right=1.2cm of v1] (v2);
			\vertex [above right=1cm of v2] (v3);
			\vertex [below right=1cm of v2] (f3) {\(\bar{\nu}\)};
			\vertex [right=1cm of v3, yshift=0.4cm] (f1) {\(\nu\)};
			\vertex [right=1cm of v3, yshift=-0.4cm] (f2) {\(Z\)};		
			\diagram* {
				(i1) -- [fermion] (v1),
				(i2) -- [anti fermion] (v1),
				(v1) -- [boson, edge label=\(Z\)] (v2),
				(v2) -- [fermion] (v3),
				(v2) -- [anti fermion] (f3),
				(v3) -- [fermion] (f1),
				(v3) -- [boson] (f2),
			};
		\end{feynman}
	\end{tikzpicture}
	\vspace{1cm}
	\begin{tikzpicture}[baseline=(current bounding box.center), >={Latex[open]}]
		\begin{feynman}
			\vertex (i1) {\(e^-\)};
			\vertex [below=2cm of i1] (i2) {\(e^+\)};
			\vertex [right=1.2cm of i1] (v1);
			\vertex [right=1.2cm of i2] (v2);
			\vertex [right=1.8cm of v1] (f1) {\(\nu_e\)};
			\vertex [right=1cm of v2] (v3);
			\vertex [right=0.8cm of v3, yshift=0.3cm] (f2) {\(\bar{\nu}_e\)};
			\vertex [right=0.8cm of v3, yshift=-0.5cm] (f3) {\(Z\)};		
			\diagram* {
				(i1) -- [fermion] (v1),
				(v1) -- [fermion] (f1),
				(i2) -- [anti fermion] (v2),
				(v2) -- [anti fermion] (v3),
				(v3) -- [anti fermion] (f2),
				(v1) -- [boson, edge label'=\(W\)] (v2),
				(v3) -- [boson] (f3),
			};
		\end{feynman}
	\end{tikzpicture}
	\hspace{0.5cm}
	\begin{tikzpicture}[baseline=(current bounding box.center), >={Latex[open]}]
		\begin{feynman}
			\vertex (i1) {\(e^-\)};
			\vertex [below=2cm of i1] (i2) {\(e^+\)};
			\vertex [right=2cm of i1] (v1);
			\vertex [right=0.8cm of i2] (v3);
			\vertex [right=1.2cm of v3] (v2);
			\vertex [right=1cm of v1] (f1) {\(\nu_e\)};
			\vertex [right=1cm of v2] (f2) {\(\bar{\nu}_e\)};
			\vertex [right=0.8cm of v3, yshift=-0.6cm] (f3) {\(Z\)};			
			\diagram* {
				(i1) -- [fermion] (v1),
				(v1) -- [fermion] (f1),
				(i2) -- [] (v3),
				(v3) -- [anti fermion] (v2),
				(v2) -- [anti fermion] (f2),
				(v1) -- [boson, edge label'=\(W\)] (v2),
				(v3) -- [boson] (f3),
			};
		\end{feynman}
	\end{tikzpicture}
	\hspace{0.5cm}
	\begin{tikzpicture}[baseline=(current bounding box.center), >={Latex[open]}]
		\begin{feynman}
			\vertex (i1) {\(e^-\)};
			\vertex [below=2cm of i1] (i2) {\(e^+\)};
			\vertex [right=1.2cm of i1] (v1);
			\vertex [right=1.2cm of i2] (v2);
			\vertex [right=1.6cm of v1] (f1) {\(\nu_e\)};
			\vertex [right=1.6cm of v2] (f2) {\(\bar{\nu}_e\)};
			\vertex [right=0.8cm of v1, yshift=-1cm] (v3);
			\vertex [right=1.2cm of v3] (f3) {\(Z\)};		
			\diagram* {
				(i1) -- [fermion] (v1),
				(v1) -- [fermion] (f1),
				(i2) -- [anti fermion] (v2),
				(v2) -- [anti fermion] (f2),
				(v1) -- [boson, edge label'=\(W\)] (v3),
				(v2) -- [boson, edge label=\(W\)] (v3),
				(v3) -- [boson] (f3),
			};
		\end{feynman}
	\end{tikzpicture}
	\caption{Representative Feynman diagrams for the irreducible SM background $e^+e^-\to \nu\bar{\nu}Z$. Top row: on-shell $ZZ$ production with one $Z$ boson decaying invisibly (left) and $s$-channel production with final-state radiation of a $Z$ from the neutrino line (right). Bottom row: $t$-channel $W$ exchange contribution for the $\nu_e\bar{\nu}_e Z$ final state; with final-state $Z$ radiation (left), initial-state $Z$ radiation (middle), and $t$-channel $WW$-fusion (right).}
	\label{fig:sm_bkg_feynman}
\end{figure}

\paragraph{Background processes:} The irreducible SM background is
$$e^+e^-\rightarrow \nu\bar{\nu} Z,$$
where $\nu$ runs over all three active neutrino flavors. The representative Feynman diagrams are shown in figure~\ref{fig:sm_bkg_feynman}. This final state receives contributions from:
\begin{itemize}
    \item $t$-channel $e^+e^- \to ZZ$, with one $Z$ boson decaying invisibly ($Z\to\nu\bar{\nu}$),
 \item $s$-channel diagrams in which an off-shell $Z$ boson produces a neutrino pair and the observed $Z$ is radiated from the outgoing neutrino or antineutrino line, contributing to the non-resonant $Z\nu\bar{\nu}$ background, and
    \item $t$-channel $W$-exchange topologies that produce a $\nu_e\bar{\nu}_e$ pair, and the visible $Z$ boson radiating from the initial-state electron/positron line or from the final-state neutrino/antineutrino line, or via $WW$-fusion.
\end{itemize}

\begin{table}[htb!]
	\centering
	\renewcommand{\arraystretch}{1.0}{
		\begin{tabular}{|>{\centering\arraybackslash}p{3cm}|
				>{\centering\arraybackslash}p{2cm}|
				>{\centering\arraybackslash}p{2cm}|
				>{\centering\arraybackslash}p{2cm}|
				>{\centering\arraybackslash}p{2cm}|
				>{\centering\arraybackslash}p{2cm}|
				>{\centering\arraybackslash}p{2cm}|}
			\hline 
			Polarization  & \multicolumn{5}{c|}{Production cross section (fb)} \\ \cline{2-6}
			$\left(P_{e^+}, P_{e^-}\right)$ &  $\mathcal{O}^{(5)}_{B\chi}$ & $\nu \overline{\nu} Z$ (SM) &  $\mathcal{O}^{(6)}_{L\phi}$ & $\mathcal{O}^{(6)}_{W\phi}$ & $\mathcal{O}^{(6)}_{B\phi}$\\ \hline
			unpolarized & 0.199 & 955.9 & 0.515 & 1.06 & 0.435\\
			$\left(+20\%, +80\%\right)$ & 0.2297 & 252.2 & 0.5971 & 0.255 & 0.524\\
			$\left(+20\%, -80\%\right)$ & 0.1255  & 2017.0 & 0.4326 & 2.292 & 0.242\\
			$\left(-20\%, +80\%\right)$ & 0.3354 & 190.7  & 0.4325 & 0.17 & 0.769\\
			$\left(-20\%, -80\%\right)$ & 0.1039 & 1349.0  & 0.5979 & 1.528 & 0.208\\ \hline
	\end{tabular}}
	\caption{Cross section for DM pair production in association with a $Z$ boson at the 1 TeV ILC, comparing various operators and the SM background. The mass of DM and $\Lambda$ for different operators are as follows. For $\mathcal{O}^{(5)}_{B\chi}$, $\Lambda=7.3$ TeV, $m_{\rm DM}=100$ GeV; for $\mathcal{O}^{(6)}_{L\phi}$, $\Lambda=2.0$ TeV, $m_{\rm DM}=100$ GeV; for $\mathcal{O}^{(6)}_{W\phi}$, $\Lambda=1.20$ TeV, $m_{\rm DM}=76$ GeV; for $\mathcal{O}^{(6)}_{B\phi}$ , $\Lambda=1.27$ TeV, $m_{\rm DM}=50$ GeV.}
	\label{tab:prodxs}
\end{table}
\paragraph{Beam polarization and operator dependence:}

The mono-$Z$ signal production rate depends strongly on the initial-state beam polarizations and is governed by the Electroweak structure of the underlying effective operator. The SM contamination also has a polarization dependence, as detailed hereunder.

The $t$-channel $W$-exchange amplitudes that produce the dominant $\nu_e\bar{\nu}_e\,Z$ background couple exclusively to left-handed electron and right-handed positron. Consequently, polarizing the electron beam to be predominantly right-handed efficiently suppresses this background. As summarized in table~\ref{tab:prodxs}, configuring the ILC beams to $(P_{e^+}, P_{e^-}) = (-20\%, +80\%)$ drastically reduces the total SM background from $955.9$~fb (unpolarized) to $190.7$~fb. Conversely, the left-handed-enhanced configuration, $(P_{e^+}, P_{e^-}) = (+20\%, -80\%)$, sharply increases the background to $2017.0$~fb.

Let us now discuss the effect of polarization on the DM-SM operators. The polarization dependence of the DM–SM operators is dictated by their Lorentz and gauge structures. The $SU(2)_L$ gauge-portal operator $\mathcal{O}^{(6)}_{W\phi}$ mediates $s$-channel annihilation that effectively projects the $W^3$ component of the neutral Electroweak current. Because weak isospin is strictly confined to left-handed chiral fields ($T_3(e_L) = -1/2$, $T_3(e_R) = 0$), this operator couples almost exclusively to left-handed electrons. Thus, its cross section is maximized under the left-handed-electron-enhanced configuration $(P_{e^+},P_{e^-})=(+20\%,-80\%)$, reaching $2.292$~fb, and drops precipitously to $0.17$~fb under the $(-20\%,+80\%)$ configuration.

Conversely, the hypercharge gauge-portal operator $\mathcal{O}^{(6)}_{B\phi}$ isolates the $U(1)_Y$ hypercharge current. The right-handed electron is an $SU(2)_L$ singlet that carries twice the hypercharge magnitude ($|Y(e_R)|$) of the left-handed doublet ($|Y(e_L)|$). Because the interaction strength scales with the square of the hypercharge, the $s$-channel $B$-field exchange couples inherently more strongly to right-handed electrons. This explains why the $\mathcal{O}^{(6)}_{B\phi}$ cross section is instead optimized by the right-handed configuration $(P_{e^+},P_{e^-})=(-20\%,+80\%)$, attaining $0.769$~fb. The hypercharge dipole operator $\mathcal{O}^{(5)}_{B\chi}$ behaves analogously for the same reason, increasing from $0.199$~fb unpolarized to $0.3354$~fb for the $(-20\%,+80\%)$ polarization configuration.

For the leptophilic scalar operator $\mathcal{O}^{(6)}_{L\phi}$, configuration $(P_{e^+},P_{e^-})=(+20\%,+80\%)$ yields the best signal-to-background ratio among those considered. This improvement arises from the combined polarization dependence of the signal amplitude and the suppression of the $t$-channel $W$-exchange background contributions by a predominantly right-handed electron beam.

\begin{figure}[htbp!]
	\centering
	\includegraphics[width=0.6\linewidth]{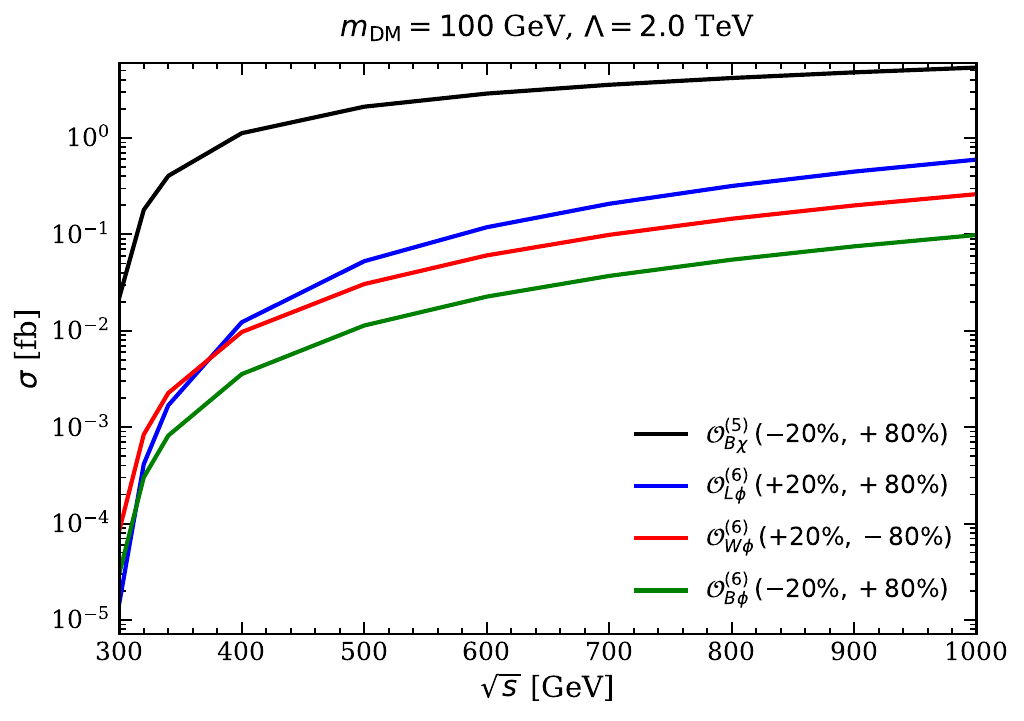}
	\caption{The variation of $e^+e^-\rightarrow {\rm DM\, DM}\, Z$ production cross sections as a function of the center-of-mass energy $\sqrt{s}$ for the DMEFT operators. The common reference point is $m_{\rm DM}=100$~GeV and $\Lambda=2.0$~TeV. The beam-polarization configuration $(P_{e^+},P_{e^-})$ used for each operator is indicated in the legend.}
	\label{fig:xsec_sqrtS}
\end{figure}
\paragraph{CM energy dependence:}
The $\sqrt{s}$ dependence of the production process $e^+e^-\rightarrow {\rm DM\, DM}\, Z$ for various DMEFT operators is shown in figure~\ref{fig:xsec_sqrtS}. We evaluate cross sections at a common reference point, $m_{\rm DM}=100$~GeV and $\Lambda=2.0$~TeV, with the beam-polarization configuration optimized separately for each operator. The dipole operator $\mathcal{O}^{(5)}_{B\chi}$ yields the largest cross section across the considered energy range and rises rapidly above threshold. The dimension-six operators $\mathcal{O}^{(6)}_{L\phi}$, $\mathcal{O}^{(6)}_{W\phi}$, and $\mathcal{O}^{(6)}_{B\phi}$ also exhibit increasing cross sections with $\sqrt{s}$, driven by the momentum dependence of the effective interactions. The different cross sections encode the distinct Lorentz and Electroweak structures of these operators.

\subsection{Event Selection}
\label{sec:selection}
We reconstruct the visible $Z$ boson in four mutually exclusive decay
channels:
\begin{equation}
	Z\rightarrow e^+e^-,
	\qquad
	Z\rightarrow \mu^+\mu^-,
	\qquad
	Z\rightarrow jj,
	\qquad
	Z\rightarrow b\bar b.
\end{equation}
The first two channels benefit from a clean reconstruction of the leptonic $Z$ decay, whereas the hadronic modes profit from the larger hadronic branching fraction. The event samples are made exclusive in order to avoid double counting between the light-jet and $b$-tagged-jet categories. 
The DMEFT operators are implemented at the Lagrangian level using \textsc{FeynRules}~\cite{Alloul:2013bka}, and the resulting Universal FeynRules Output (UFO)~\cite{Degrande:2011ua} model is imported into \textsc{MadGraph5\_aMC@NLO}~\cite{Alwall:2014hca}. Signal and background events are generated with \textsc{MadGraph5\_aMC@NLO}, showered and hadronized using \textsc{Pythia\,8}~\cite{Sjostrand:2014zea}, and subsequently passed through a fast detector simulation based on \textsc{Delphes\,3}~\cite{deFavereau:2013fsa}. The default \textsc{Delphes} card supplied with \textsc{MadGraph5\_aMC@NLO} is used to simulate detector resolution, object reconstruction, and tagging efficiencies. Jets are reconstructed with the anti-$k_T$~\cite{Cacciari:2008gp} algorithm using a radius parameter $R=0.4$ and a minimum transverse momentum $p_T^{\text{jet}} > 20~\text{GeV}$, and events containing isolated photons are vetoed.
\begin{table}[htb!]
\centering
\begin{tabular}{l l}
\hline
Channel & Selection requirements \\ \hline
Di-electron ($ee$) & $N_{e} = 2, N_{\mu} = 0, N_{jet} = 0, N_{\gamma} = 0$ \\
Di-muon ($\mu\mu$) & $N_{e} = 0, N_{\mu} = 2, N_{jet} = 0, N_{\gamma} = 0$ \\
Light jets ($jj$) & $N_{e} = 0, N_{\mu} = 0, N_{jet} = 2, N_{btag} = 0, N_{\gamma} = 0$ \\
Heavy jets ($bb$) & $N_{e} = 0, N_{\mu} = 0, N_{jet} = 2, N_{btag} = 2, N_{\gamma} = 0 $ \\ \hline
Kinematic cuts & value \\ \hline
Missing transverse momentum & $p_{T}^{miss} > 25$ GeV \\
Angular separation & $\Delta R_{12} < 3.0$ \\
Invariant mass & $80\,\text{GeV}\, < M_{12} < 100$ GeV \\ \hline
\end{tabular}
\caption{Basic event selection criteria for mono-$Z+\text{missing energy}$ topology. $M_{12}$ and $\Delta R_{12}$ are the invariant mass of the two reconstructed objects from the $Z$-boson decay and their angular separation in the $\eta$–$\phi$ plane, respectively.}
\label{tab:selection_criteria}
\end{table}
The basic object-multiplicity requirements and the kinematic selection are given in table~\ref{tab:selection_criteria}. In the leptonic channels, the selected same-flavor opposite-sign lepton pair is used to reconstruct the $Z$ candidate. In the hadronic channels, the two selected jets are required to reconstruct the $Z$ boson. The requirement of invariant mass of two visible objects (either a jet pair or a lepton pair)
\begin{equation*}
	80~{\rm GeV}<M_{12}<100~{\rm GeV},
\end{equation*}
suppresses non-resonant backgrounds and ensures that the visible system is consistent with an on-shell $Z$ boson. 
The requirement of missing transverse momentum $p_T^{\rm miss}>25$~GeV further suppresses SM backgrounds that lack an intrinsic source of missing energy. The angular separation $\Delta R_{12} = \sqrt{(\Delta\eta)^2 + (\Delta\phi)^2}$, where $\Delta\eta$ and $\Delta\phi$ are the differences in pseudorapidity and azimuthal angle between the two reconstructed objects, is required to satisfy $\Delta R_{12}<3.0$, ensuring that the two reconstructed objects are compatible with the $Z$ decay while rejecting widely separated pairs.

\subsection{Scalar and Dirac Fermion Operators}
\label{subsec:sensitivity}
Among the complete set of scalar and Dirac fermion DMEFT operators illustrated in figure~\ref{fig:dm1} and table~\ref{tab:ops}, several are already ruled out by non-collider observations. As established in Sec.~\ref{sec:dpheno}, the Dirac operators $\mathcal{O}_{D\chi}^{(6)}$ and $\mathcal{O}_{e\chi}^{(6)}$ are entirely excluded by spin-independent direct detection limits. The leptophilic fermion operator $\mathcal{O}_{\ell\chi}^{(6)}$ remains cosmologically viable only for DM masses $m_{\rm DM} > 520$~GeV, rendering pair production kinematically inaccessible at the $\sqrt{s} = 1$~TeV ILC. Consequently, our collider analysis focus on the remaining viable operators: the fermion dipole operator $\mathcal{O}^{(5)}_{B\chi}$, the scalar leptophilic operator $\mathcal{O}^{(6)}_{L\phi}$, and the gauge-portal operators $\mathcal{O}^{(6)}_{W\phi}$ and $\mathcal{O}^{(6)}_{B\phi}$. The collider analysis for the operators $\mathcal{O}^{(6)}_{D\phi}$ and $\mathcal{O}^{(6)}_{DX}$ are given in Subsec.~\ref{colli-Higgs}.

The signal and background distributions for $\mathcal{O}^{(5)}_{B\chi}$ operator are shown in figure~\ref{fig:dists2}, while the corresponding distributions for $\mathcal{O}^{(6)}_{L\phi}$ and $\mathcal{O}^{(6)}_{B\phi}$ are displayed in figures~\ref{fig:lphi7} and \ref{fig:Bphi}, respectively. These distributions clearly demonstrate that a universal selection strategy for these operators is inadequate for isolating the mono-$Z$ signal from the SM background. Instead, the optimal kinematic cuts depend on the operator structure, and the beam-polarization configuration. The DMEFT signals deviate from the SM background in several distinct ways. The missing energy ($\slashed{E}$), missing transverse momentum ($p_T^{\rm miss}$), recoil mass ($m_{\rm rec}$), and $Z$-boson pseudorapidity ($\eta_Z$) provide effective discrimination between the DM signal and the SM background. The recoil mass of the invisible system against the reconstructed $Z$-boson is defined in equation~\ref{eq:recoil-mass}.

For the SM background, the missing energy distribution exhibits a prominent peak near $\slashed{E}\simeq 500$~GeV, which originates from $ZZ$ production, $e^+e^-\to ZZ$, with one $Z$-boson decaying invisibly, $Z\to\nu\bar{\nu}$. When both $Z$ bosons are approximately on-shell, they share the available collision energy, so the invisibly decaying $Z$ carries an energy close to $\sqrt{s}/2$, giving rise to this peak when $\sqrt{s}=1$ TeV. In addition, neutral-current diagrams in which an off-shell $Z$ produces a neutrino pair and the visible $Z$ is radiated from the outgoing neutrino or antineutrino line contribute to the non-resonant $Z\nu\bar{\nu}$ continuum.

The high missing energy tail, $800\lesssim\slashed{E}\lesssim900$~GeV, is predominantly populated by the charged-current $t$-channel $W$-exchange contribution to $e^+e^-\to\nu_e\bar{\nu}_e Z$. This gauge-invariant set includes diagrams with the visible $Z$ radiated from the incoming electron or positron line, from the outgoing neutrino or antineutrino line, and from the internal $W$ propagator via the $WWZ$ gauge vertex. The $t$-channel $W$-exchange topologies favor a relatively soft visible $Z$, allowing the neutrino pair to carry a large fraction of the beam energy, and thus dominate the large-$\slashed{E}$ region. We discuss next the collider sensitivity of each DM operator that survives cosmological constraints. 

\FloatBarrier
\subsubsection{Fermion Dipole Operator}
\label{subsub:BChi}
\begin{figure}[htbp!]
	\centering
	\includegraphics[width=0.475\linewidth]{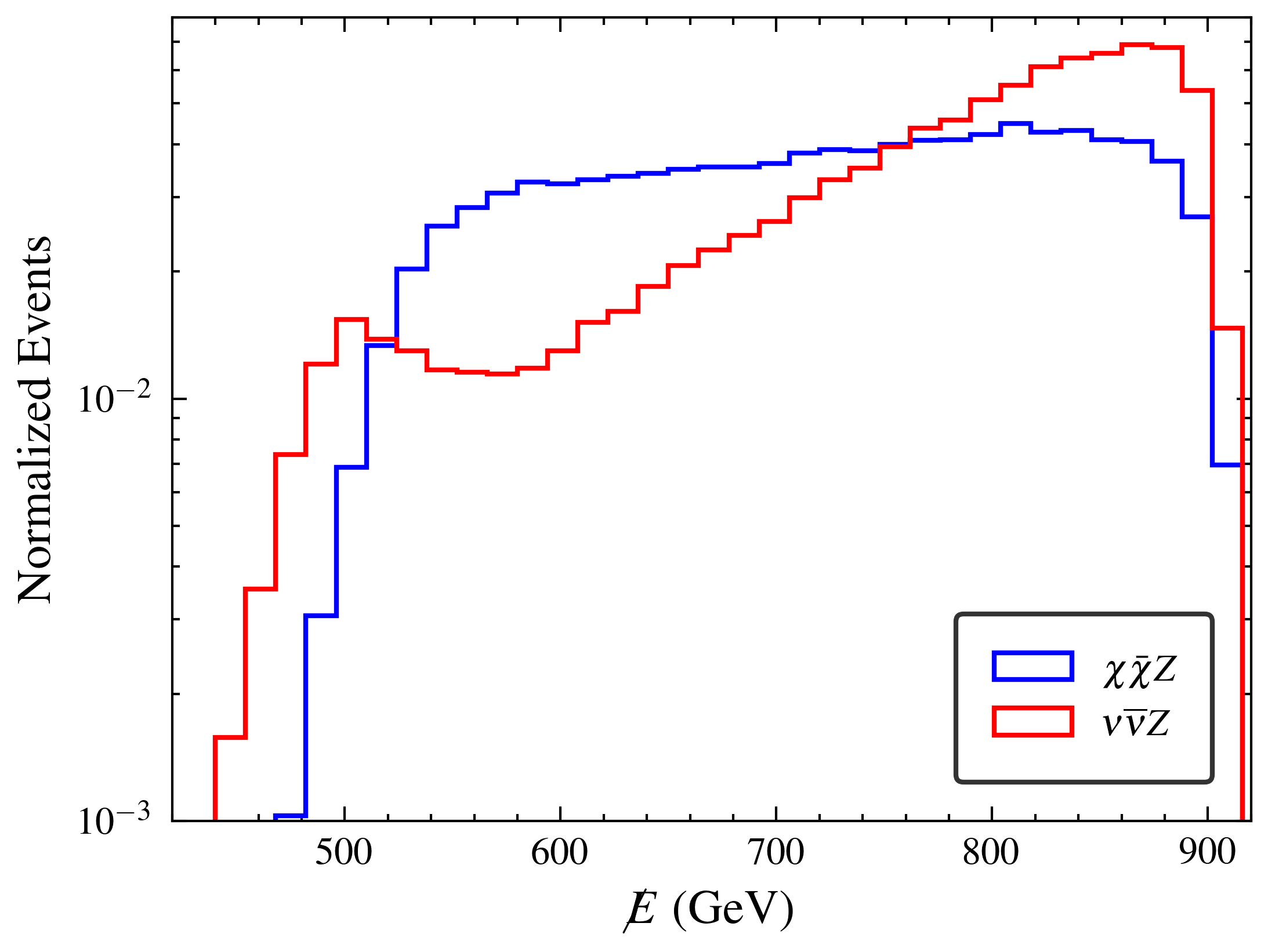}
	\includegraphics[width=0.475\linewidth]{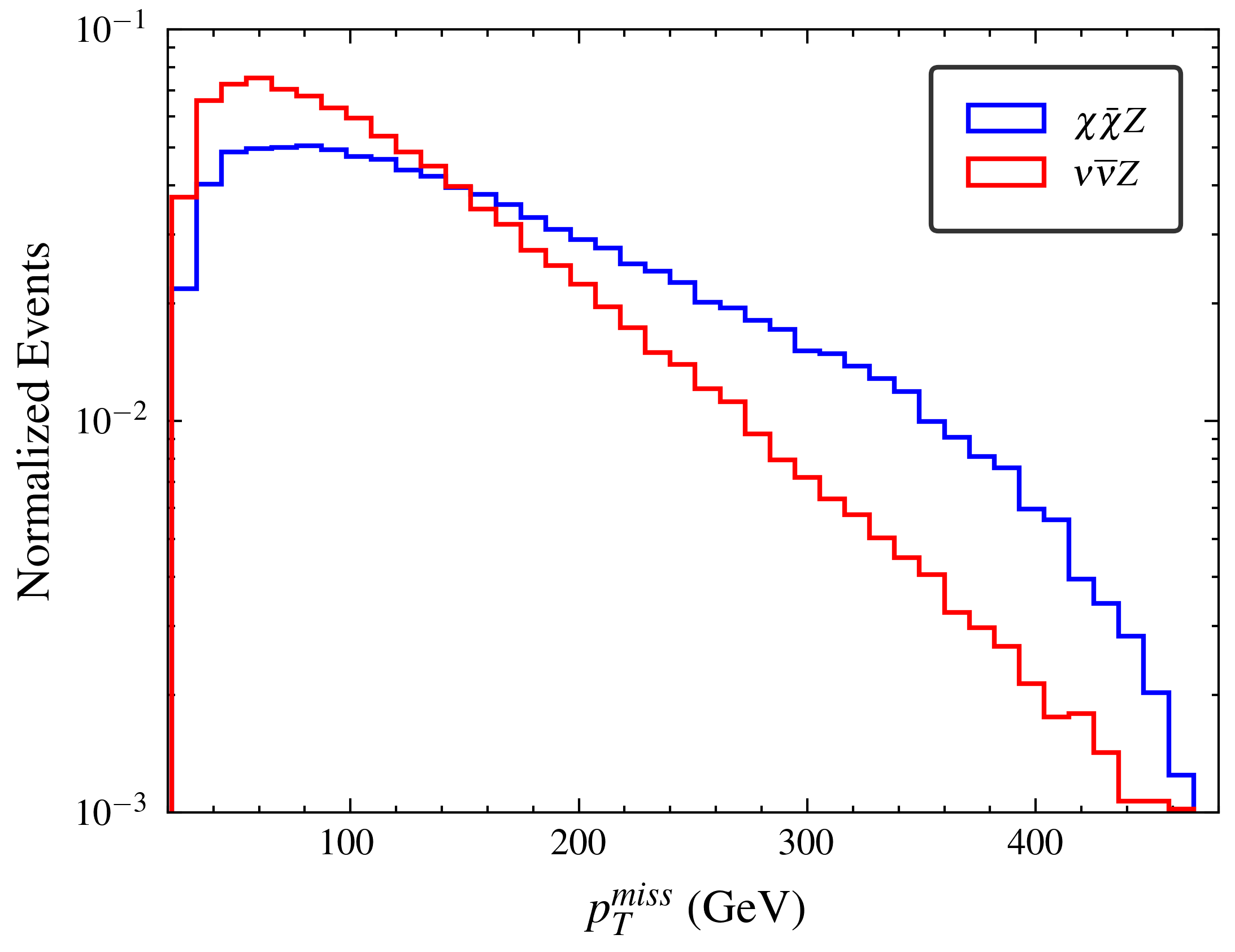}
	\includegraphics[width=0.475\linewidth]{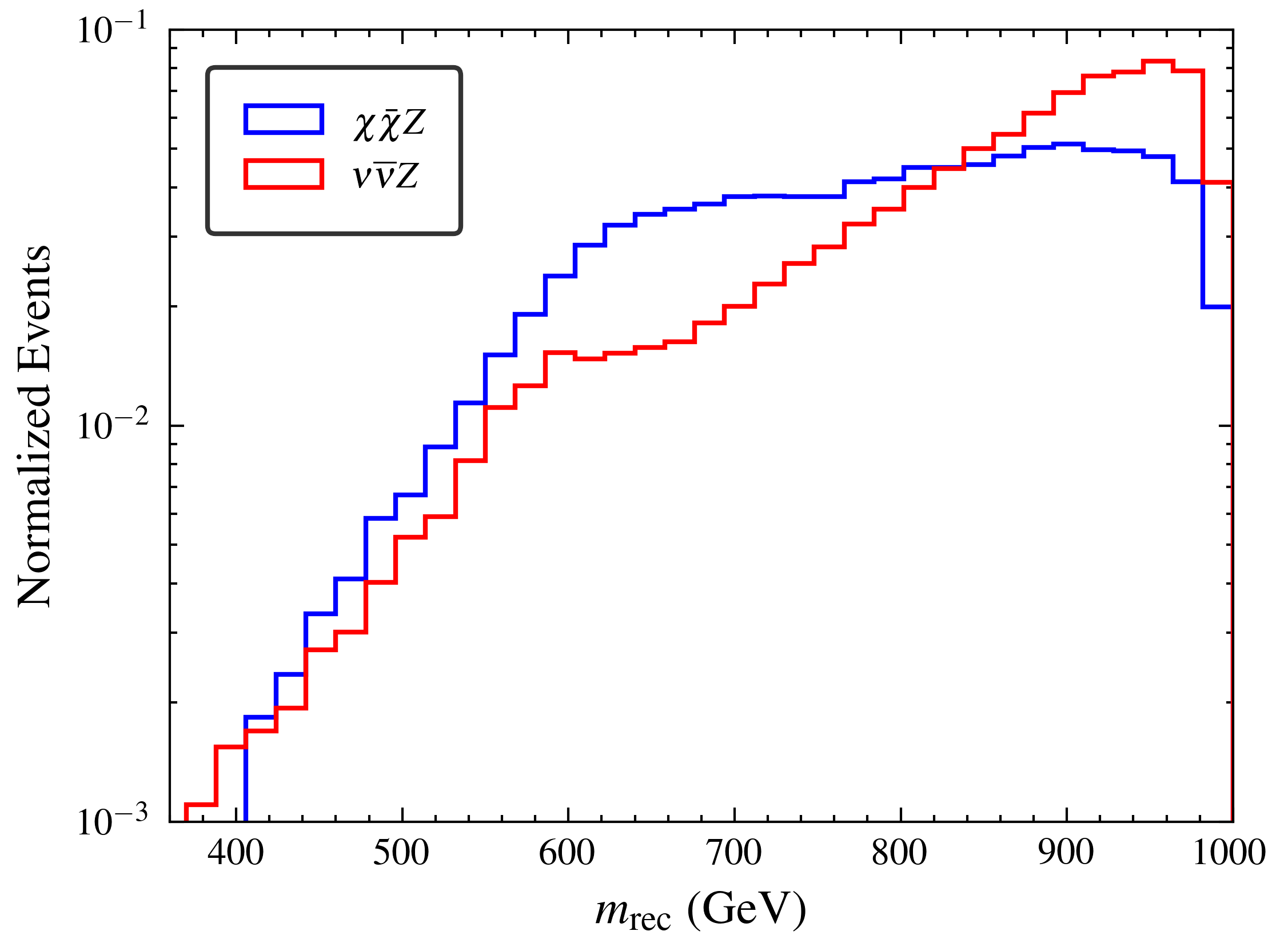}
	\includegraphics[width=0.475\linewidth]{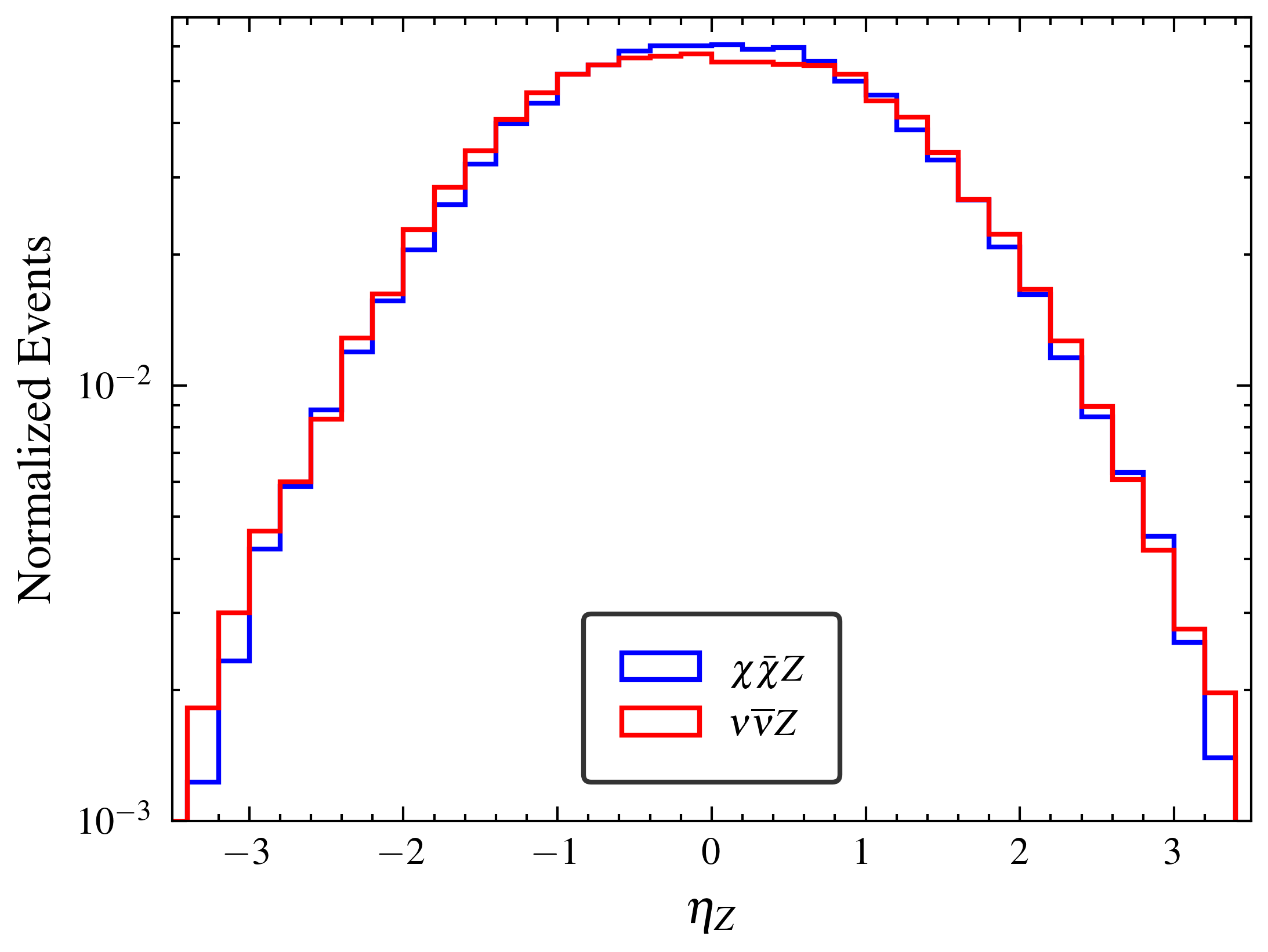}
	\caption{Kinematic distributions of the signal ($\mathcal{O}^{(5)}_{B\chi}$, $\Lambda=7.3$~TeV, $m_{\rm DM}=100$~GeV) and background processes at a 1~TeV $e^+ e^-$ collider with unpolarized beams, after applying the event selection criteria described in Subsec.~\ref{sec:selection}. The signal benchmark is compatible with cosmological constraints.}
	\label{fig:dists2}
\end{figure}
\begin{table}[htbp!]
	\centering
	\renewcommand{\arraystretch}{1.0}{
		\begin{tabular}{|>{\centering\arraybackslash}p{2.6cm}|
				>{\centering\arraybackslash}p{1.8cm}|
				>{\centering\arraybackslash}p{1.8cm}|
				>{\centering\arraybackslash}p{1.5cm}|
				>{\centering\arraybackslash}p{1.8cm}|
				>{\centering\arraybackslash}p{1.8cm}|
				>{\centering\arraybackslash}p{1.5cm}|}
			\hline 
			\multirow{3}*{Cuts} & \multicolumn{6}{c|}{Number of events} \\ \cline{2-7}
			& \multicolumn{3}{c|}{Unpolarized} & \multicolumn{3}{c|}{$\left(P_{e^+}, P_{e^-}= -20\%, +80\%\right)$} \\ \cline{2-7}
			& Signal ($S$) & Background ($B$) & $\frac{S}{\sqrt{B}}$ & Signal ($S$) & Background ($B$) & $\frac{S}{\sqrt{B}}$ \\ \hline
			Basic event  & 379.8 & $1.61 \times 10^{6}$& \textbf{0.299} & 639.44 & 313225 & \textbf{1.143}\\
			selection         &  [100\%] & [100\%] & &[100\%] & [100\%] &\\
			\hline
			\multirow{2}*{$\slashed{E}< 800$ GeV} & 267.8 & 859239 & 0.289 & 452.73 & 193667 & 1.029 \\
			&  [70.51\%] & [53.33\%] & &[70.8\%] & [61.83\%] & \\
			\hline
			\multirow{2}*{$p_{T}^{\rm miss} > 100$ GeV} & 216.7 & 633341 & 0.272 & 365.64 & 142983 & 0.967 \\
			&  [57.07\%] & [39.31\%] &  &[57.18\%] & [45.65\%] &\\
			\hline
			\multirow{2}*{$m_{\rm rec} < 850$ GeV} & 197.4 & 542894 & 0.268 & 333.44 & 128223 & 0.931 \\
			&  [51.97\%] & [33.69\%] & &[52.14\%] & [40.94\%] &\\
			\hline
			Significance, $\mathcal{Z}$ & \multicolumn{3}{c|}{\textbf{0.299}} & \multicolumn{3}{c|}{\textbf{1.143}} \\ \hline
	\end{tabular}}
	\caption{Cut-flow table for signal ($\mathcal{O}^{(5)}_{B\chi}$, $\Lambda=7.3$~TeV, $m_{\rm DM}=100$~GeV) and background, together with the statistical significance $S/\sqrt{B}$, at a 1~TeV ILC with $\mathcal{L}_{\rm int}=8$~ab$^{-1}$, for unpolarized and polarized $(-20\%, +80\%)$ beams. The signal benchmark is compatible with cosmological constraints.}
	\label{tab:cutflow}
\end{table}

The dimension-5 hypercharge dipole operator, $\mathcal{O}^{(5)}_{B\chi} = (\overline{\chi} \sigma_{\mu\nu}\chi) B^{\mu\nu}$, contributes to the mono-$Z$ signature through both initial-state and final-state radiation topologies. For collider simulation, we consider the benchmark point compatible with the cosmological constraint:
$$ m_\chi = 100\text{ GeV}, \qquad \Lambda = 7.3\text{ TeV}. $$
To ensure the theoretical consistency of the effective field theory, we restrict the collider analysis to regions where the new physics scale satisfies $\Lambda > \sqrt{s} = 1$~TeV. Applying the baseline event selection described in Subsec.~\ref{sec:selection}, the normalized kinematic distributions for the signal (blue) and background (red) are displayed in figure~\ref{fig:dists2} for unpolarized beams. The kinematic observables:
\begin{itemize}
\item \textbf{Pseudorapidity ($\eta_Z$):} The lower-right panel of figure~\ref{fig:dists2} demonstrates that the $Z$-boson pseudorapidity distributions for the signal and background are remarkably similar, both peaking symmetrically in the central region.
\item \textbf{Missing energy ($\slashed{E}$) and recoil mass ($m_{\rm rec}$):} The invariant mass of the invisible system is kinematically bounded from below by $m_{\rm rec} \ge 2m_\chi = 200$~GeV. Consequently, both the recoil mass and missing energy spectra for the signal start at a higher threshold than the massless neutrino background. The SM background exhibits a sharp resonance peak at $\slashed{E} \approx 500$~GeV from on-shell $ZZ$ production, followed by a steep rise toward the kinematic boundary ($\sim 850$~GeV) driven by the $t$-channel $W$-exchange contribution. In contrast, the signal distributions in both $\slashed{E}$ and $m_{\rm rec}$ are nearly flat over an extended interval, $\slashed{E} \sim 550\text{--}870$~GeV and $m_{\rm rec} \sim 650\text{--}900$~GeV.
	
This broad, nearly flat plateau in the missing energy and recoil mass distributions reflects the interplay between the momentum-dependent dipole vertex and the non-resonant three-body phase space. Because the operator couples through the hypercharge field-strength tensor, the scattering amplitude grows with energy and preferentially populates configurations with a large invariant mass of the invisible $\chi\bar{\chi}$ system. At the same time, the three-body kinematics of $e^+e^- \to Z\chi\bar{\chi}$ at $\sqrt{s}=1$~TeV suppresses configurations in which the $Z$ boson carries most of the beam energy, since these correspond to a small invariant mass of the DM pair and lie close to the kinematic endpoint. The combination of these two yields an approximately uniform distribution in recoil mass and missing energy over a broad range, rather than a peaked structure.
\item Missing transverse momentum ($p_T^{\rm miss}$) for most of the events in the signal and background is relatively soft.

\end{itemize}

The detailed cut-flow after applying the basic event selection (see table~\ref{tab:selection_criteria}) and subsequent kinematic cuts is presented in table~\ref{tab:cutflow}. The left and right side of table~\ref{tab:cutflow} are corresponding to the unpolarized and polarized $(P_{e^+}, P_{e^-}) = (-20\%, +80\%)$ beam configurations respectively. This specific polarization directly exploits the Electroweak structure of the dipole operator to simultaneously enhance the signal cross section and suppress the dominant $t$-channel $W$-exchange background, immediately elevating the statistical significance to $\mathcal{Z} = 1.143$ at an integrated luminosity $\mathcal{L}_{\rm int} = 8~{\rm ab}^{-1}$. For naive estimation, the statistical significance is defined as
\begin{equation}
	\mathcal{Z} = \frac{S}{\sqrt{B}},
\end{equation}
where $S$ and $B$ denote the number of signal and background events surviving selection, respectively. The resulting projected sensitivity contours in the $(m_\chi, \Lambda)$ plane are shown in the left panel of figure~\ref{fig:excl1} for the polarized configuration $(P_{e^+}, P_{e^-}) = (-20\%, +80\%)$. The red, blue, and green dash-dotted curves denote the $2\sigma$ exclusion, $3\sigma$ evidence, and $5\sigma$ discovery reaches, respectively. 

The projected sensitivity covers a remarkably broad range of DM masses. For a light DM candidate with $m_\chi = 10$~GeV, the $2\sigma$ exclusion reaches a cutoff scale of $\Lambda \simeq 5.57$~TeV, while the $3\sigma$ evidence and $5\sigma$ discovery reach extend to $\Lambda \simeq 4.55$~TeV and $\Lambda \simeq 3.53$~TeV, respectively. At $m_\chi = 300$~GeV, the corresponding limits are $\Lambda \simeq 4.99$~TeV ($2\sigma$), $4.07$~TeV ($3\sigma$), and $3.15$~TeV ($5\sigma$). For a heavier mass of $m_\chi = 400$~GeV, the sensitivity remains substantial, with $\Lambda \simeq 4.32$~TeV ($2\sigma$), $3.53$~TeV ($3\sigma$), and $2.73$~TeV ($5\sigma$). As the DM mass approaches the kinematic boundary, $m_\chi \to \sqrt{s}/2 = 500$~GeV, the available phase space becomes highly suppressed, the production rate drops sharply, and the sensitivity contours fall steeply. It is noteworthy that, from the LHC recasting analysis in Sec.~\ref{sec:lhc-recast}, the $2\sigma$ exclusion for $\mathcal{O}_{B\chi}^{(5)}$ extends only to $\Lambda \approx 460$~GeV and $420$~GeV for DM masses of $100$~GeV and $200$~GeV, respectively. Thus, the future lepton collider provides a significantly improved reach for this operator.

\FloatBarrier
\subsubsection{Scalar Leptophilic Operator}
\label{subsub:LPhi}
\begin{figure}[htbp!]
	\centering
	\includegraphics[width=0.475\linewidth]{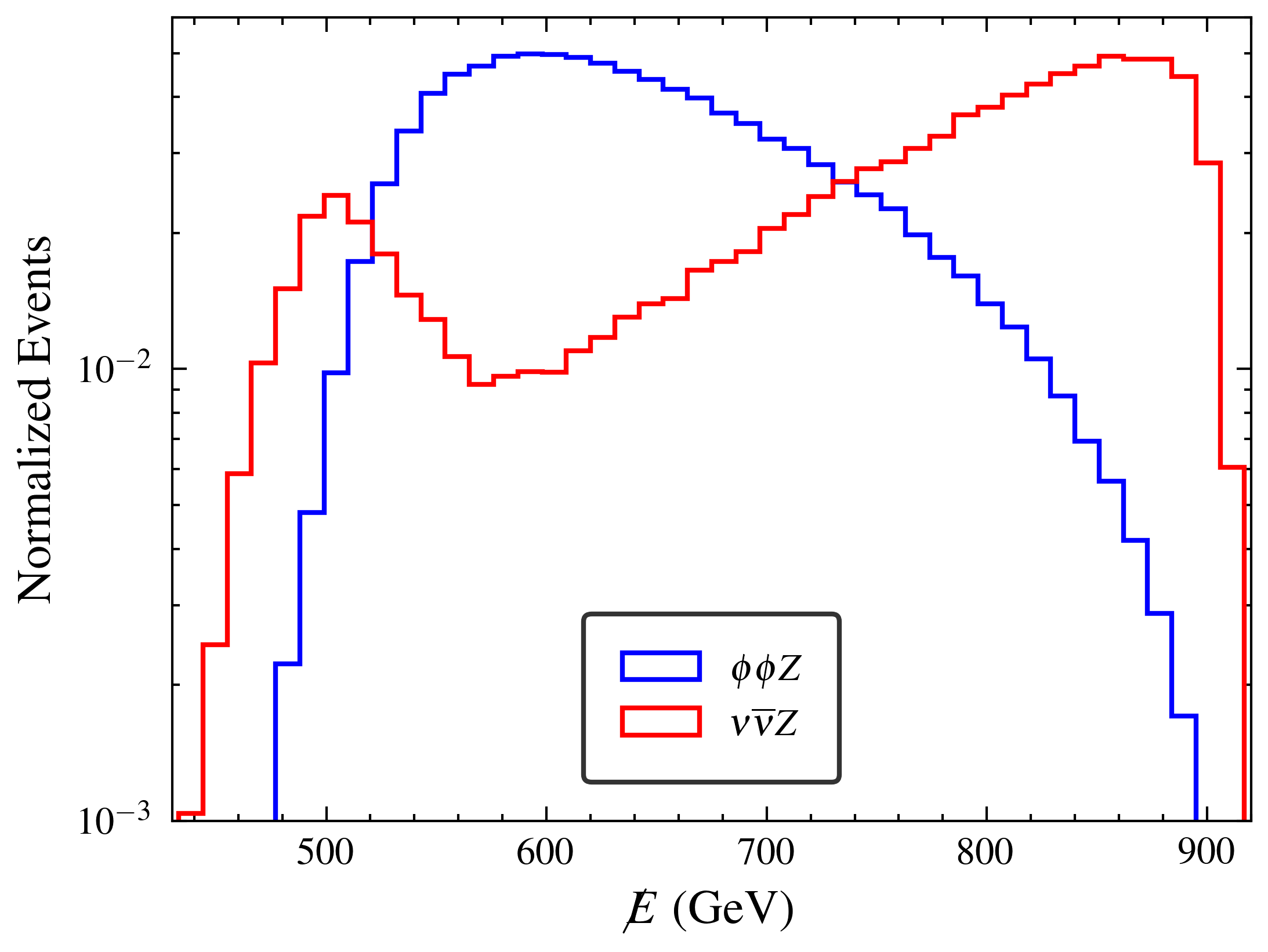}
	\includegraphics[width=0.475\linewidth]{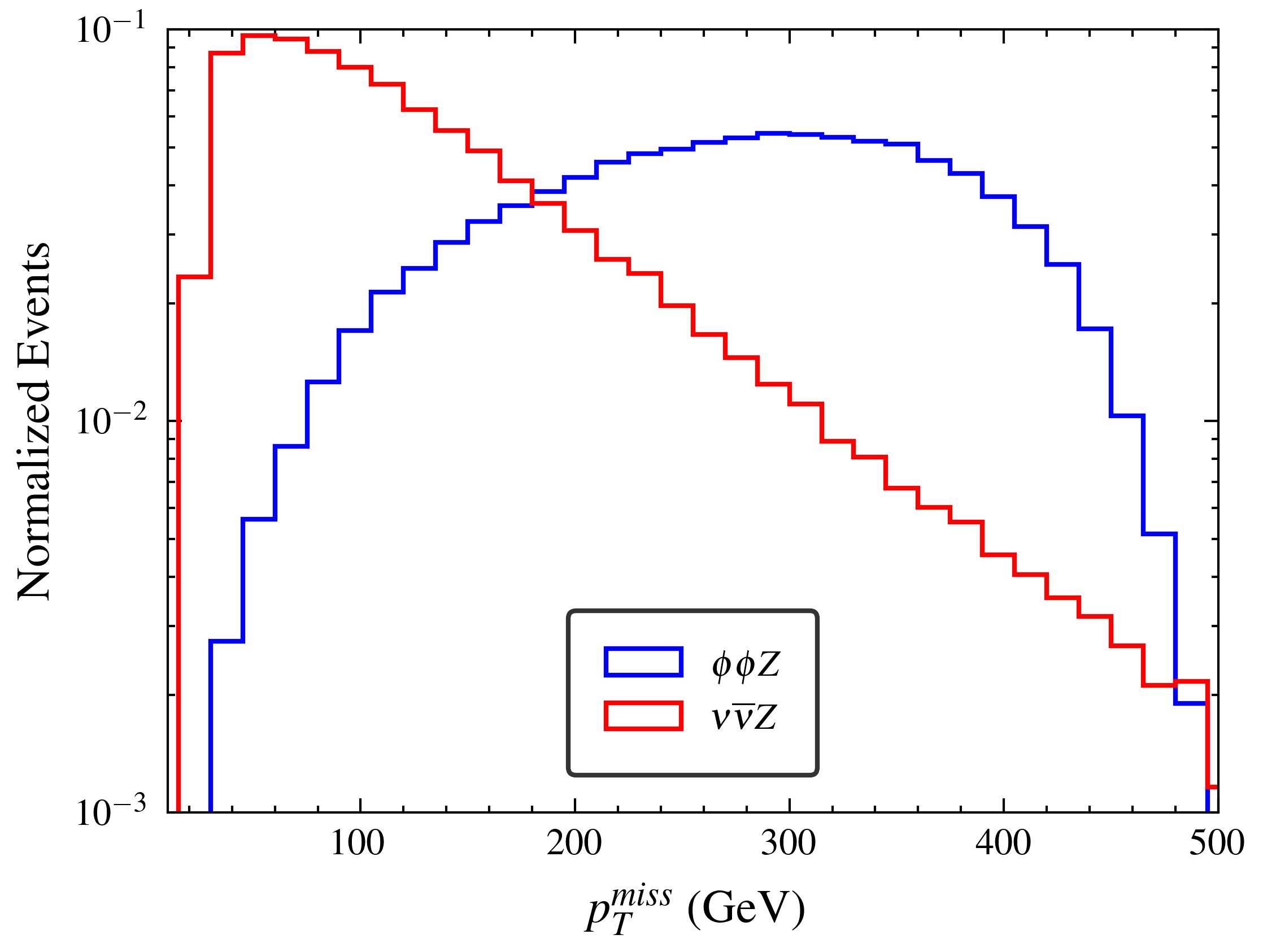}
	\includegraphics[width=0.475\linewidth]{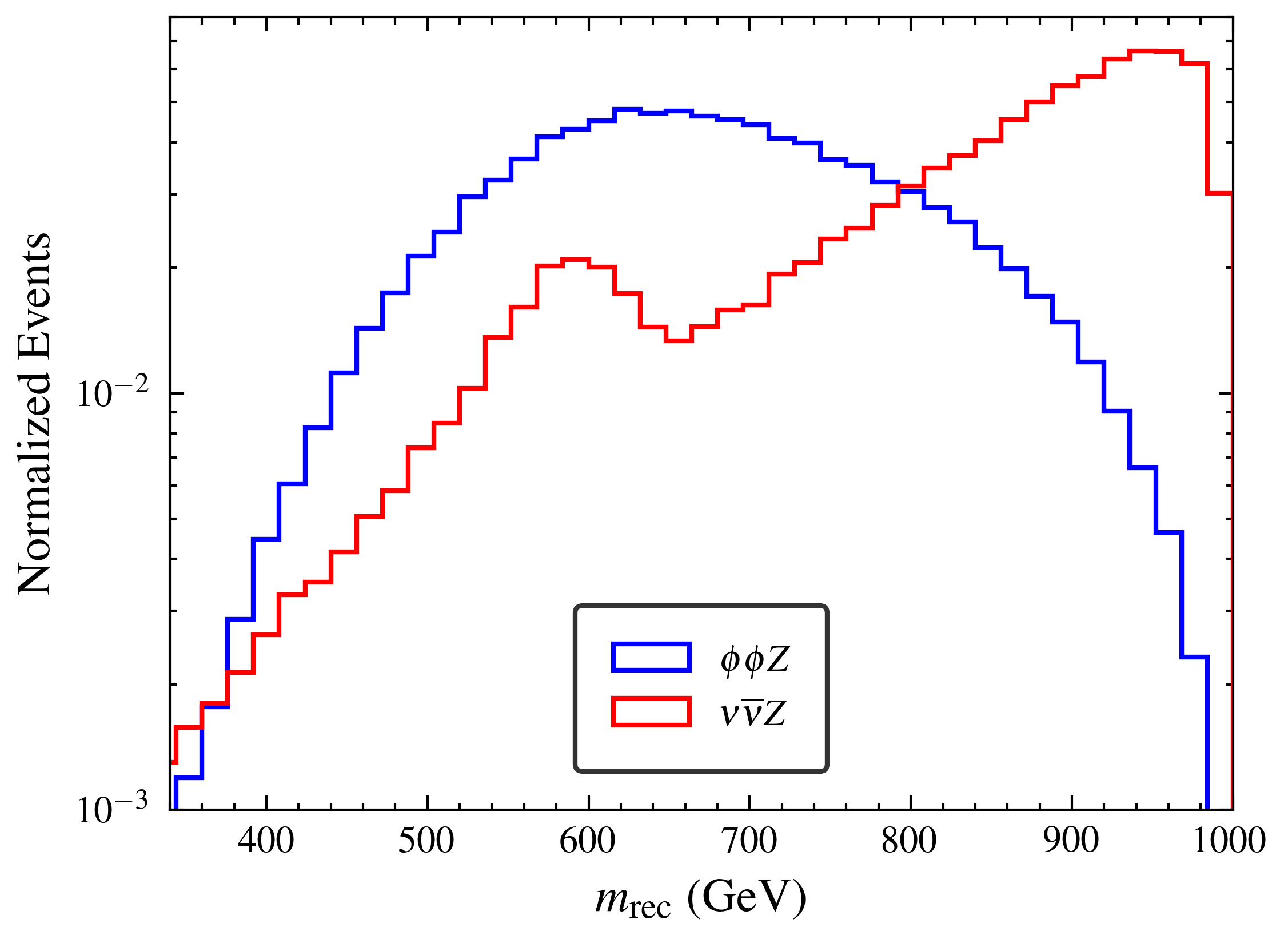}
	\includegraphics[width=0.475\linewidth]{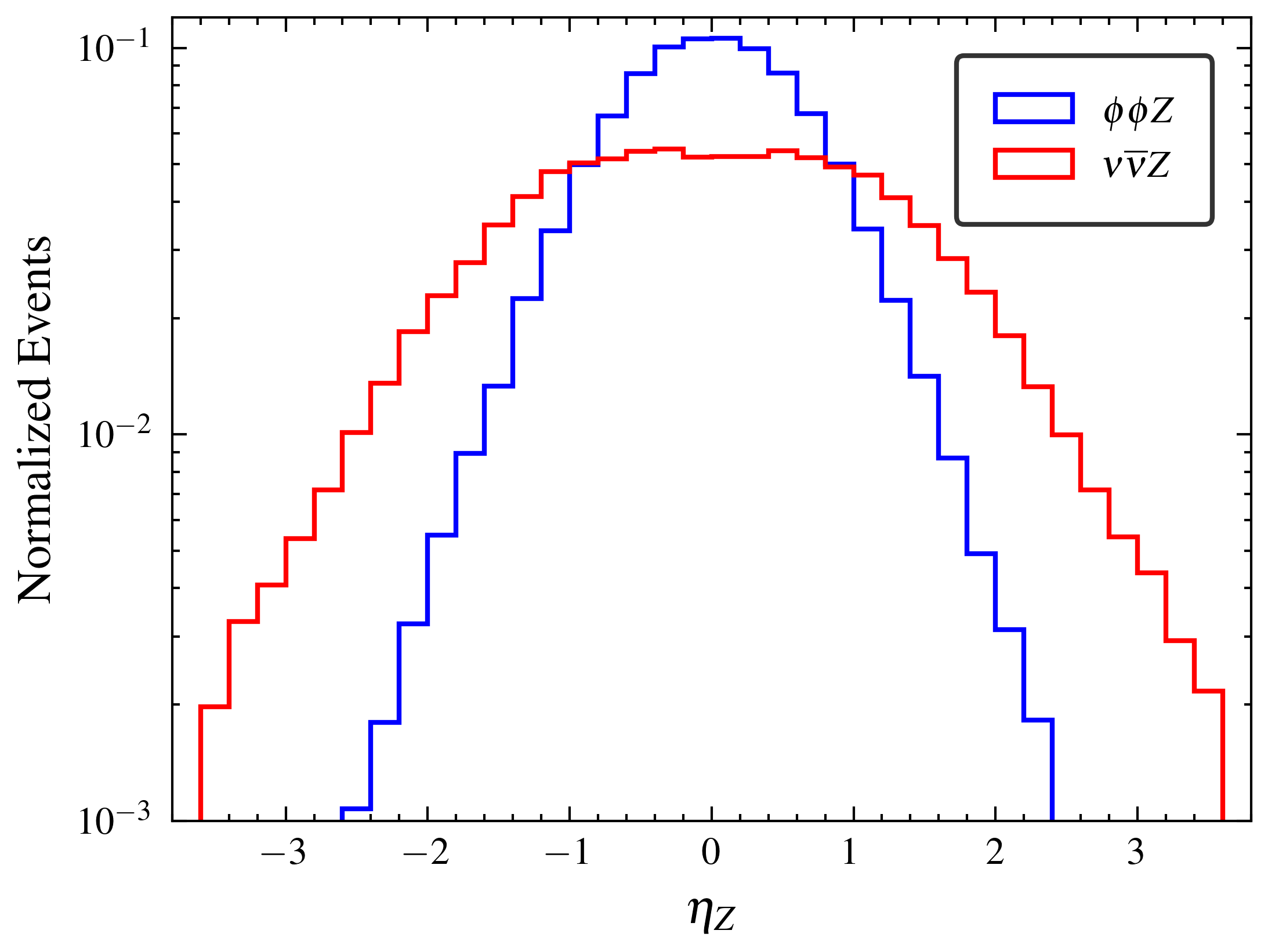}
	\caption{Kinematic distributions of the signal ($\mathcal{O}^{(6)}_{L\phi}$, $\Lambda=2.0$ TeV, $m_{\rm DM}=100$ GeV) and background processes at a 1~TeV $e^+ e^-$ collider with polarized beam configurations, $P_{e^+}, P_{e^-}= +20\%, +80\%$, after applying the event selection criteria described in Subsec.~\ref{sec:selection}. The signal benchmark is compatible with cosmological constraints.}
	\label{fig:lphi7}
\end{figure}
\begin{table}[htbp!]
	\centering
	\renewcommand{\arraystretch}{1.0}{
		\begin{tabular}{|>{\centering\arraybackslash}p{2.6cm}|
				>{\centering\arraybackslash}p{1.8cm}|
				>{\centering\arraybackslash}p{1.8cm}|
				>{\centering\arraybackslash}p{1.5cm}|
				>{\centering\arraybackslash}p{1.8cm}|
				>{\centering\arraybackslash}p{1.8cm}|
				>{\centering\arraybackslash}p{1.5cm}|}
			\hline 
			\multirow{3}*{Cuts} & \multicolumn{6}{c|}{Number of events} \\ \cline{2-7}
			& \multicolumn{3}{c|}{Unpolarized} & \multicolumn{3}{c|}{$P_{e^+}, P_{e^-}= +20\%, +80\%$} \\ \cline{2-7}
			& Signal ($S$) & Background ($B$) & $\frac{S}{\sqrt{B}}$ & Signal ($S$) & Background ($B$) & $\frac{S}{\sqrt{B}}$ \\ \hline
			Basic event  & 1331.11 & $1.61 \times 10^{6}$ & 1.05 & 1538.61 & 419540.0 & 2.37 \\
			selection         &  [100\%] & [100\%] & & [100\%] & [100\%] &\\
			\hline
			\multirow{2}*{$\slashed{E}< 740$ GeV} & 1103.86 & 561687.0 & 1.47 & 1276.75 & 169269.0 & 3.10 \\
			&  [82.93\%] & [34.86\%] & & [82.98\%] & [40.34\%] & \\
			\hline
			\multirow{2}*{$p_{T}^{\rm miss} > 160$ GeV} & 1026.14 & 311853.0  & 1.84 & 1185.57 & 94944.2  & 3.84 \\
			&  [77.09\%] & [19.35\%] &  & [77.05\%] & [22.63\%] &\\
			\hline
			\multirow{2}*{$m_{\rm rec} < 750$ GeV} & 905.64 & 231328.0 & 1.88 & 1047.62 & 74909.5 & 3.83\\
			&  [68.04\%] & [14.36\%] & & [68.08\%]& [17.85\%] &\\
			\hline
			\multirow{2}*{$|\eta_Z|< 1.0$} & 784.3 & 143117.0 & 2.073 & 907.61 & 42785.2 & 4.39\\
			&  [58.92\%] & [8.88\%] & & [58.99\%] & [10.19\%] &\\
			\hline
			Significance, $\mathcal{Z}$ & \multicolumn{3}{c|}{2.073} & \multicolumn{3}{c|}{4.39} \\ \hline
	\end{tabular}}
	\caption{Same as table~\ref{tab:cutflow} with unpolarized beam configurations (left), and polarized $P_{e^+}, P_{e^-}= +20\%, +80\%$ beam comfiguration (right), for the operator $\mathcal{O}^{(6)}_{L\phi}$ with $\Lambda=2.0$ TeV, $m_{\rm DM}=100$ GeV, compatible with cosmological constraints.}
	\label{tab:cutflow-2}
\end{table}

The scalar leptophilic operator, $\mathcal{O}^{(6)}_{L\phi}=(\overline{\ell}_L H e_R)\phi^2+\mathrm{h.c.}$, induces the mono-$Z$ signature through a contact interaction between the initial-state leptons and the scalar DM pair, with the visible $Z$ boson radiated from the incoming lepton line. We consider the benchmark point $$m_\phi=100~{\rm GeV}, \qquad \Lambda=2.0~{\rm TeV}.$$ This benchmark satisfies the DM constraints discussed in Sec.~\ref{sec:dpheno} and reproduces the observed relic abundance. The analysis is performed at $\sqrt{s}=1$~TeV with an integrated luminosity of $\mathcal{L}_{\rm int}=8~{\rm ab}^{-1}$. For this operator, the beam-polarization configuration $(P_{e^+},P_{e^-})=(+20\%,+80\%)$ yields the most favorable signal-to-background ratio among the configurations considered. Although the signal cross section is only moderately enhanced relative to the unpolarized case, the predominantly right-handed electron beam strongly suppresses the dominant $t$-channel $W$-exchange background. The normalized distributions of the signal and the SM background for this polarized configuration are shown in figure~\ref{fig:lphi7}.

The kinematic distributions of the $\mathcal{O}^{(6)}_{L\phi}$ signal differ significantly from those of the SM background. The $t$-channel $W$-exchange background topologies preferentially produce a relatively soft $Z$ boson, while the neutrino pair carries a large fraction of the collision energy. As a result, the background is enhanced near the kinematic endpoints, $\slashed{E}\gtrsim800$~GeV and $m_{\rm rec}\gtrsim900$~GeV. Consequently, the requirements $\slashed{E} < 740\text{ GeV}$ and $m_{\rm rec} < 750\text{ GeV}$ efficiently eliminate this high-energy invisible component with minimal signal loss. Furthermore, because the signal exhibits a significantly harder missing transverse momentum spectrum than the background (figure~\ref{fig:lphi7}, top right), imposing $p_T^{\rm miss} > 160\text{ GeV}$ suppresses softer background events. Finally, the cut $\vert{}\eta_Z\vert{} < 1.0$ suppresses the background more than it suppresses the more central signal events.

The cut-flow analysis for the unpolarized and polarized configurations is presented in table~\ref{tab:cutflow-2}. For the preferred polarized configuration, $(P_{e^+},P_{e^-})=(+20\%,+80\%)$, the basic event selection yields $1538$ signal events over a background of $419540$ events, corresponding to a statistical significance of $\mathcal{Z}=2.37$. After applying the full sequence of kinematic selections, the signal yield is reduced to $907$ events, while the background decreases to $42785$ events. The resulting statistical significance, $\mathcal{Z}=S/\sqrt{B}=4.39$, is substantially larger than the corresponding unpolarized result, $\mathcal{Z}=2.07$. This improvement arises from the combined effect of beam polarization and kinematic selections that exploit the harder missing transverse momentum spectrum and more central angular distribution of the signal relative to the background.

Although the recoil mass distributions of the signal and background are visibly different, additional selection $m_{\rm rec}<750$~GeV does not improve the statistical significance. This is because $m_{\rm rec}$ is correlated with the missing energy observable, as is evident from equation~\ref{eq:recoil-mass}. The preceding requirement $\slashed{E}<740$~GeV already suppresses the high recoil mass region dominated by the $t$-channel $W$-exchange background, leaving limited additional discrimination from the $m_{\rm rec}$ selection.

\begin{figure}[htbp!]
	\centering
	\includegraphics[width=0.45\linewidth]{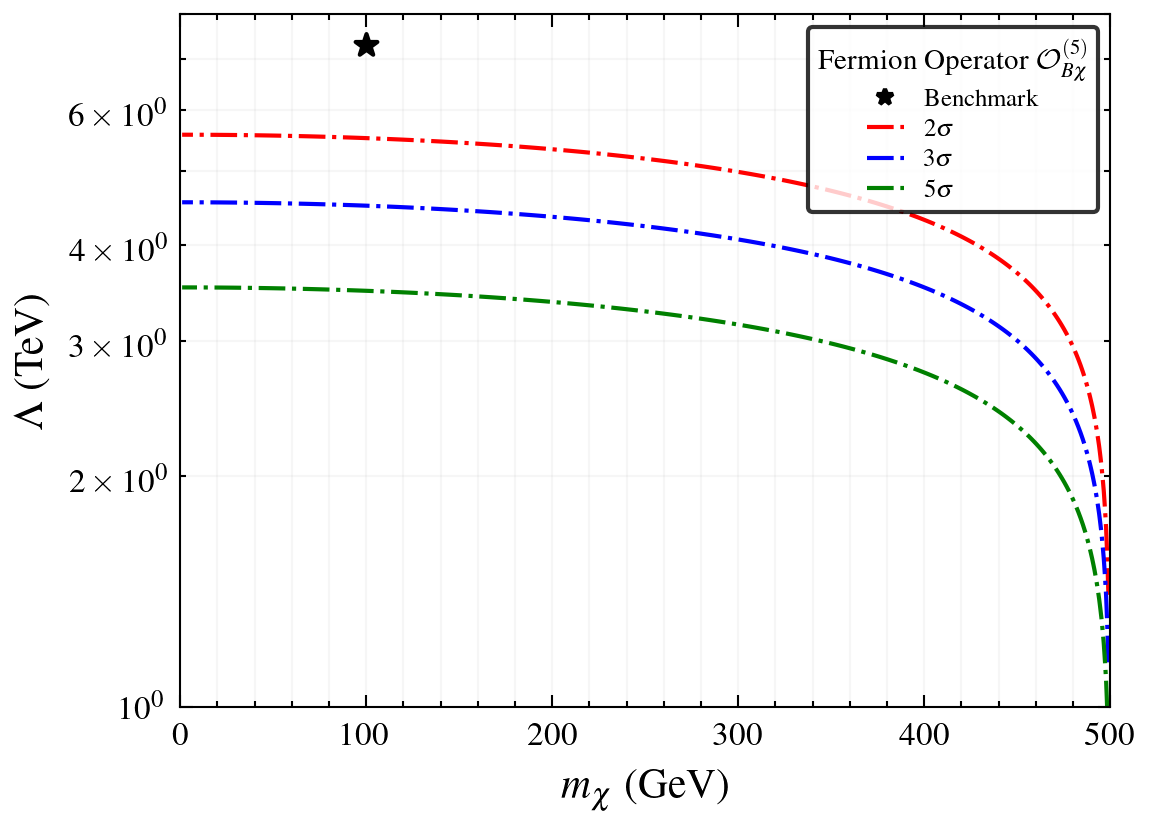}
	\includegraphics[width=0.45\linewidth]{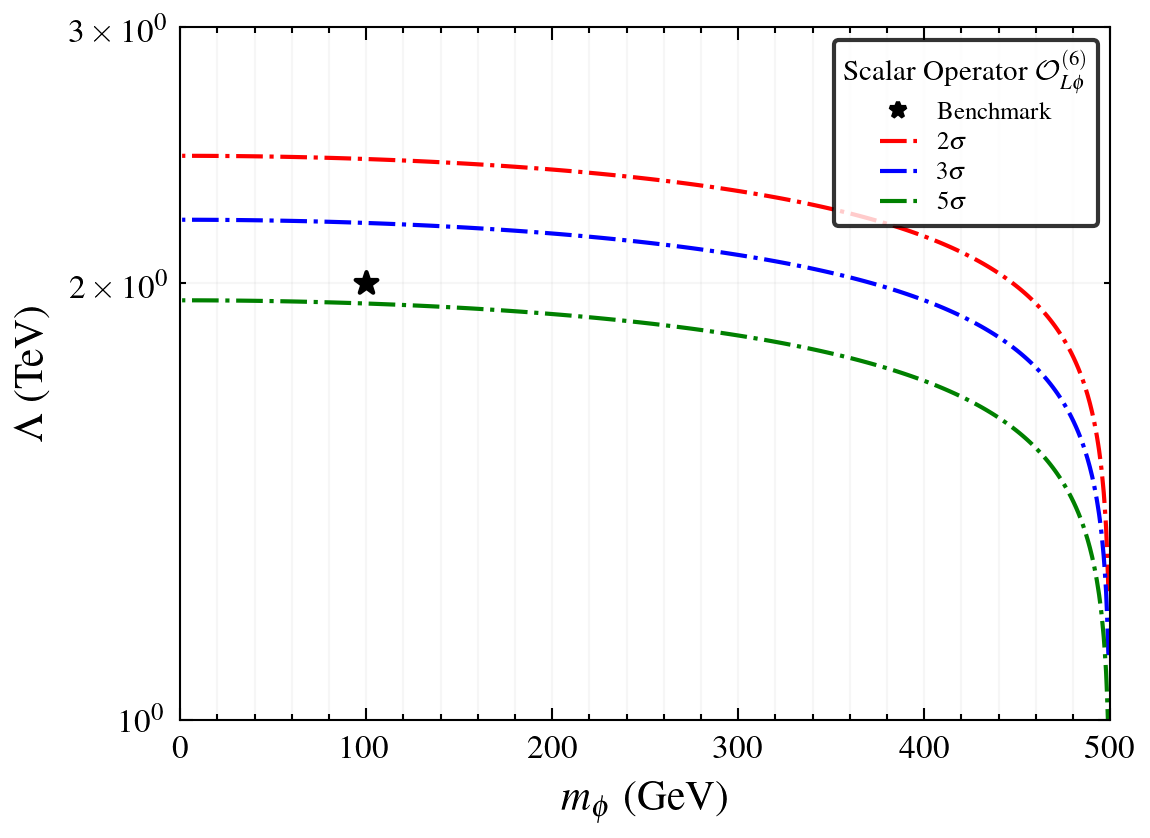}
	\caption{Statistical significance of the signal over the SM background is presented in the DM mass vs. $\Lambda$ plane at the $\sqrt{s} = 1$ TeV ILC with $\mathcal{L}_{\rm int}=8~{\rm ab}^{-1}$. Left: Operator $\mathcal{O}^{(5)}_{B\chi}$ with beam polarization $(P_{e^+}, P_{e^-}) = (-20\%, +80\%)$ and benchmark $(\Lambda = 7.3\text{ TeV}, m_{\chi} = 100\text{ GeV})$. Right: Operator $\mathcal{O}^{(6)}_{L\phi}$ with $(P_{e^+}, P_{e^-}) = (+20\%, +80\%)$ and benchmark $(\Lambda = 2.0\text{ TeV}, m_{\phi} = 100\text{ GeV})$. Both benchmarks are compatible with the cosmological bound.
    }
	\label{fig:excl1}
\end{figure}

The projected sensitivity in the $(m_\phi,\Lambda)$ plane is shown in the right panel of figure~\ref{fig:excl1}. The sensitivity contours are obtained for $(P_{e^+},P_{e^-})=(+20\%,+80\%)$ with $\mathcal{L}_{\rm int}=8~{\rm ab}^{-1}$. For scalar DM, $m_\phi\lesssim 300$~GeV, the $5\sigma$ discovery reach extends up to $\Lambda\simeq 1.84$~TeV, while the $3\sigma$ evidence and $2\sigma$ exclusion reaches extend up to $\Lambda\simeq 2.1$~TeV and $\Lambda\simeq 2.3$~TeV, respectively. The sensitivity decreases with increasing $\Lambda$, as expected from the $\Lambda^{-2}$ suppression of the dimension-six amplitude, while the mass dependence is governed primarily by the available phase space for producing the invisible scalar DM pair along with the $Z$ boson.

\FloatBarrier
\subsubsection{Gauge-portal Operators}
\label{subsub:WBphi}
\begin{figure}[htbp!]
	\centering
	\includegraphics[width=0.475\linewidth]{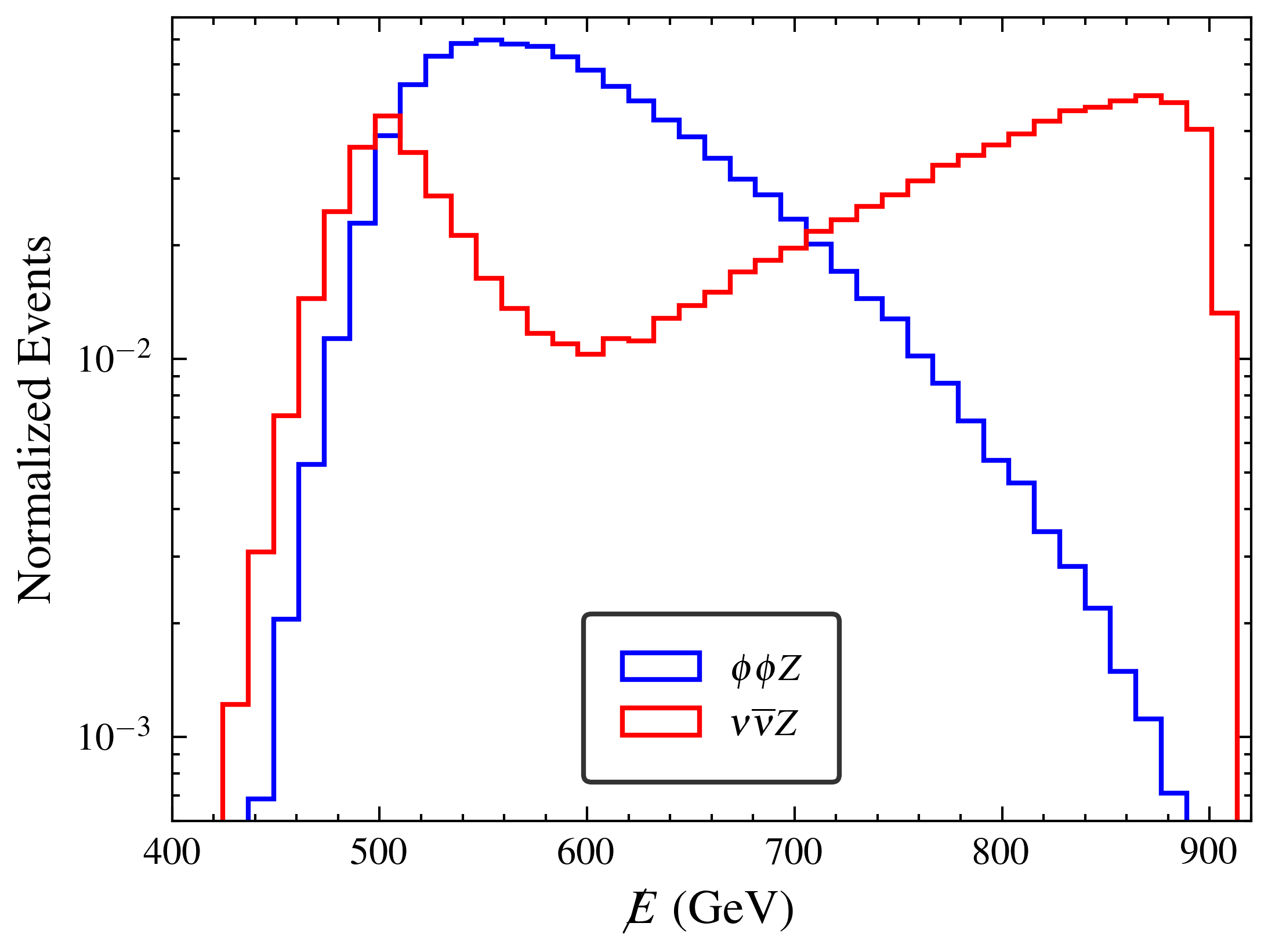}
	\includegraphics[width=0.475\linewidth]{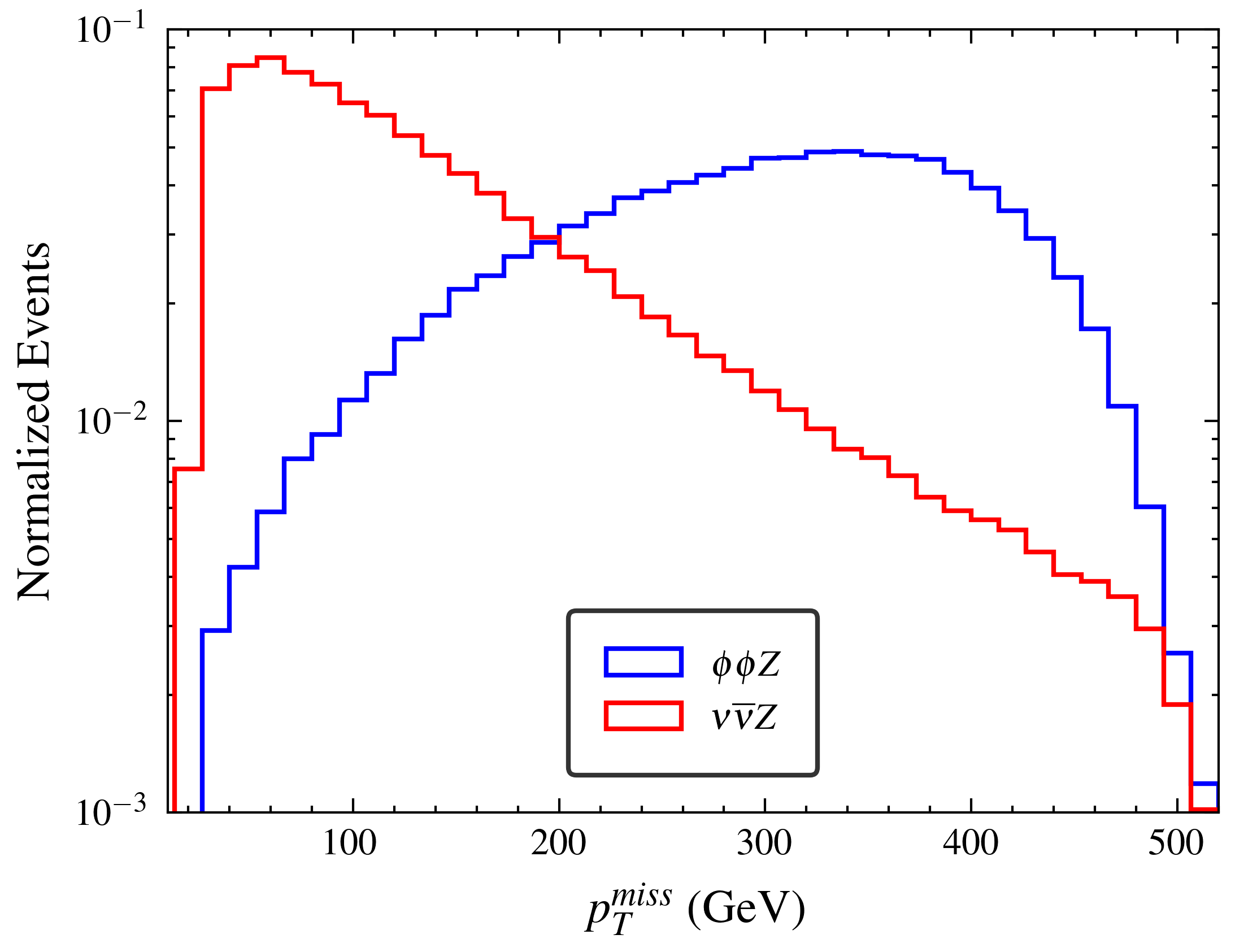}
	\includegraphics[width=0.475\linewidth]{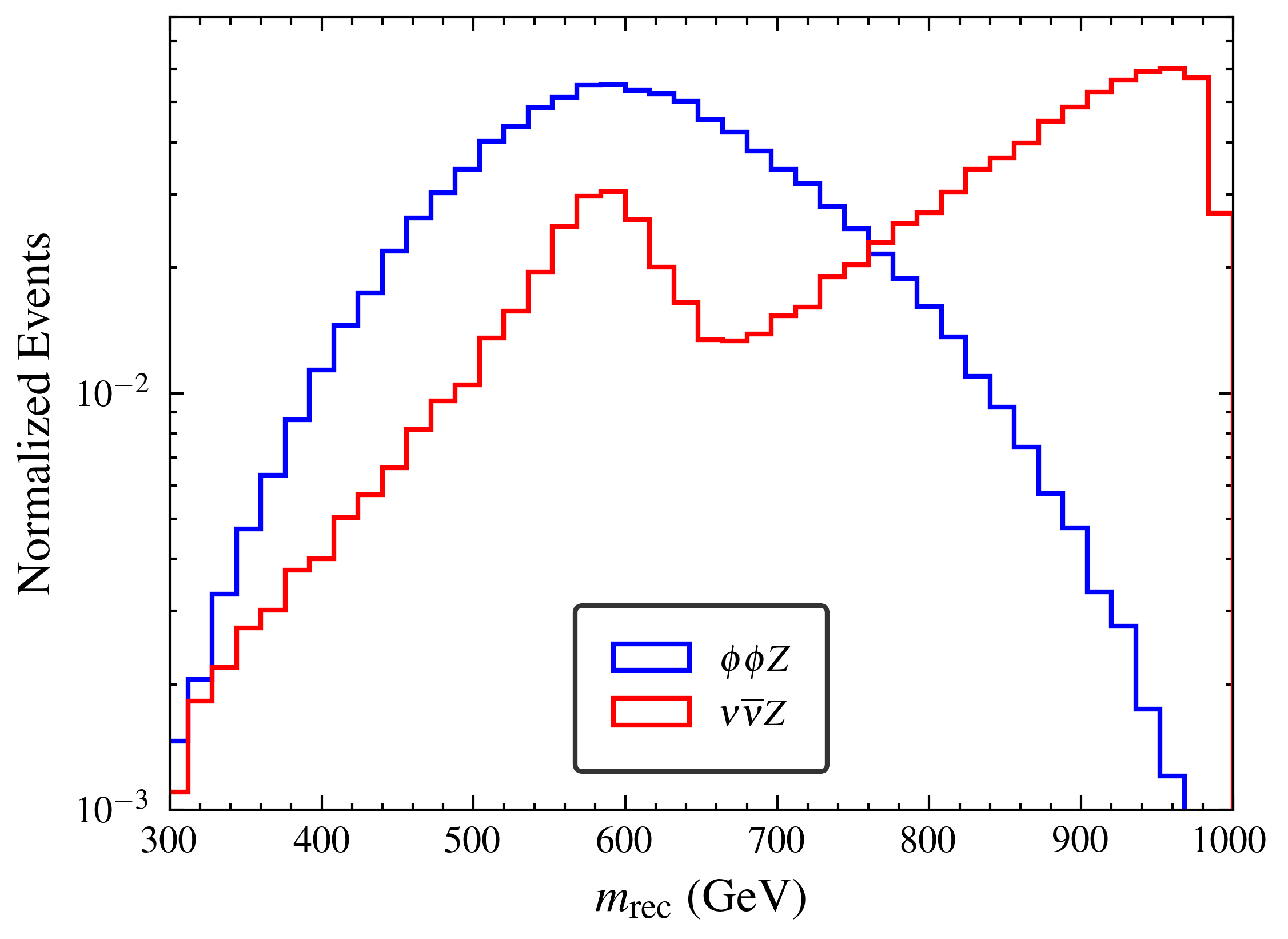}
	\includegraphics[width=0.475\linewidth]{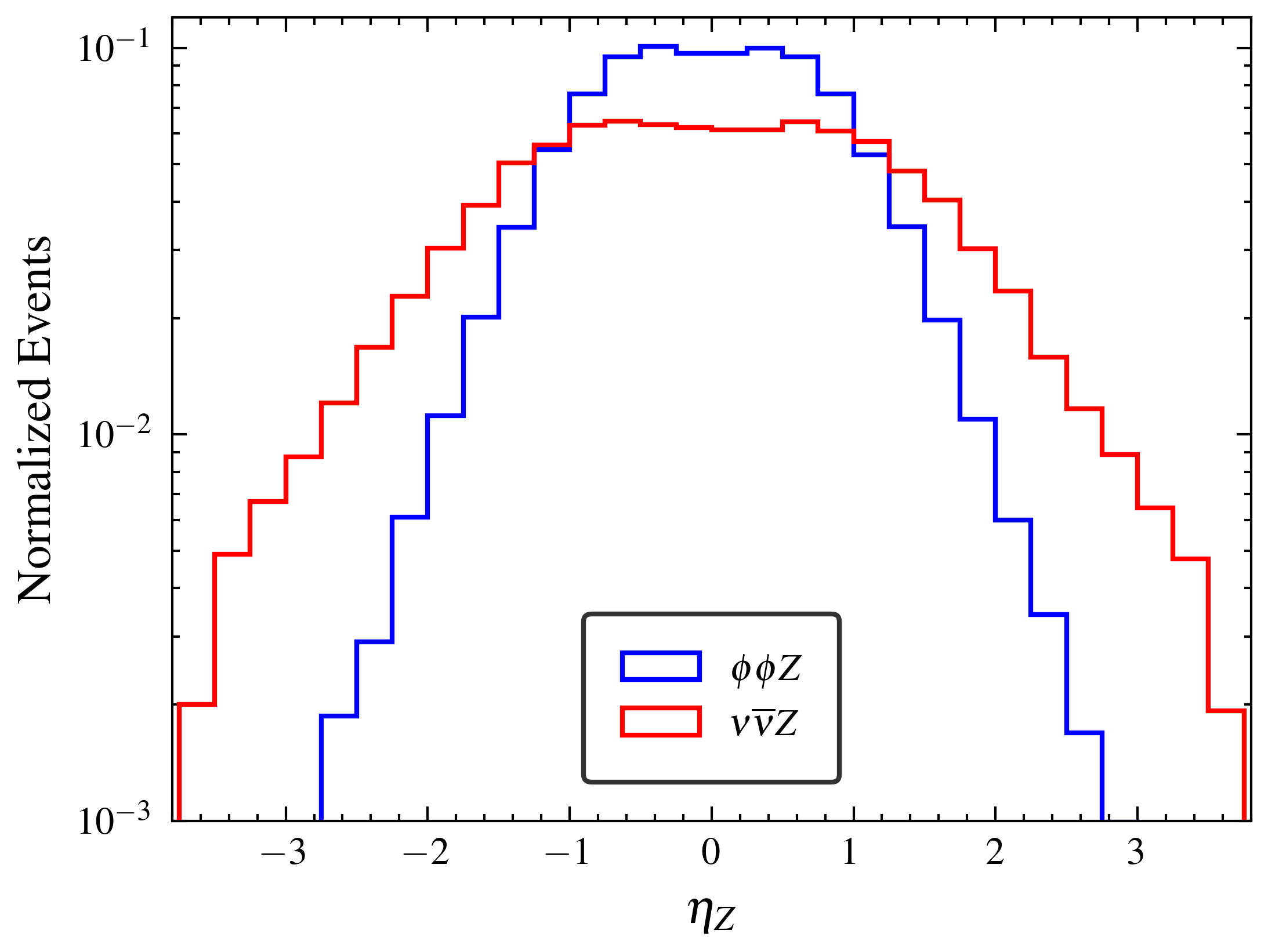}
	\caption{Same as figure~\ref{fig:lphi7} with polarized beam configurations, $P_{e^+}, P_{e^-}= -20\%, +80\%$ for the operator $\mathcal{O}^{(6)}_{B\phi}$ with $\Lambda=1.27$ TeV, $m_{\phi}=50$ GeV, compatible with cosmological constraints.}
	\label{fig:Bphi}
\end{figure}

The dimension-six gauge-portal operators $\mathcal{O}^{(6)}_{W\phi}=(W^I_{\mu\nu}W^{I\mu\nu})\phi^2$ and $\mathcal{O}^{(6)}_{B\phi}=(B_{\mu\nu}B^{\mu\nu})\phi^2$ couple the scalar DM to the Electroweak gauge fields through the $SU(2)_L$ and $U(1)_Y$ field-strength tensors, respectively. They induce the mono-$Z$ process $e^+e^-\to V^\ast\to Z\phi\phi$, with $V=\gamma/Z$, through the effective gauge-boson contact diagram shown in figure~\ref{fig:c}. This is different from the leptophilic case, in which the observed $Z$ boson is radiated from the incoming lepton line. 

The different Electroweak structures of the gauge-portal operators result in different polarization dependence, as discussed in second paragraph of Sec.~\ref{sec:cpheno}. Specifically, the $\mathcal{O}^{(6)}_{W\phi}$ signal is maximized for $(P_{e^+}, P_{e^-}) = (+20\%, -80\%)$, whereas the hypercharge operator $\mathcal{O}^{(6)}_{B\phi}$ is most effectively probed for $(-20\%, +80\%)$. We consider the benchmark points
\begin{equation*}
 \begin{alignedat}{3}
  m_\phi &= 76\text{ GeV}, \qquad & \Lambda &= 1.20\text{ TeV}, \qquad &&\text{for } \mathcal{O}^{(6)}_{W\phi}, \text{ and} \\
  m_\phi &= 50\text{ GeV}, \qquad & \Lambda &= 1.27\text{ TeV}, \qquad &&\text{for } \mathcal{O}^{(6)}_{B\phi}.
 \end{alignedat}
\end{equation*}
Both benchmarks satisfy the relic density and direct detection constraints discussed in Sec.~\ref{sec:dpheno}. The normalized kinematic distributions for $\mathcal{O}^{(6)}_{B\phi}$ are shown in figure~\ref{fig:Bphi}~\footnote{The corresponding distributions for $\mathcal{O}^{(6)}_{W\phi}$ have similar shapes and not shown separately.}. The signal and background distributions show clear separation. In particular, the signal peaks near 540~GeV in both missing energy and recoil mass, and features a harder $p_T^{\rm miss}$ spectrum and a more central $Z$-boson pseudorapidity than the background.

\begin{table}[htbp!]
	\centering
	\renewcommand{\arraystretch}{1.0}{
		\begin{tabular}{|>{\centering\arraybackslash}p{2.6cm}|
				>{\centering\arraybackslash}p{1.8cm}|
				>{\centering\arraybackslash}p{1.8cm}|
				>{\centering\arraybackslash}p{1.5cm}|
				>{\centering\arraybackslash}p{1.8cm}|
				>{\centering\arraybackslash}p{1.8cm}|
				>{\centering\arraybackslash}p{1.5cm}|}
			\hline 
			\multirow{3}*{Cuts} & \multicolumn{6}{c|}{Number of events} \\ \cline{2-7}
			& \multicolumn{3}{c|}{$\mathcal{O}^{(6)}_{W\phi}\,(P_{e^+}, P_{e^-}= +20\%, -80\%)$)} & \multicolumn{3}{c|}{$\mathcal{O}^{(6)}_{B\phi}\,(P_{e^+}, P_{e^-}= -20\%, +80\%)$)} \\ \cline{2-7}
			& Signal ($S$) & Background ($B$) & $\frac{S}{\sqrt{B}}$ & Signal ($S$) & Background ($B$) & $\frac{S}{\sqrt{B}}$ \\ \hline
			Basic event  & 5907.0 & $ 3.39\times 10^{6}$ & 3.20 & 1973.7 & 315207.0  & 3.52 \\
			selection         &  [100\%] & [100\%] & & [100\%] & [100\%] &\\
			\hline
			\multirow{2}*{$\slashed{E}< 720$ GeV} & 5292.7 & 972000.0 & 5.37 & 1797.8 & 133212.0 & 4.93  \\
			&  [89.59\%] & [28.65\%] & & [91.08\%] & [42.26\%] & \\
			\hline
			\multirow{2}*{$p_{T}^{\rm miss} > 180$ GeV} & 4695.0 & 482208.0  & 6.76 & 1600.7 & 70150.1  & 6.04 \\
			&  [79.48\%] & [14.21\%] &  & [81.10\%] & [22.25\%] &\\
			\hline
			\multirow{2}*{$m_{\rm rec} < 780$ GeV} & 4685.2 & 474592.0 & 6.80 & 1597.9  & 69549.1 & 6.06\\
			&  [79.32\%] & [13.99\%] & & [80.96\%]& [22.06\%] &\\
			\hline
			\multirow{2}*{$|\eta_Z|< 1.2$} & 4352.1  & 405401.0  & 6.84 & 1469.8 & 53414.3 & 6.36\\
			&  [73.68\%] & [11.95\%] & & [74.47\%] & [16.94\%] &\\
			\hline
			Significance, $\mathcal{Z}$ & \multicolumn{3}{c|}{6.84} & \multicolumn{3}{c|}{6.36} \\ \hline
	\end{tabular}}
	\caption{Same as table~\ref{tab:cutflow}. Left: for the operator $\mathcal{O}^{(6)}_{W\phi}$, with $\Lambda=1.20$ TeV, $m_{\phi}=76$ GeV; Right: for $\mathcal{O}^{(6)}_{B\phi}$ with $\Lambda=1.27$ TeV, $m_{\phi}=50$ GeV, compatible with cosmological constraints.}
	\label{tab:cutflow-WBphi}
\end{table}

We therefore impose the selections $$\slashed{E}<720~{\rm GeV}, \qquad p_T^{\rm miss}>180~{\rm GeV}, \qquad m_{\rm rec}<780~{\rm GeV}, \qquad |\eta_Z|<1.2.$$ The cut-flow results are summarized in table~\ref{tab:cutflow-WBphi}. The upper missing energy requirement efficiently suppresses the dominant $W$-exchange background, whereas the $p_T^{\rm miss}$ and centrality requirements reject the low transverse momentum and forward background events. The recoil mass selection gives only a modest additional improvement because it is correlated with the missing energy observable.

For $\mathcal{O}^{(6)}_{W\phi}$, the basic selection yields $5907$ signal events over $3.39\times10^6$ background events corresponding to $\mathcal{Z}=3.20$. After the full selection, $4352$ signal events remain over a background of $405401$ events, yielding $$\mathcal{Z}=\frac{S}{\sqrt{B}}=6.84.$$ For $\mathcal{O}^{(6)}_{B\phi}$, the preferred right-handed electron polarization gives $1973$ signal events over $315207$ background events after the basic selection. The final selection retains $1469$ signal events over $53414$ background events, corresponding to $$\mathcal{Z}=\frac{S}{\sqrt{B}}=6.36.$$ Thus, despite its smaller inclusive production rate, $\mathcal{O}^{(6)}_{B\phi}$ achieves a sensitivity comparable to that of $\mathcal{O}^{(6)}_{W\phi}$ because its preferred polarization strongly suppresses the $W$-exchange background.
\begin{figure}[htbp!]
	\centering
	\includegraphics[width=0.45\linewidth]{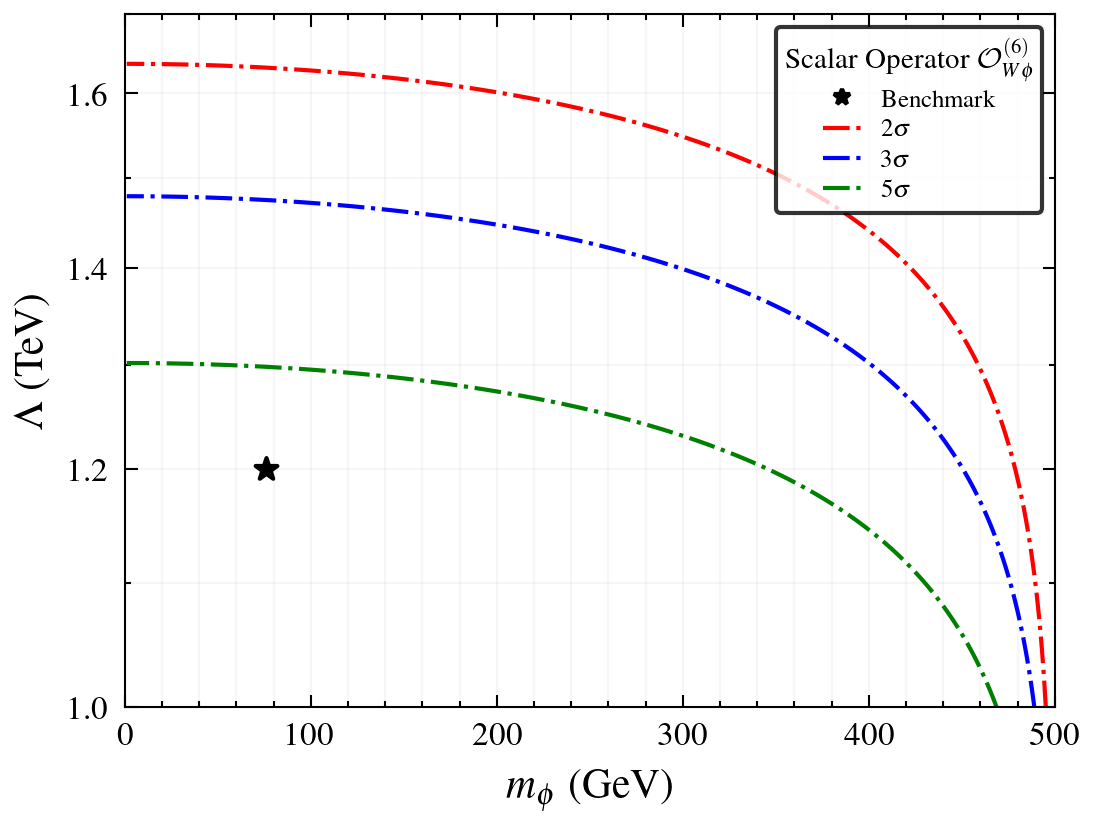}
	\includegraphics[width=0.45\linewidth]{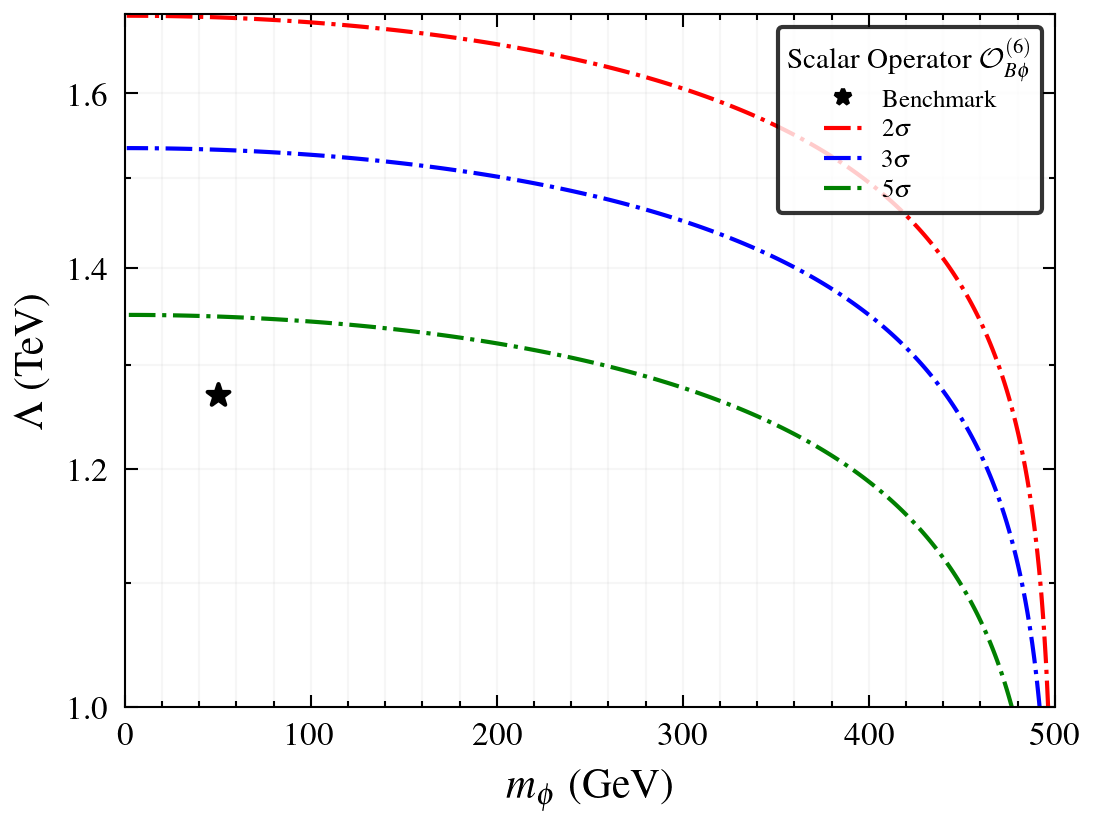}
	\caption{Same as figure~\ref{fig:excl1}. Left: for operator $\mathcal{O}^{(6)}_{W\phi}$, with polarized ($P_{e^+}, P_{e^-}= +20\%, -80\%$) beam configurations, and benchmark ($m_\phi = 76\text{ GeV},\Lambda = 1.20\text{ TeV}$); Right: for $\mathcal{O}^{(6)}_{B\phi}$ operator with ($P_{e^+}, P_{e^-}= -20\%, +80\%$), and benchmark ($m_\phi = 50\text{ GeV}, \Lambda = 1.27\text{ TeV}$). Both benchmarks are compatible with the cosmological bound.}
	\label{fig:exclWphi}
\end{figure}

The projected sensitivities in the $(m_\phi,\Lambda)$ plane are shown in figure~\ref{fig:exclWphi}. For a light scalar DM candidate with $m_\phi=10$~GeV, the $\mathcal{O}^{(6)}_{W\phi}$ search reaches $\Lambda\simeq 1.64$~TeV at the $2\sigma$ level and $\Lambda\simeq 1.30$~TeV at the $5\sigma$ level. The corresponding reach for $\mathcal{O}^{(6)}_{B\phi}$ is slightly stronger, extending to $\Lambda\simeq 1.70$~TeV at $2\sigma$ and $\Lambda\simeq 1.35$~TeV at $5\sigma$. For $m_\phi=200$~GeV, the sensitivity to $\mathcal{O}^{(6)}_{W\phi}$ decreases to $\Lambda\simeq 1.60$~TeV and $\Lambda\simeq 1.27$~TeV at the $2\sigma$ and $5\sigma$ levels, respectively. The corresponding reach for $\mathcal{O}^{(6)}_{B\phi}$ is $\Lambda\simeq 1.66$~TeV at $2\sigma$ and $\Lambda\simeq 1.32$~TeV at $5\sigma$. The sensitivity remains approximately stable across the low-to-intermediate mass region, where phase-space suppression is mild, and decreases rapidly only when $m_\phi$ approaches the kinematic limit $m_\phi\simeq\sqrt{s}/2$. From the LHC recasting analysis in Sec.~\ref{sec:lhc-recast}, the $2\sigma$ exclusion for a DM mass $m_\phi=100$~GeV ($200$~GeV) extends only to $\Lambda\approx230$~GeV ($195$~GeV) for the $\mathcal{O}_{B\phi}^{(6)}$ operator, and to $\Lambda\approx345$~GeV ($315$~GeV) for $\mathcal{O}_{W\phi}^{(6)}$. These results show that the mono-$Z$ channel at a polarized 1~TeV $e^+e^-$ collider provides a sensitive probe of viable scalar gauge-portal interactions.

\subsection{Higgs-portal Operators}   
\label{colli-Higgs}
\begin{table}[htbp!]
    \centering
    \renewcommand{\arraystretch}{1.0}{
    \begin{tabular}{|>{\centering\arraybackslash}p{3cm}|
                     >{\centering\arraybackslash}p{2cm}|
                     >{\centering\arraybackslash}p{2cm}|
                     >{\centering\arraybackslash}p{2cm}|
                     }
    \hline 
    polarization  & \multicolumn{3}{c|}{production cross section $\mathrm{[fb]}$} \\ \cline{2-4}
    $\left(P_{e^+}, P_{e^-}\right)$ &  $\mathcal{O}^{(6)}_{D\phi}$ &  $\mathcal{O}^{(6)}_{DX}$ & $\nu \overline{\nu} Z$ (SM) \\ \hline
    unpolarized & 0.091 & 0.949 & 502.852\\
    $\left(+20\%, +80\%\right)$ & 0.069 & 0.723 & 276.608\\
    $\left(+20\%, -80\%\right)$ & 0.118 & 1.225 & 827.293\\
    $\left(-20\%, +80\%\right)$ & 0.094 & 0.978  & 340.480\\
    $\left(-20\%, -80\%\right)$ & 0.084 & 0.869 & 570.203\\ \hline
    \end{tabular}}
    \caption{Cross section for scalar and vector DM pair-production corresponding to the operators $\mathcal{O}_{D\phi}^{(6)}$ and $\mathcal{O}_{DX}^{(6)}$ respectively along with cross section for the SM background. For both of the signal operators, $\Lambda = 2.935~\mathrm{TeV}$, $m_{DM} = 59.16~\mathrm{GeV}$.}
    \label{tab:prodH}
\end{table}
\begin{figure}[htbp!]
	\centering
    {\label{mrec_lep}\includegraphics[width=0.6\textwidth]{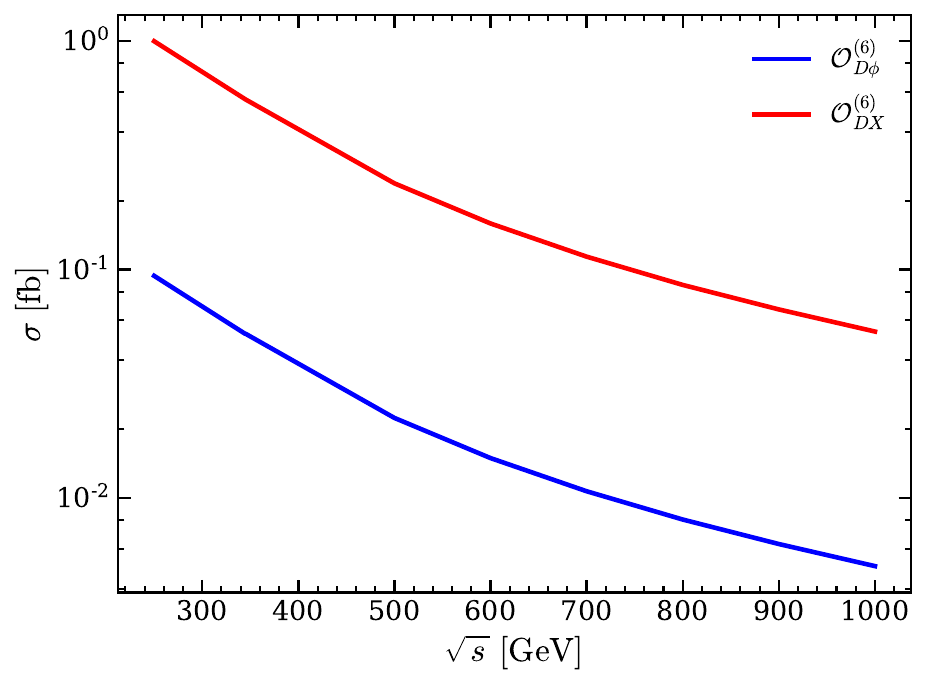}}
	\caption{Variation of signal cross section corresponding to $\mathcal{O}_{D\phi}^{(6)}$ and $\mathcal{O}_{DX}^{(6)}$ with center-of-mass energy, $\sqrt{S}$, for an $e^{+}e^{-}$ collider with polarized $(-20\%,+80\%)$ beams. For both of the signal operators, $\Lambda = 2.935~\mathrm{TeV}$, $m_{DM} = 59.16~\mathrm{GeV}$.} 
	\label{fig:prodH}
\end{figure} 
\begin{figure}[htbp!]
	\centering
    \subfloat[]{\label{mrec_lep}\includegraphics[width=0.48\textwidth]{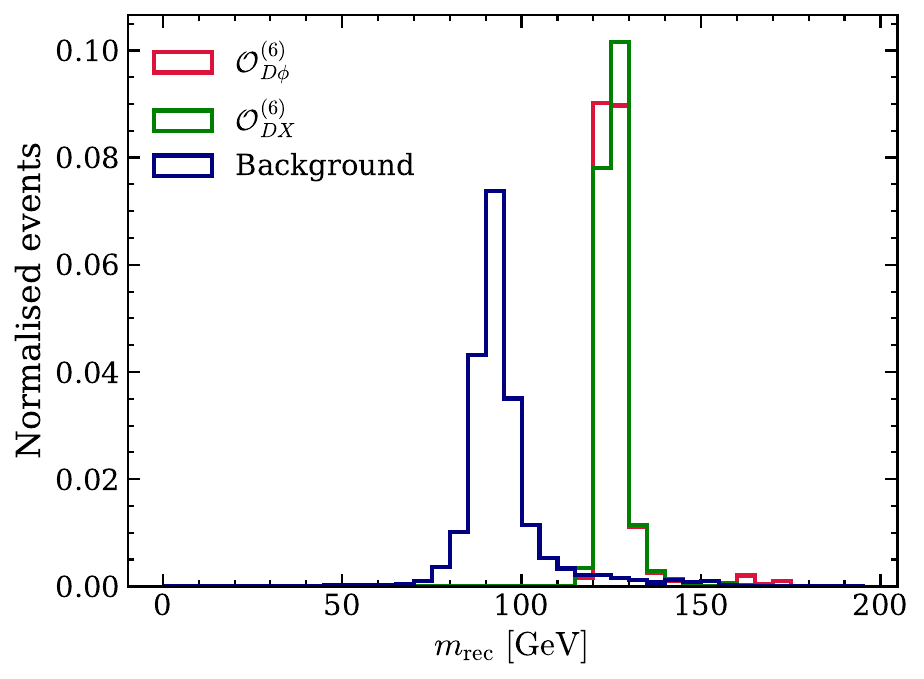}}
	\subfloat[]{\label{D6_met}\includegraphics[width=0.48\textwidth]{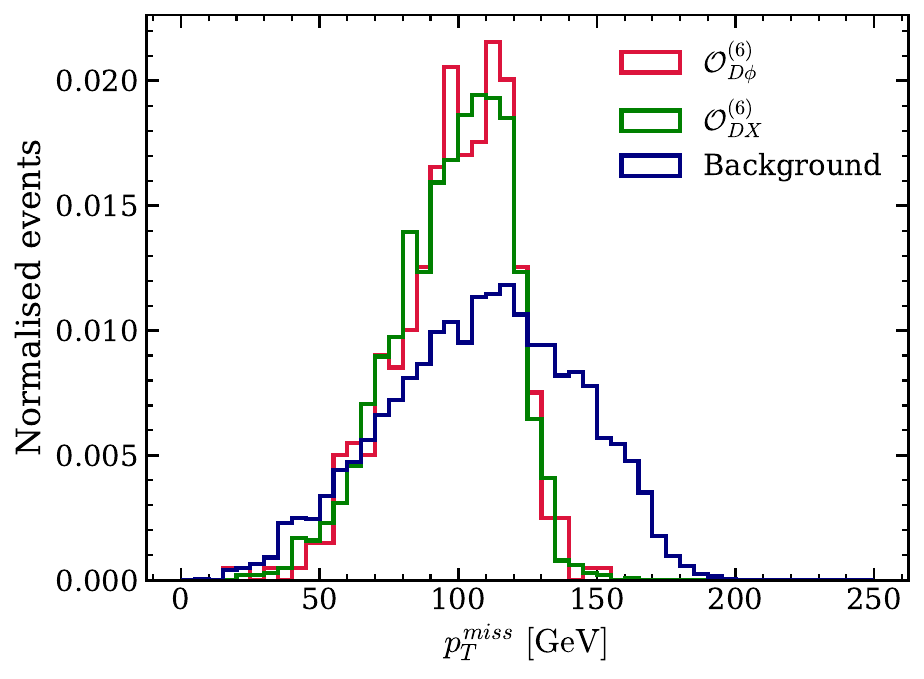}}\\
    {\label{D6_me}\includegraphics[width=0.48\textwidth]{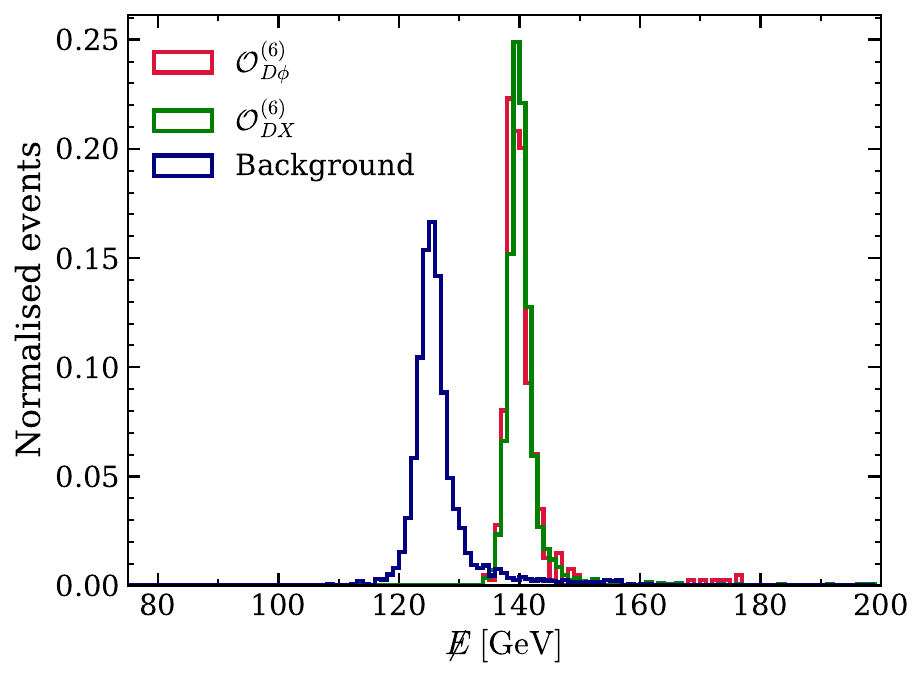}}
    {\label{D6_drl}\includegraphics[width=0.48\textwidth]{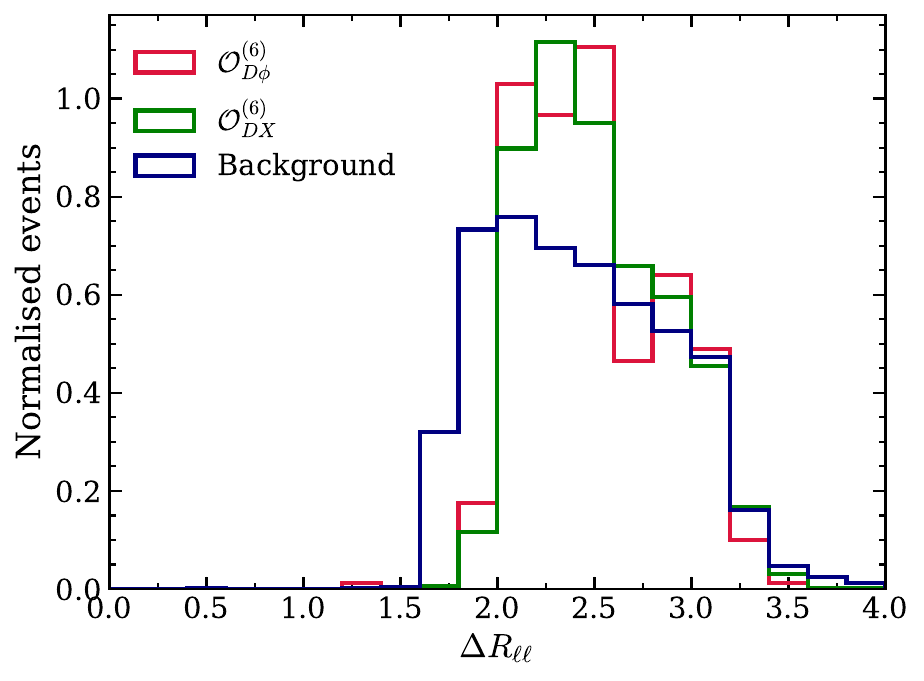}}
	\caption{Kinematic distributions(corresponding to leptonic final state) for operators $\mathcal{O}^{(6)}_{DX}$ and $\mathcal{O}^{(6)}_{D\phi}$ together with the SM background. The signal corresponds to the benchmark point $\Lambda=2.935~\text{TeV}$, $m_{\rm DM}=59.16~\text{GeV}$. $\Delta R_{ll}$ is the angular separation of the dilepton in the $\eta-\phi$ plane.} 
	\label{fig:scalar_lep}
\end{figure} 
\begin{figure}[htbp!]
	\centering
    \subfloat[]{\label{mrec_lep}\includegraphics[width=0.48\textwidth]{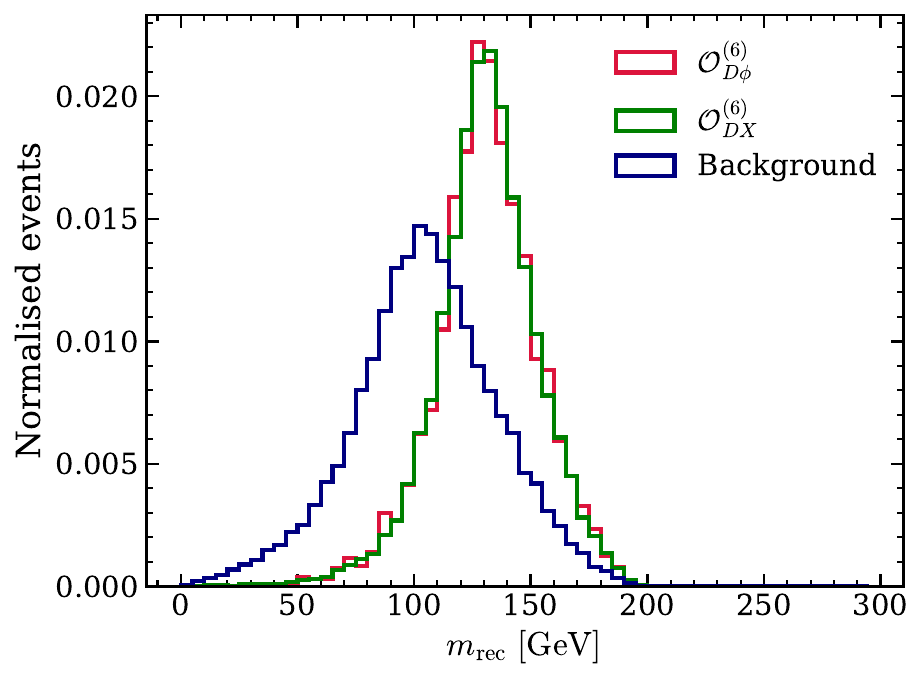}}
	\subfloat[]{\label{D6_met}\includegraphics[width=0.48\textwidth]{MET_polbeam_59_2_all_e2.pdf}}\\
    \subfloat[]
    {\label{D6_me}\includegraphics[width=0.48\textwidth]{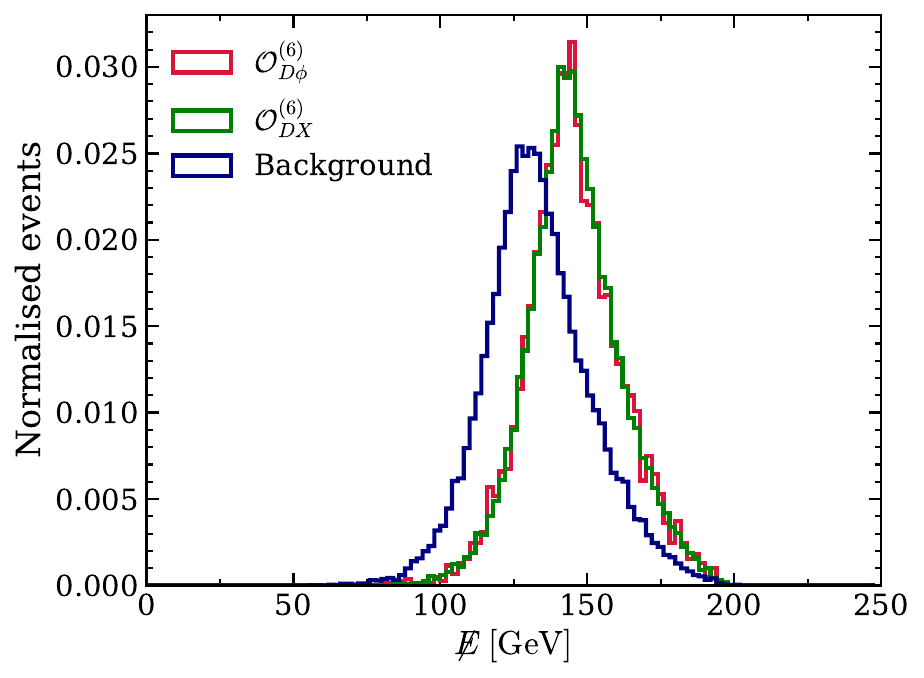}}
    {\label{D6_drl}\includegraphics[width=0.48\textwidth]{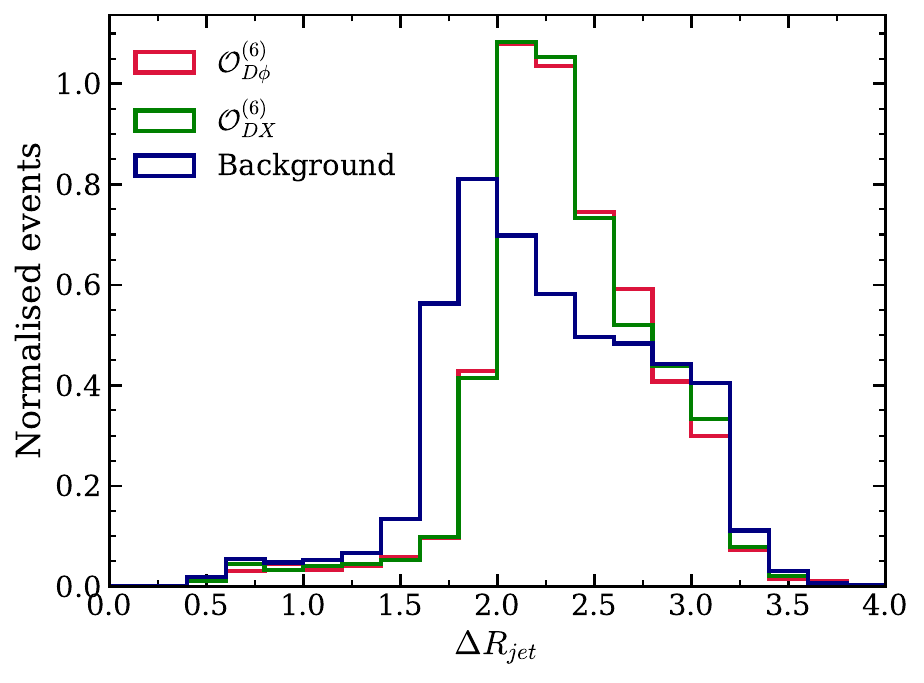}} 
	\caption{Same as figure~\ref{fig:scalar_lep}, but corresponding to hadronic analysis. $\Delta R_{jj}$ is the angular separation of the dijet in the $\eta-\phi$ plane.} 
	\label{fig:scalar_had}
\end{figure} 

As evident from figures~\ref{fig:relic} and \ref{fig:ID_scalar}, for the $\mathcal{O}_{D\phi}^{(6)}$ operator the only region of the parameter space that simultaneously yields the observed DM relic density and remains consistent with the current direct-detection constraints is located in the vicinity of the Higgs-boson resonance. Hence for $\mathcal{O}_{D\phi}^{(6)}$ the benchmark point (BP) considered in this analysis corresponds to a DM mass of $m_{\mathrm{DM}} = 59.16~\mathrm{GeV}$ and a new-physics scale of $\Lambda = 2.935~\mathrm{TeV}$. We have considered this benchmark for $\mathcal{O}_{DX}^{(6)}$ operator as well.

The production cross sections for signals and background are tabulated in Table \ref{tab:prodH} for 4 different polarizations of the beams. For the beam polarization of $(-20\%,+80\%)$ the background is reduced remarkably while keeping the signal almost fixed to the value corresponding to the unpolarized one. On the other hand though the polarization configuration of $(+20\%,+80\%)$ cuts down the background even better, it reduces the signal as well. Hence we have studied the prospect of these operators for the beam polarization of $(-20\%,+80\%)$.

Fig,~\ref{fig:prodH} presents the dependence of the production cross section for the $\mathcal{O}_{D\phi}^{(6)}$ and $\mathcal{O}_{DX}^{(6)}$ operators on the center-of-mass energy, $\sqrt{s}$, of the $e^+e^-$ collider. It can be observed that the production cross section decreases rapidly as the center-of-mass energy increases. Therefore, in order to maximize the signal production rate and improve the sensitivity to these operators, we restrict our collider analysis to $\sqrt{s}=250~\mathrm{GeV}$ for these operators.

We have studied both the leptonic and hadronic decay modes of $Z$ boson and the corresponding kinematic distributions are presented separately for the dilepton and dijet final states in figure~\ref{fig:scalar_lep} and figure~\ref{fig:scalar_had}, respectively. It is evident from the figures that leptonic final state is more efficient to distinguishing signal from background. For the leptonic channel, as shown in figure~\ref{fig:scalar_lep}, the recoil mass, $m_{\mathrm{rec}}$, and the missing energy, $\slashed{E}$, emerge as the most effective discriminating variables between the signal and the Standard Model background. For the hadronic channel, these variables remain important discriminants, as expected; however, their separation power is comparatively reduced with respect to the leptonic channel. This degradation can primarily be attributed to the poorer reconstruction efficiency  associated with the hadronic final state. Hence, in the cut-based analysis as summarized in Table~\ref{tab:cutflow-DphiX}, we have taken only the leptonic final state into account. 

\begin{table}[htbp!]
    \centering
    \renewcommand{\arraystretch}{1.0}{
    \begin{tabular}{|>{\centering\arraybackslash}p{2.6cm}|
                     >{\centering\arraybackslash}p{1.8cm}|
                     >{\centering\arraybackslash}p{1.8cm}|
                     >{\centering\arraybackslash}p{1.5cm}|
                     >{\centering\arraybackslash}p{1.8cm}|
                     >{\centering\arraybackslash}p{1.8cm}|
                     >{\centering\arraybackslash}p{1.5cm}|}
    \hline 
    \multirow{3}*{Cuts} & \multicolumn{6}{c|}{Number of events} \\ 
    \cline{2-7}
    & \multicolumn{3}{c|}{$\mathcal{O}^{(6)}_{D\phi}\,(P_{e^+}, P_{e^-}= -20\%, +80\%)$)} & \multicolumn{3}{c|}{$\mathcal{O}^{(6)}_{DX}\,(P_{e^+}, P_{e^-}= -20\%, +80\%)$)} \\ \cline{2-7}
    & Signal ($S$) & Background ($B$) & $\frac{S}{\sqrt{B}}$ & Signal ($S$) & Background ($B$) & $\frac{S}{\sqrt{B}}$ \\ \hline
    Basic event  & 25.0 & 85120.0 & 0.086 & 259.1 & 85120.0 & 0.888\\
     selection (leptonic)        &  [100\%] & [100\%] & & [100\%] & [100\%] &\\
    \hline
    \multirow{2}*{$\slashed{E}> 135$ GeV}
 & 24.9 & 3677.2 & 0.411 & 258.3 & 3677.2 & 4.259\\
 & [99.6\%] & [4.3\%] & & [99.7\%] & [4.3\%] & \\
    &  & & & & & \\
    \hline
    {$120~\mathrm{GeV} < m_{rec} < 140$ GeV} & 24.8 & 1225.7 & 0.708 & 254.1 & 1225.7 &  7.258 \\
    & [99.2\%] & [1.4\%]  &  & [98.1\%] & [1.4\%] &\\
    & & & & & & \\
    \hline
    Significance, $\mathcal{Z}$ & \multicolumn{3}{c|}{0.708} & \multicolumn{3}{c|}{7.258} \\ \hline
    \end{tabular}}
    \caption{Cut-flow table and signal significance for signal and background process at the $e^+ e^-$ collider with $\sqrt{s} = 250$ GeV, $\mathcal{L}_{\rm int}=8$ ab$^{-1}$, with polarized $\left(-20\%, +80\%\right)$ beam configurations. The DM signal corresponds to the operators $\mathcal{O}^{(6)}_{D\phi}$ and $\mathcal{O}^{(6)}_{DX}$ with $\Lambda=2.935$ TeV, $m_{\rm DM}=59.16$ GeV.}
    \label{tab:cutflow-DphiX}
\end{table}

At $\sqrt{s}=250$~GeV with $\mathcal{L}_{\rm int}=8$~ab$^{-1}$, the leptonic mono-$Z$ searches yields $\mathcal{Z}\simeq0.71$ for $\mathcal{O}_{D\phi}^{(6)}$ and $\mathcal{Z}\simeq7.3$ for $\mathcal{O}_{DX}^{(6)}$ at the Higgs-resonance benchmark $(m_{\rm DM},\Lambda)=(59.16~\mathrm{GeV},\,2.935~\mathrm{TeV})$. This implies that $\mathcal{O}_{DX}^{(6)}$ is accessible with high significance at the first ILC stage, whereas $\mathcal{O}_{D\phi}^{(6)}$ remains below the $2\sigma$ evidence threshold for the same parameters. The substantially larger significance for $\mathcal{O}_{DX}^{(6)}$ follows from its higher production cross section and more favorable signal-to-background ratio under the chosen polarization.

\subsection{Discovery Prospects at the ILC}
\label{subsec:summary_discovery}
\begin{table}[htbp!]
    \centering
    \renewcommand{\arraystretch}{1.25}
    \begin{tabular}{lccccc}
        \hline\hline
        Operator & $\sqrt{s}$ [TeV] & polarization ($P_{e^+}, P_{e^-}$) & $m_{\rm DM}$ [GeV] & $\Lambda$ [TeV] & $\mathcal{L}_{5\sigma}$ [ab$^{-1}$] \\
        \hline
        $\mathcal{O}^{(5)}_{B\chi}$
        & 1
        & $(-20\%,+80\%)$
        & 100 & 7.3  & 153.0  \\
        \hline
        \multirow{3}*{$\mathcal{O}^{(6)}_{L\phi}$}
        & \multirow{3}*{1}
        & \multirow{3}*{$(+20\%,+80\%)$}
        &   50 & 2.0  & 10.0  \\
        & & & 100 & 2.0   & 10.37 \\
        & & & 300 & 2.0   & 15.56 \\
        \hline
        \multirow{2}*{$\mathcal{O}^{(6)}_{W\phi}$}
        & \multirow{2}*{1}
        & \multirow{2}*{$(+20\%,-80\%)$}
        &  73 & 1.1   & 2.12  \\
        & & &  80 & 1.36   & 11.66  \\
        \hline
        \multirow{3}*{$\mathcal{O}^{(6)}_{B\phi}$}
        & \multirow{3}*{1}
        & \multirow{3}*{$(-20\%,+80\%)$}
        &  37 & 1.07   & 1.25  \\
        & & &  50 & 1.27   & 4.94  \\
        & & & 60 & 1.4   & 10.83  \\
        \hline
        \multirow{1}*{$\mathcal{O}^{(6)}_{D\phi}$}
        & \multirow{1}*{0.25}
        & \multirow{1}*{$(-20\%,+80\%)$}
        &  59.16 & 2.935   & 399.0  \\
        \hline
        \multirow{1}*{$\mathcal{O}^{(6)}_{DX}$}
        & \multirow{1}*{0.25}
        & \multirow{1}*{$(-20\%,+80\%)$}
        &  59.16 & 2.935   & 3.79  \\
        \hline\hline
    \end{tabular}
    \caption{Estimated integrated luminosity required for a $5\sigma$ discovery at the future $e^+e^-$ collider for a few benchmark points of various operators consistent with the cosmological bounds. The CM energy ($\sqrt{s}$) and optimized beam-polarization configuration are listed in the second and third columns, respectively.}
    \label{tab:luminosity-5sigma}
\end{table}

To quantify the discovery prospects for the individual DMEFT operators consistent with cosmological constraints, we estimate the integrated luminosity required for a $5\sigma$ observation of the mono-$Z$ signal at the ILC. The results are summarized in table~\ref{tab:luminosity-5sigma} for representative benchmark points at $\sqrt{s} = 250$~GeV and $1$~TeV.

At $\sqrt{s} = 1$~TeV, the discovery reach is governed by the interplay between operator structure and cosmological constraints. For the fermionic dipole operator $\mathcal{O}^{(5)}_{B\chi}$, relic density and direct detection limits push the viable cutoff scale well into the multi-TeV regime, severely suppressing the production rate and rendering a $5\sigma$ discovery unattainable within realistic ILC operating scenarios. In contrast, the bosonic operators $\mathcal{O}^{(6)}_{B\phi}$ and $\mathcal{O}^{(6)}_{W\phi}$ exhibit promising discovery potential: their lower-mass benchmarks are accessible within the baseline luminosity, whereas heavier DM mass requires higher exposure as the viable cutoff scale increases. The leptophilic operator $\mathcal{O}^{(6)}_{L\phi}$ on the other hand, demonstrates an intermediate reach, where discovery across the mass spectrum is achievable but preferably at high-luminosity run.

The derivative Higgs-portal operators survive the cosmological constraints only in the Higgs resonance region. At $\sqrt{s} = 250$~GeV, the production cross section (as shown in table~\ref{tab:prodH}) of the vector boson DM operator $\mathcal{O}^{(6)}_{DX}$ yields a rate an order of magnitude larger than its scalar counterpart $\mathcal{O}^{(6)}_{D\phi}$. Consequently, $\mathcal{O}^{(6)}_{D\phi}$ requires an unfeasible large luminosity and remains out of reach, whereas the vector DM operator $\mathcal{O}^{(6)}_{DX}$ is comfortably discoverable within the nominal luminosity with the optimal $(-20\%, +80\%)$ polarization configuration. 

\section{Summary and Conclusion}
\label{sec:conc}
We have investigated the mono-$Z$ signature as a probe of DM interactions at the future electron--positron colliders within the framework of DMEFT. Mono-$Z$ signal can probe the Electroweak structure of the DMEFT operators, which mono $\gamma$ signal is often blind to. Secondly, having a strong constraint coming from the non-observation of DM in spin independent direct search via nuclear recoil, motivates DM to interact primarily with the leptonic sector. We consider gauge-invariant operators up to dimension six involving scalar, Dirac/Majorana-fermion and vector-DM candidates, classify them according to the underlying production topology, consider their cosmological constraints involving relic density, direct and indirect searches, choose suitable benchmark points allowed by them, and finally analyse their collider detection prospect. 

We additionally adhere to the constraint from the Higgs invisible branching fraction, and recast the ATLAS mono-$Z$ plus missing energy search bound at LHC. These constraints restrict the parameter space available for collider searches, however a large parameter space remains uncovered. In particular, over the DM mass range considered, the Dirac operators $\mathcal{O}^{(6)}_{D\chi}$ and $\mathcal{O}^{(6)}_{e\chi}$ are excluded by current spin-independent direct-detection limits, whereas $\mathcal{O}^{(6)}_{\ell\chi}$ remains viable only for $m_\chi \gtrsim 520~\mathrm{GeV}$, beyond the kinematic reach of mono-$Z$ production at a $1$~TeV collider. The remaining operators are therefore relevant for future collider searches.

A mono $Z$ signal at future lepton collider can arise from $Z$-boson recoiling against an invisible dark-sector pair. Such signal can arise majorly in four different ways: (i) through Higgsstrahlung process followed by an invisible Higgs decay, (ii) initial-state radiation from leptonic contact interactions, (iii) gauge-field strength induced production, or (iv) a combination of initial- and final-state radiation. Although these possibilities lead to the same visible final state, their distinct Lorentz and Electroweak structures leave characteristic imprints on the production rate, beam-polarization dependence, and kinematic distributions.
 
We perform a detailed search for the mono-$Z$ plus missing energy channel at the $\sqrt{s} = 250$~GeV and $1$~TeV ILC, incorporating detector effects. Kinematic selection criteria and beam-polarization configurations are optimized individually for each operator. The SM background, $e^+e^-\to Z\nu\bar{\nu}$, receives contributions from $ZZ$ production, non-resonant neutral-current diagrams, and $t$-channel $W$ exchange topologies. The latter contribution is particularly important because of its large rate and its strong chiral dependence. Beam polarization therefore plays an important role in the analysis: a predominantly right-handed electron beam suppresses the $W$-exchange background, while the signal rate depends on the Lorentz and Electroweak structures of the effective operators.

 Missing energy, missing transverse momentum, recoil mass, and $Z$-boson pseudorapidity provide good handles to discriminate the signal from the background. While the recoil mass directly reconstructs the invariant mass of the invisible particles, it is highly correlated with the missing energy in the mono-$Z$ topology and offers limited additional separation power once missing energy requirements are applied. Consequently, the operator-specific selections on missing energy, missing transverse momentum, and pseudorapidity efficiently isolate the signal from the background.

Our discovery projections underscore how DMEFT sensitivity is fundamentally shaped by the interplay of operator structures and cosmological constraints. At $\sqrt{s} = 1$~TeV, stringent relic-density and direct-detection constraints push the dipole operator for Dirac-fermion DM, $\mathcal{O}^{(5)}_{B\chi}$, beyond realistic collider reach, whereas the gauge portals ($\mathcal{O}^{(6)}_{B\phi}, \mathcal{O}^{(6)}_{W\phi}$) and the leptophilic scalar $\mathcal{O}^{(6)}_{L\phi}$ offer viable discovery channels when paired with appropriate beam polarization, though higher masses demand large luminosity. At $\sqrt{s} = 250$~GeV, the derivative Higgs portal operators exhibit their spin-dependence: the vector DM operator $\mathcal{O}^{(6)}_{DX}$ is readily discoverable within the nominal luminosity, whereas its scalar counterpart $\mathcal{O}^{(6)}_{D\phi}$ remains unavailable to the future lepton collider. These results establish that collider sensitivity in the mono-$Z$ channel is governed not merely by DM mass and cutoff scale, but significantly by the spin of DM,  Lorentz and Electroweak structure of the DMEFT operators.
Our results demonstrate that polarized mono-$Z$ searches at future lepton colliders provide a powerful and complementary probe of the dark sector. The precisely known initial state, missing energy, missing transverse momentum, and operator-dependent beam polarization allow the $e^+ e^-$ collider to probe viable DMEFT interactions at scales well beyond the reach of the present LHC mono-$Z$ recast. 

Finally, we emphasize that the collider projections presented here rely on the standard assumption of thermal dark matter freeze-out during a radiation-dominated epoch. Non-standard cosmological scenarios, such as freeze-out occurring during the reheating era, involve entropy dilution that can significantly alter the  DM relic-density and broaden the cosmologically viable parameter space~\cite{Bhattacharya:2025wef}, which we leave for future investigation.

In this study, we consider one effective operator at a time. A more comprehensive treatment including operator interference, systematic uncertainties, and higher-order corrections would further refine the projected sensitivities. Nevertheless, our analysis establishes the mono-$Z$ channel as a robust and complementary probe of effective dark sector interactions at future electron--positron colliders, summarizing almost all the possibilities, which can be a useful guideline for future DM searches.

\section*{Acknowledgments}
The authors sincerely thank the Indian Institute of Technology Hyderabad for their excellent hospitality during WHEPP-2025, where this work was initiated during the working group sessions. The authors also thank Radhika Vinze and Tanumoy Mandal for their involvement in the very initial stages of the work. SB acknowledges the ANRF grant CRG/2023/000580. RS acknowledges partial support from the Anusandhan National Research Foundation (ANRF), Government of India, through Grant No. CRG/2023/008234. P.K.P. acknowledges the Ministry of Education, Government of India, for providing financial support for his research via the Prime Minister’s Research Fellowship (PMRF) scheme. 
\appendix
\section{Vector and Majorana Operators}
\label{app:vector}
\begin{figure}[htbp!]
    \centering
    \includegraphics[width=0.55\textwidth]{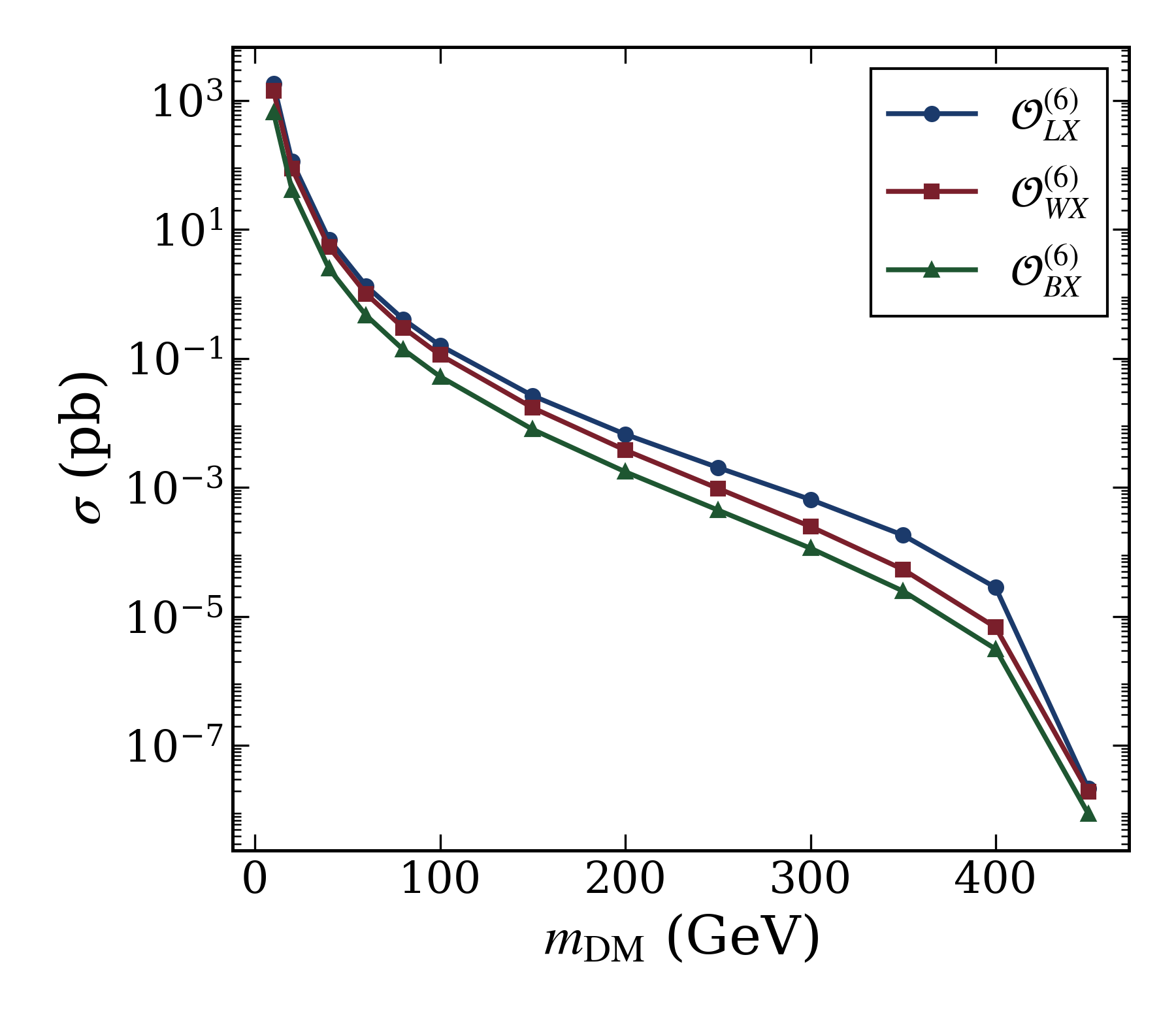}
    \caption{Production cross sections for the process $e^+e^- \to Z X X$ with an unpolarized beam configuration as a function of the vector DM mass for the operators $\mathcal{O}_{LX}^{(6)}$, $\mathcal{O}_{WX}^{(6)}$, and $\mathcal{O}_{BX}^{(6)}$ at the $\sqrt{s} = 1 $~TeV ILC. The effective new physics scale is fixed at $\Lambda = 2$~TeV. }
    \label{fig:vector-cs}
\end{figure}

In figure~\ref{fig:vector-cs}, we shown the variation of production cross section for the process $e^+e^- \to Z X X$ with an unpolarized beam configuration as a function of the vector DM mass for the operators $\mathcal{O}_{LX}^{(6)}$, $\mathcal{O}_{WX}^{(6)}$, and $\mathcal{O}_{BX}^{(6)}$ at the $\sqrt{s} = 1 $~TeV ILC. The effective new physics scale is fixed at $\Lambda = 2$~TeV. For Higgs-portal vector DM operator, $\mathcal{O}_{DX}^{(6)}$, the details are discussed in section~\ref{colli-Higgs}.

\begin{figure}[htbp!]
    \centering
    \begin{subfigure}[b]{0.48\textwidth}
        \centering
        \includegraphics[width=\textwidth]{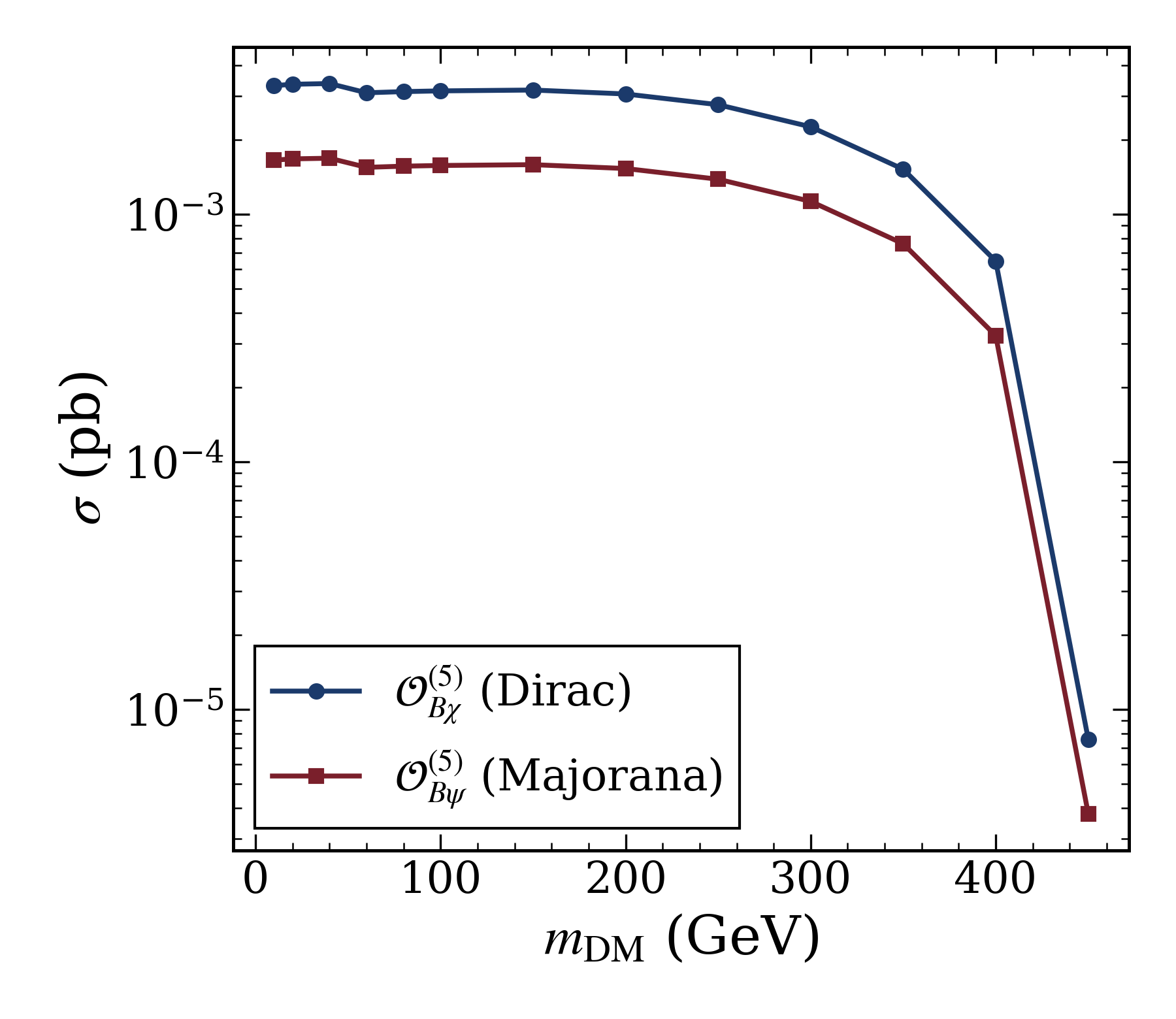}
        \caption{$\mathcal{O}_{B\psi}^{(5)}$ and $\mathcal{O}_{B\chi}^{(5)}$}
        \label{fig:Bpsi-dirac-vs-maj}
    \end{subfigure}
    \hfill
    \begin{subfigure}[b]{0.48\textwidth}
        \centering
        \includegraphics[width=\textwidth]{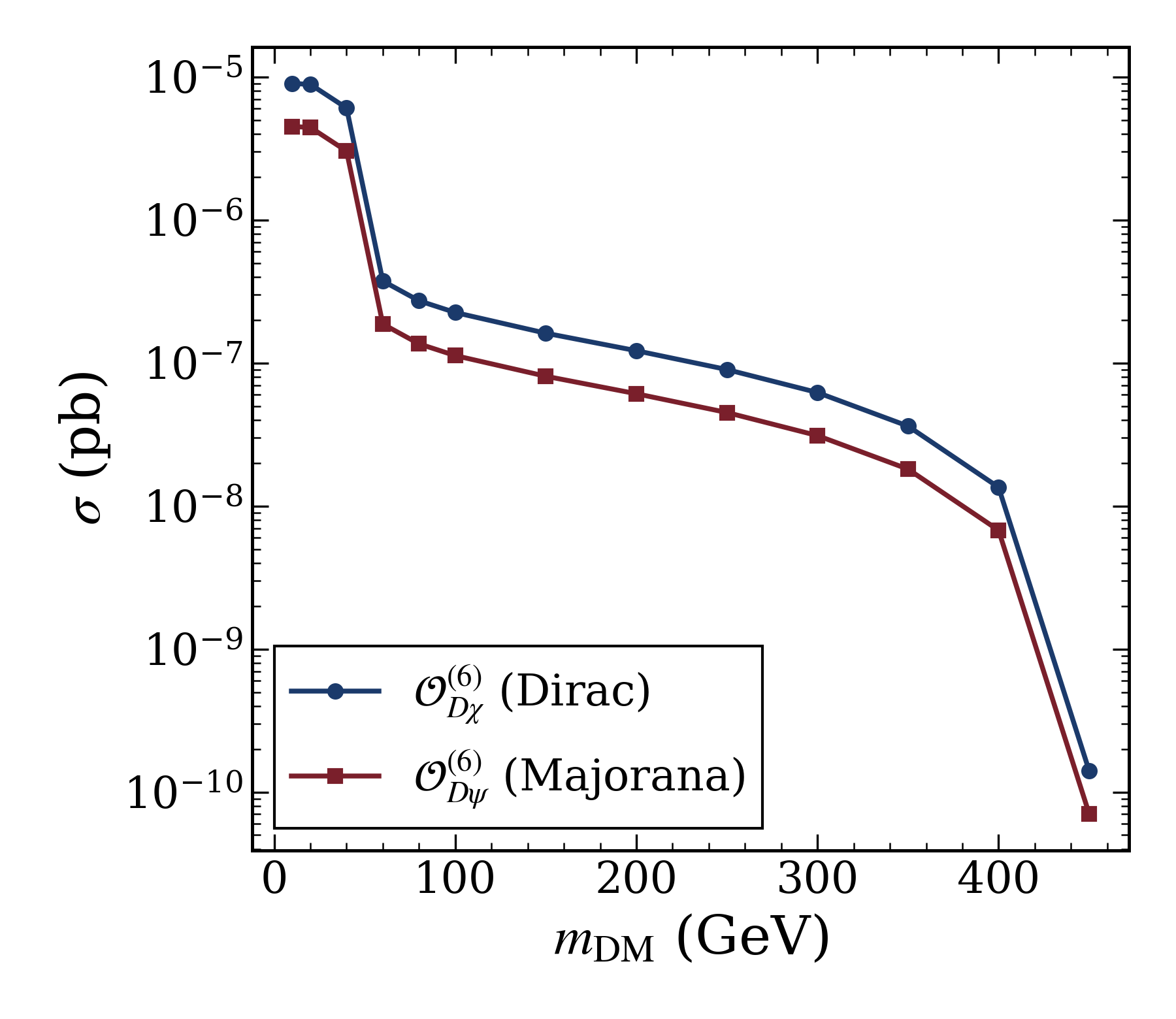}
        \caption{$\mathcal{O}_{D\psi}^{(6)}$ and $\mathcal{O}_{D\chi}^{(6)}$}
        \label{fig:Dpsi-dirac-vs-maj}
    \end{subfigure}
    \caption{Comparison between the production cross section for Dirac and Majorana DM of the processes $e^+e^- \to Z\, \text{DM\,DM}$, for the (a) $\mathcal{O}_{B\psi}^{(5)}$ and $\mathcal{O}_{B\chi}^{(5)}$;  (b) $\mathcal{O}_{D\psi}^{(6)}$ and $\mathcal{O}_{D\chi}^{(6)}$ operators, at a $1$~TeV $e^+e^-$ collider with $\Lambda=2$~TeV for the unpolarized beam configuration.}
    \label{fig:psi-dirac-vs-majorana}
\end{figure}

In figure~\ref{fig:psi-dirac-vs-majorana}, we provide the comparison between the production cross section for Dirac and Majorana DM of the processes $e^+e^- \to Z\, \text{DM\,DM}$, for the (a) $\mathcal{O}_{B\psi}^{(5)}$ and $\mathcal{O}_{B\chi}^{(5)}$ (figure~\ref{fig:Bpsi-dirac-vs-maj});  (b) $\mathcal{O}_{D\psi}^{(6)}$ and $\mathcal{O}_{D\chi}^{(6)}$ (figure~\ref{fig:Dpsi-dirac-vs-maj}) operators, at a $1$~TeV $e^+e^-$ collider with $\Lambda=2$~TeV for the unpolarized beam configuration.

\bibliographystyle{JHEP}
\bibliography{biblio.bib}
\end{document}